\documentclass[ejs]{imsart_arxiv}

\RequirePackage{amsthm,amsmath,amsfonts,amssymb}
\RequirePackage[authoryear]{natbib}
\RequirePackage[colorlinks,citecolor=blue,urlcolor=blue]{hyperref}
\RequirePackage{graphicx}
\usepackage{subcaption}
\usepackage{caption}
\usepackage{float}
\usepackage{times}
\usepackage{bm}
\allowdisplaybreaks
\usepackage[boxruled,scleft]{algorithm2e}
\usepackage{url}

\startlocaldefs
\theoremstyle{plain}

\newtheorem{theorem}{Theorem}[section]

\newtheorem{proposition}[theorem]{Proposition}
\theoremstyle{definition}

\theoremstyle{remark}

\endlocaldefs

\begin{document}
\begin{frontmatter}
\title{Testing for the Network Small-World Property}
\runtitle{Testing Small-World}

\begin{aug}
\author[A]{\fnms{Kartik}~\snm{Lovekar}\ead[label=e1]{lovekar.1@osu.edu}},
\author[B]{\fnms{Srijan}~\snm{Sengupta}\ead[label=e2]{ssengup2@ncsu.edu}}
\and
\author[A]{\fnms{Subhadeep}~\snm{Paul}\ead[label=e3]{paul.963@osu.edu}}
\address[A]{Department of Statistics,
The Ohio State University\printead[presep={,\ }]{e1,e3}}

\address[B]{Department of Statistics,
North Carolina State University\printead[presep={,\ }]{e2}}
\runauthor{Lovekar, Sengupta, Paul}
\end{aug}

\begin{abstract}
Researchers have long observed that the ``small-world" property, which combines the concepts of high transitivity or clustering with a low average path length, is ubiquitous for networks obtained from a variety of disciplines, including social sciences, biology, neuroscience, and ecology. However, we find several shortcomings of the currently prevalent definition and detection methods rendering the concept less powerful. First, the widely used \textit{small world coefficient} metric combines high transitivity with a low average path length in a single measure that confounds the two separate aspects. We find that the value of the metric is dominated by transitivity, and in several cases, networks get flagged as ``small world" solely because of their high transitivity. Second, the detection methods lack a formal statistical inference. Third, the comparison is typically performed against simplistic random graph models as the baseline, ignoring well-known network characteristics and risks confounding the small world property with other network properties. We decouple the properties of high transitivity and low average path length as separate events to test for. Then we define the property as a statistical test between a suitable null hypothesis and a superimposed alternative hypothesis. We propose a parametric bootstrap test with several null hypothesis models to allow a wide range of background structures in the network. 
In addition to the bootstrap tests, we also propose an asymptotic test under the Erd\"{o}s-Ren\'{y}i null model for which we provide theoretical guarantees on the asymptotic level and power. Our theoretical results include asymptotic distributions of clustering coefficient for various asymptotic growth rates on the probability of an edge. Applying the proposed methods to a large number of network datasets, we uncover new insights about their small-world property.
\end{abstract}


\begin{keyword}
\kwd{Network }
\kwd{Small-world property}
\kwd{Testing}
\kwd{Dependent data}
\end{keyword}

\end{frontmatter}

\section{Introduction}
The ``small-world property'' is one of the most widely observed properties of complex networks encountered in applications across a range of disciplines \citep{watts1998collective, amaral2000classes,humphries2008network,bassett2006small}. The idea of a ``small-world'' in networks was first conceived and experimentally validated in the context of social networks by \cite{milgram1967small} and was formulated in its currently used form in the seminal work by \cite{watts1998collective}. Roughly, the small-world property consists of ``segregation" of vertices into small tightly knit groups that leads to high local clustering in an otherwise sparse network, while at the same time the network has a small average path length that ``integrates'' the network.  Over the last two decades,  ``small-world'' networks have been observed in anatomical and functional brain networks \citep{bassett2006small,bullmore09,rubinov10,bassett2017small}, metabolic networks \citep{jeong2000large,wagner2001small}, protein-protein interaction networks \citep{jeong2001lethality}, gene co-expression network \citep{van2004yeast}, the internet \citep{albert2002statistical}, ecological networks and food webs \cite{montoya2002small,sole2001complexity}, scientific collaboration networks \cite{newman2001structure}, and air transportation networks \cite{guimera2005worldwide}. Some authors have also wondered if the small-world property is ``ubiquitous" \citep{telesford2011ubiquity} or even ``nearly-universal" for networks \cite{bassett2017small}.

The term ``small-world" is traditionally used to mean the average distance between pairs of vertices, $L$, is small (typically $O(\log n)$ or less as $n$ increases). Therefore, even for large networks, the average number of hops needed to reach one vertex from another is quite small. The authors in \cite{watts1998collective} showed that a large number of real-life networks have small $L$ (comparable to a random graph) but at the same time high global clustering coefficient $C$ (comparable to a ring lattice). In the modern use of the term small world, the term refers to both of these properties simultaneously holding (See definition and discussions in \cite{amaral2000classes,humphries2008network,humphries2005brainstem,telesford2011ubiquity,bassett2006small}). A popular tool for detecting the small-world property is through the \textit{small-world coefficient} defined as, $\sigma = \frac{\hat{C}/C_R}{\hat{L}/L_R}$, where $\hat{C}$ and $\hat{L}$ are observed global clustering coefficient and average path length respectively, while $C_R$ and $L_R$ are the expected values of the same quantities in a Erd\"{o}s-Ren\'{y}i random graph (ER) of equivalent density \citep{humphries2005brainstem,humphries2008network,bassett2008hierarchical,bullmore09,guye2010graph}.

Despite decades of empirical and methodological work on the property, several aspects of the measure and the methods popularly used to detect small-world property have been criticized in the literature  \citep{papo2016beware,bialonski2010brain,hlinka2017small,muldoon2016small}.
For example, it is rather surprising that little work exists on the quantification of uncertainty and statistical significance of the measure. The small-world coefficient in \cite{humphries2008network} can be thought of as measuring the ratio of $C/L$ between the observed network against what one would expect from an Erd\"{o}s-Ren\'{y}i baseline model. However, the methods lack the statistical theory of a formal test and usually do not come with a p-value to quantify the significance of the ratio. Therefore, it is not clear that what value of the coefficient should be considered high enough to call a network small world \citep{telesford2011ubiquity,hilgetag2016brain}. 
The coefficient further suffers from an issue of linear scaling, whereby larger networks are more likely to have a higher small-world coefficient than smaller networks purely due to their size \citep{humphries2008network,telesford2011ubiquity}. 
In addition, we find in our analysis in this paper that the small world coefficient measure is heavily influenced by the clustering coefficient. In most cases, using this measure leads to the same decision as \textit{just using the clustering coefficient}.

In an attempt to remedy the shortcomings of the coefficient, two alternative formulations of the small world coefficient have also been considered in the literature that compares $\hat{L}$ with $L$ expected from a random graph and $\hat{C}$ with $C$ obtained from (deterministic) ring lattice \citep{telesford2011ubiquity,muldoon2016small}. The paper \cite{muldoon2016small} defines a ``small-world propensity" as $1-\sqrt{\frac{\Delta_C^2 + \Delta_L^2}{2}}$. Here $\Delta_C$ is the ratio of the difference between $\hat{C}$ and $C$ of ring lattice and the difference between $C$ of ring lattice and expected $C$ of ER graph. Similarly, $\Delta_L$ represents the ratio of the difference between $\hat{L}$ and $L$ of ER model and the difference between $L$ of the ring lattice and expected $L$ of ER model. The paper \cite{telesford2011ubiquity} defines a metric $\omega = \frac{L_R}{\hat{L}} - \frac{\hat{C}}{C_{lattice}}$. Therefore, $\hat{L}$ is compared to expected $L$ from the ER model, and $\hat{C}$ is compared to the $C$ of ring lattice, and the difference between the ratios makes the metric. However, neither approach includes a test of statistical significance of observed values of the new small-world metrics. 

The use of ER random graph model and the regular ring lattice model as two ``extremes" can be justified by comparing their properties in terms of $L$ and $C$ values from Propositions \ref{prop1} and \ref{prop2}. Briefly, for ring lattice, $C$ is 3/4 (deterministic). For the ER random graph model with $n$ vertices and probability of an edge $p=\frac{\delta}{n}$, expected $C$ grows asymptotically as $p$, i.e., the average density of the graph. For sparse graphs, therefore, expected $C$ goes to 0. Hence, in terms of $C$, typically, the ring lattice has high values of $C$ while the ER random graph has a low value of $C$. On the other hand, for $L$, the ring lattice on $n$ vertices with $\delta$ degrees has a deterministic $L = \frac{n}{2\delta}$, which is quite high in sparse graphs since $\delta$ typically grows much slower than $n$. For the ER model, the expected $L$ is $\frac{\log n}{\log (\delta)}$, which only grows as $\log n$ and hence is small. Therefore, in terms of $L$, ER random graph is characterized by small $L$, while the regular ring lattice has a high $L$. 

Both the classical procedure and the modifications proposed in the literature ignore the presence of \textit{community structure and presence of hub structure due to degree heterogeneity}, both of which are capable of generating networks with substantial small-world properties. It has been shown that highly modular networks are segregated and tends to produce a high clustering coefficient \citep{pan2009modularity,meunier2010modular}. On the other hand, networks with high-degree hub nodes facilitate communication between modules leading to a short average path length. Yet, the small-world property is not just the presence of a modular organization. Highly modular networks usually do not have small path lengths \cite{gallos2012small}. Therefore, the popular small-world coefficient metric might incorrectly determine them as small-world due to the high influence of the clustering coefficient on the metric. Hence it seems relevant for scientific discovery, link prediction, and estimating effects of treatment interventions to understand if the small-world property is a manifestation beyond the simultaneous presence of community and hub structure. For example, if a network is small-world, then a network model that fits the small-world property better might be preferred for more accurate link prediction due to its ability to better predict transitivity. Similarly, a network that exhibits significant small-world property will likely have a higher network interference or spillover of treatments due to units being relatively well connected due to small path length.  After accounting for the additional variations, perhaps the near universality will give away to the specialty of small-world networks, making the small-world property more useful. Finally, in studies of small-world property, the alternative model is rarely fully specified, and consequently, the statistical powers of small-world detection procedures are unknown.  

In view of these serious limitations of the classical definition of small-world coefficient and the procedure of estimating it, we propose a different strategy. The key components of our approach are as follows.
\begin{enumerate}
    \item We replace the quantitative measure small world coefficient with an \textit{intersection criteria} as a mechanism for determining if a given network is a small world network. The intersection criteria decouples high clustering coefficient and low average path length into two separate criteria that can be tested separately. Accordingly we implement the \textit{intersection test} as a combination of two tests whose simultaneous rejection determines if an observed network is \textit{significantly} more small world than a posed null model. We will show that this approach avoids the undesirable weight of the clustering coefficient in determining small-worldness using the small world coefficient and is, in spirit, closer to the small world property defined in the works of \cite{watts1998collective} and \cite{newman1999scaling}.

    \item The proposed intersection test involves testing if the expected $C$ and $L$ for the population from which the given graph is drawn is respectively greater and not appreciably smaller than expected from a reference (null) population. We propose to use a number of different random graph models as \textit{null models} and define a class of \textit{alternative superimposed} small world models. The expansion of null models include the Chung-Lu random graph model, the stochastic block model and the degree corrected stochastic block model. Together these models explain a variety of observed network characteristics including degree heterogeneity, modular organization, and hub structure. 

    \item We develop an \textit{asymptotic test} with Erd\"{o}s-Ren\'{y}i random graph being the null model and theoretically study the asymptotic level and power of the resulting test. This is the first detection method for small world property that \textit{does not require comparison with simulated networks} from a benchmark model and hence is computationally efficient for large networks. Our asymptotic framework considers a wide range of rates at which the probability of an edge $p$ converges to 0 as number of vertices $n$ increases.
    As a byproduct of our analysis we also characterize the asymptotic distribution of  clustering coefficient in Erd\"{o}s-Ren\'{y}i random graph for various growth rates on $p$.
   
   \end{enumerate} 
    
    A key shortcoming of the existing small-world toolbox is that networks are compared with the simplistic Erd\"{o}s-Ren\'{y}i model.
    It is well-known that, in contrast to the Erd\"{o}s-Ren\'{y}i model, real-world networks exhibit a range of properties such as degree heterogeneity, preferential attachment, and community structure \citep{aiello2000random,vazquez2003growing,bickel2009nonparametric,fortunato2010community,sengupta2018block,girvan2002community}.
    Crucially, some of these properties can be confounded with the small-world property when compared to the Erd\"{o}s-Ren\'{y}i model.
    For example, a network with community structure  is likely to exhibit significantly higher clustering than Erd\"{o}s-Ren\'{y}i graphs. We seek to avoid this confounding of network properties by testing networks against more general null models which account for such properties.
    
    In terms of statistical principle, this approach is similar to any other statistical hypothesis test. We want to test the population quantities $E[C]$ is larger than a certain user-provided value (say $C_0$) and $E[L]$ is smaller than a certain user-provided value (say $L_0$). However, it is not obvious what those user-provided values should be to declare a network as a small world. Hence we resort to the null models to obtain those user-provided values.  The varied choices of the null models provide meaning to the small-world property as a phenomenon beyond what could be explained by other well-known graph properties, including degree-heterogeneity and community structure.

    The above formulation is somewhat  changing the meaning normally attached to a network being ``small-world". In particular, we are interpreting the property not in isolation, but in relation to other salient properties of networks. Indeed, our intention is to view small-world property as something beyond degree heterogeneity and community structure. It is certainly possible (and expected, as we show through simulation in Figure \ref{CLdist}) that community structure increases the clustering coefficient and does not increase the average path length - the core definition of small-world property. Our tests will determine a network to be small-world in comparison to what would be predicted from certain reasonable models of networks that otherwise account for well-known network properties. Such relative characterization is relevant in applications to predict the behavior of the network under interventions or to predict connections with a new vertex. Elsewhere in the network science literature, such an approach of testing against various null hypothesis models is taken when testing for the presence of community structure \citep{bickel2016hypothesis,gao2017testing}.
    
When the null hypothesis model is the Erd\"{o}s-Ren\'{y}i model, we develop an asymptotic test, as discussed earlier. For the null hypothesis models more general than Erd\"{o}s-Ren\'{y}i model, we further develop a \textit{bootstrap test}. Our bootstrap detection method involves computing a p-value for the statistical significance of the observed deviation of $C$ and $L$ from a suitable random graph model denoting the null hypothesis. For four different null network models described previously, we derive procedures to compute p-values of the test statistic using parametric bootstrap. 

The methods have been implemented in the software R and the package is available to download freely from the Github repository  \url{https://github.com/KartikSL/SWTest}. The code to reproduce simulations and results on real-world data is available on \url{https://github.com/KartikSL/SWsims}.

\section{The null and the alternative hypotheses and superimposed Newman-Watts type models}
To formalize the notion of small-worldness in terms of a statistical hypothesis, we consider the Newman-Watts (NW) small-world model \citep{newman1999scaling}. The Newman-Watts model is a modification of the original Watts-Strogatz model of \cite{watts1998collective}, and is also based on an interpolation between (or mixture of) a regular ring lattice and an Erd\"{o}s-Ren\'{y}i (ER) random graph. The Newman-Watts model fixes problems in the Watts-Strogatz model related to having a finite probability of the lattice becoming detached and non-uniformity of the distribution of the shortcuts, and makes the model suitable for an analytic treatment \citep{newman1999scaling}. 
The particular variation of the model that we consider in this paper is parameterized by three quantities --- the number of vertices $n$, the expected degree $2\delta$, and the mixing proportion $\beta \in [0,1]$.
A network from this model is generated as follows:
\begin{enumerate}
    \item Construct a $ \left \lceil{2\delta\beta}\right \rceil  $-regular ring lattice of $n$ nodes, where $\lceil \cdot \rceil$ denotes the ceiling function.
    To do this, first construct a cycle of $n$ nodes, which is a 2-regular ring lattice.
    Then, connect each node to its neighbors that are two hops away, thereby forming a 4-regular ring lattice, and continue until $\left \lceil{\delta\beta}\right \rceil$ hops.
    \item Next, for each of the ${n \choose 2}$ node pairs, randomly add an edge connecting them with probability $p=\frac{2\delta-\left \lfloor{2\delta\beta}\right \rfloor}{n-1}$, where $\lfloor \cdot \rfloor$ denotes the floor function.
\end{enumerate}
We assume $\delta \rightarrow \infty$, as $n\rightarrow \infty$ and $\beta$ is a constant not dependent on $n$. 
Clearly for $\beta=1$, the random graph portion of the mixture is an empty graph and this model yields a $ \left \lceil{2\delta}\right \rceil  $ regular ring lattice with high global clustering coefficient and high average path length.
On the other hand, for $\beta=0$, the ring lattice portion of the mixture is an empty graph, and this model yields a pure Erd\"{o}s-Ren\'{y}i random graph with $n$ nodes and $p=\frac{2\delta}{n-1}$.
For $0 < \beta < 1$, this model yields small-world networks with high global clustering coefficient and low average path length.

Under this model, we define the detection of small world property as the test of a statistical hypothesis.
The \textit{null hypothesis} asserts that expected $C$ and $L$ of the graph is same as expected $C$ and $L$ for ER random graph with parameters $(n,\frac{2 \delta}{n-1})$ or pure ring lattice graph with parameters $(n,2\delta)$. 
That is, the null hypothesis is given by $H_0: \beta \in \{0,1\}$. The \textit{alternative hypothesis} states that expected $C$ is higher than expected under the null hypothesis and the expected $L$ is similar to that expected under the null hypothesis, i.e., $H_1: 0<\beta <1$. In Section 3.2 we show that the $E[C]$ and $E[L]$ under the NW-ER $(n,2\delta,\beta)$ model with mixing parameter $0<\beta<1$ have the required properties of the alternative hypothesis. Therefore NW-ER $(n,2\delta,\beta)$ model with mixing parameter $0<\beta<1$ can be thought of as an alternative hypothesis model.

The model can be viewed as a mixture or superimposed model similar to the superimposed stochastic block model proposed in \citep{paul2023higher}. We note that the model will produce some multi-edges, however, the number of such multi-edges is small compared to total number of edges. Also, when considering higher order structures, e.g., connected triples or triangles, which are required to define small-world property, the edges from the two components of the mixture will interact with each other to produce certain ``incidental" higher order structures in addition to model component generated structures \citep{paul2023higher}.

\subsection{Extension to SBM, CL and DCSBM null models}
The above superimposed alternative model framework can be extended to include other null models we might be interested in. We consider three such models, namely the Degree Corrected Stochastic Block Model (DCSBM), the Stochastic Blockmodel (SBM), and the Chung-Lu (CL) model, all part of the larger family of inhomogeneous random graph models.

The SBM exhibits community structure, the CL model exhibits degree heterogeneity, and the DCSBM exhibits both degree heterogeneity and community structure.
Therefore, by using the DCSBM as the null model, we can test whether a network has small-world property after accounting for both properties.
Similarly, by using the SBM as the null model, we can test for small-world property after accounting for community structure only, and by using the CL model as the null model, we can test for small-world property after accounting for degree heterogeneity only.

Let us consider the case of Degree Corrected Stochastic Block Model (DCSBM) since the other two are special cases of this model. The alternative model can be described as follows:

\begin{enumerate}
\item Fix, the following quantities. (a) The number of communities $k$ and $n$ dimensional community assignment vector $z$. (b) The $n$ dimensional vector of degree parameters $\theta$, such that $\sum_{i:z_i =q}\theta_i =1$, (c) a $k \times k$ matrix of parameters $B$ such that $\sum_{q,l}B_{ql} = 1.$ (d) A constant $0 \leq \beta \leq 1$.
    \item Construct a $ \left \lceil{2\delta\beta}\right \rceil$-regular ring lattice of $n$ nodes as before.
    \item Next, for each of the ${n \choose 2}$ node pairs, add an edge connecting them according to the outcome of the Bernoulli trial:  
    $A_{ij} \sim Bernoulli(n(2\delta-\left \lfloor{2\delta\beta}\right \rfloor)\theta_i \theta_j B_{z_iz_j}).$
\end{enumerate}
Note $\theta,z,B$ are not dependent on $\beta$. Essentially, for this alternative model, as we decrease the value of $\beta$ away from 1, three key properties of the DCSBM null model, namely, average density, degree heterogeneity and community memberships are approximately being preserved. The only quantity that changes with $\beta$ is how much information is available regarding the DCSBM portion of the model.
For $\beta=1$, the DCSBM portion of the mixture yields an empty graph and this model yields a $2 \delta$ regular ring lattice, while for $\beta=0$, the ring lattice is an empty graph, and this model yields a graph from DCSBM model.
 
The \textit{hypothesis} that we will test is as before. The null hypothesis now asserts that the expected $C$ and $L$ of the graph is same as expected $C$ and $L$ for DCSBM random graph with parameters $(n,2 \delta, \theta)$ or pure ring lattice graph with parameters $(n,2\delta)$. The alternative hypothesis states that expected $C$ is higher than expected under the null and the expected $L$ is similar to that expected under the null, or, equivalently as those expected from a NW-DCBM $(n,2\delta,\theta, \beta)$ model with mixing parameter $0<\beta<1$.\\

\textit{\textbf{Remark: mixed membership models.}} We want to emphasize that the above superimposed framework is general and can accommodate many other random graph models as well. These include the mixed-membership stochastic block model (MMSBM) and its degree-corrected version (DCMMSBM), as well as the random dot product graph (RDPG) model. While we do not formally include these models in our simulation and real data results, the framework can be similarly extended to these models. The mixed membership models can be thought of as a flexible way of modeling the community structure (by allowing a vertex to be simultaneously member of multiple communities).

\section{The intersection criteria and testing procedure}
Define $E_{ij},V_{ijk},T_{ijk}$ as the indicator variables denoting an edge between node pair $i,j$, the number of open triples or $V$ structures (2 of the 3 possible edges exist and the third one does not exist, also called 2-armed stars) among the node triple $i,j,k$, and the number of closed triples or triangle structures among the node triple $i,j,k$ respectively. Then,
\[
V_{ijk} = E_{ij} E_{ik}(1-E_{jk}) + E_{ik} E_{jk}(1-E_{ij}) + E_{ij} E_{jk}(1-E_{ik}),
\]
and 
\[
T_{ijk} = E_{ij} E_{ik} E_{kj}.
\]
Let $S_{ijk}=3 T_{i,j,k} + V_{i,j,k}$. Further, define
\begin{align*}
    E & =\sum_{1 \leq i <j \leq n} E_{ij}/\dbinom{n}{2}, \\
    V& =  \sum_{1 \leq i <j <k \leq n} V_{ijk}/\dbinom{n}{3},\\ 
    T& =  \sum_{1 \leq i <j <k \leq n} T_{ijk}/\dbinom{n}{3},
\end{align*}
 and 
 \[
 S= 3T+V = \sum_{1 \leq i <j <k \leq n} (E_{ij} E_{ik} + E_{ik} E_{jk} + E_{ij} E_{jk})/ \dbinom{n}{3}.
 \]
 With these notations, we define $C=\frac{3T}{3T+V} = \frac{3T}{S}$ as the \textit{clustering coefficient or transitivity} of the graph. 
 
 Further, consider any pair of nodes $(i,j)$, and denote $L_{ij}$ as the length of the shortest path length between them. Then $$L=
 \frac{1}{\dbinom{n}{2}}\sum_{1\leq i<j \leq n}L_{ij}$$ is the \textit{average shortest path length} of the graph.

We propose a multiple testing procedure with two test statistics $[C,L]$, which we call the \textit{intersection test}. In particular we reject the null hypothesis if, 
\begin{equation}
      \{C> K_1 \} \cap  \{L < K_2 \},
      \label{test_statistic}    
\end{equation}

for suitable choices of $K_1$ and $K_2$.
The cutoff $K_1$ is chosen to be a high quantile of the null hypothesis sampling distribution, while $K_2$ is set to a value that is high in comparison to the null hypothesis sampling distribution. 
      
      To provide an intuition for the test consider the case when the null model is the ER model. Therefore the quantities $K_1$ and $K_2$ are chosen according to the distribution of $C$ and $L$ under the ER model. When the observed network is actually generated from the ER model, the first event is low-probability, and the second event is high-probability, which means the intersection event is low-probability. Therefore we fail to reject. On the other hand, when the observed network is generated from the Newman-Watts model with mixing parameter $\beta \neq \{0,1\}$, the first event is high-probability (since clustering is greater), the second event is high-probability (distances are still small), which means the intersection event is high-probability and therefore we reject. Finally, when the observed network is generated from a pure lattice, the first event is high-probability (since clustering is greater), the second event is low-probability (distances are much higher than ER), which means the intersection event is low-probability and therefore we fail to reject. Thus the intersection test ensures we have the correct decision under all three scenarios (Table \ref{tab:intuition}).

      \begin{table}[h]
	\centering
      \resizebox{0.9\textwidth}{!}{
	\begin{tabular}{|r|r|r|r|r|}
		\hline
		Network Model &  SW & $C> K_1$   & $L < K_2$   & Intersection Test\\ 
		\hline
		\hline	
		Pure Erd\"{o}s-Ren\'{y}i ($\beta=0$) & No & No & Yes & Not rejected\\ \hline
		Newman-Watts ($0 < \beta < 1$) & Yes & Yes & Yes & Rejected\\ \hline
		Pure Ring Lattice ($\beta=1$) & No & Yes & No & Not rejected\\
				\hline
		\hline
	\end{tabular}	
	}	
	\caption{Heuristics of the Intersection Test} 
	\label{tab:intuition}
\end{table}

\subsection{Relating the test, the small-world property, and the superimposed small-world models as alternative hypotheses}

An important question is how the small-world \textit{property} relates to the \textit{test}, the \textit{models}, and the null \textit{hypotheses} described above. First, the test closely mimics the definition of the property since we call a network small-world if observed $C$ is \textit{higher than} some reference value $K_1$ and observed $L$ is \textit{not greater} than another reference value $K_2$. The various null models are used to determine these reference values and serve as null hypotheses. However, it is not immediately clear how the alternative hypothesis is  related to the superimposed small-world models posed or how the test statistic can distinguish between the null and the alternative hypotheses. We provide some results and discussions below. First, we show that the population quantity $E[C]$ is an order of magnitude higher for any $0 < \beta <1$ compared to at $\beta=0$ for all models under mild conditions. More specifically, the ratio of $E[C]$ between $0 < \beta <1$ to $\beta=0$ goes to $\infty$ as $n \to \infty$.

We first consider the superimposition of the ER and the ring lattice model as described earlier (NW-ER model). We assume $\delta \to \infty$ while $\frac{\delta}{n} \to 0$ as $n \to \infty $, and $\beta$ is a constant not dependent on $n$. We first determine the asymptotic limit of $E[3T]$ and $E[S]$ as $n \to \infty$ under the model. The calculations are similar to \cite{barrat2000properties}, however, the result we present is under the Newman-Watts superimposed model while the result in \cite{barrat2000properties} was under the Watts-Strogatz model and hence the result is slightly different. For two functions $f(n)$ and $g(n)$, we use the notation $f(n) \asymp g(n)$ to mean that the functions are asymptotically equivalent, i.e., there exists constants $c_1, c_2$ and a number $n_0$ such that $c_1 g(n) \leq f(n) \leq c_2 g(n) $ for all $n>n_0$. In the same vein we will use the notation $f(n) \lesssim g(n)$ to mean that there exists a constant $c_3$ and a number $n_1$ such that $f(n) \leq c_3 g(n)$ for all $n>n_1$. We will further use the notations $f(n) >> g(n)$ to mean $\lim_{n\to \infty} \frac{f(n)}{g(n)} =\infty $, and $f(n) << g(n)$ to mean $\lim_{n\to \infty} \frac{f(n)}{g(n)} =0 $.

\begin{proposition}
    Consider the superimposed NW-ER model with mixing proportion $\beta \in [0,1]$. Assume $\delta \to \infty$ while $\frac{\delta}{n} \to 0$ as $n \to \infty $. Then as $n \to \infty$ we have the following results: 
\begin{enumerate}
\item For $\beta=0$, $E[C] \asymp \frac{2\delta}{n-1} \to 0$.
\item For $\beta=1$, $C=\frac{3}{4}$ (deterministic). 
    \item  For $0<\beta<1$, $
\dbinom{n}{3}E[T] \asymp \frac{n \delta^2\beta^2}{2} + \frac{8 \delta^3 (1-\beta)^3}{6} + 4 \delta^3\beta^2(1-\beta) + 8 \delta^3 \beta(1-\beta)^2,$ \hspace{10pt} $ \dbinom{n}{3}E[S] \asymp  2n\delta^2,$ and consequently, 
$
E \left[C\right]  \to \frac{3}{4}\beta^2.
$
\end{enumerate}

\label{prop1}
\end{proposition}
 From this proposition we can see the ratio of $E[C]$ for any $\beta \neq 0$ to that for $\beta=0$ converges to $\infty$. Note while this conclusion holds for any constant $\beta$, it also holds when $\beta$ is dependent on $n$ as long as $\beta >> \sqrt{\frac{\delta}{n}}$.  The proof of this proposition along with the proofs of other propositions, lemmas, and theorems are given in the Appendix. 

Next, we generalize the above calculations to inhomogeneous random graphs to show that the expected global clustering coefficient $C$ is asymptotically an order higher for the superimposed models than the corresponding null models  SBM, CL, and DCSBM under certain conditions. Consider the inhomogeneous random graph model that independently generates an edge between a pair of nodes $(i,j)$ with probability $p_{ij}$. The CL, SBM, and DCSBM are special cases of this model. Let $p_{\max}=\max_{i,j} p_{ij}$ and $p_{\min}=\min_{i,j} p_{ij}$ be the maximum and minimum probability of an edge. Let $\delta_{\max} = \frac{(n-1)}{2}p_{\max}$ and $\delta_{\min} = \frac{(n-1)}{2}p_{\min}$, and $\bar{p} =  \sum_{1 \leq i<j \leq n} p_{ij}/\dbinom{n}{2}$. Therefore note that $\bar{p} = \frac{2\delta}{n-1}.$

\begin{proposition}
    Consider a NW superimposed inhomogeneous random graph on $n$ nodes, mixing proportion $\beta \in [0,1]$ and overall expected degree $2\delta$. 
 
    \begin{enumerate}
        \item At $\beta=0$, when the graph is a purely inhomogeneous random graph, $E[3T] \leq 3\bar{p}p_{\max}^2 $, and $E[S] \geq 3\bar{p}p_{\min}$, and consequently $E[C] \lesssim \left(\frac{\delta_{\max}}{\delta_{\min}}\right)\frac{2\delta_{\max}}{n}$ as $n \to \infty$.
        \item At $\beta=1$, $C=\frac{3}{4}$.
\item  For $0<\beta<1$, we have, $\dbinom{n}{3}E[3T] \gtrsim \frac{3n \delta \delta_{\min} \beta^2}{2}$ and $\dbinom{n}{3}E[S] \lesssim 2n \delta \delta_{\max} $ and consequently, $E[C] \gtrsim \frac{3}{4}\left(\frac{\delta_{\min}}{\delta_{\max}}\right)\beta^2$ as $n \to \infty$.
    \end{enumerate}
    \label{prop2}
\end{proposition}
We note that the ratio of $E[C]$ for $0<\beta<1$ to $E[C]$ for $\beta=0$ goes to $\infty$ for all constant $\beta$, as long as $n\frac{\delta_{\min}^2}{\delta_{\max}^3} \to \infty$. Therefore we make an assumption that $n\frac{\delta_{\min}^2}{\delta_{\max}^3} \to \infty$, as $n \to \infty$. This assumption holds true for a wide range of parameter values for sparse graphs in SBM, DCSBM and CL, and is only likely to be violated under either severe degree heterogeneity in DCSBM and CL models, or very tightly bound community structure in SBM and DCSBM models.  For example, this condition holds if $\delta_{\min} \asymp \delta_{\max} \asymp \delta$ and if $\frac{\delta}{n} \to 0$ at any rate.  However, it is possible to have a severe degree heterogeneity in DCSBM graphs. For example, \cite{jin2021improvements} considered the case when $\frac{\delta_{\min}}{\delta_{\max}} \to 0$ in DCSBM graphs in the context of community detection. In that case, the above condition will require $\frac{\delta_{\min}}{\delta_{\max}} >> \sqrt{\frac{\delta_{\max}}{n}}$. As \cite{jin2021improvements} discusses in the most interesting regimes, $\delta_{\max}$ is asymptotically between $\log n$ and $\sqrt{n}$. Therefore the requirement is satisfied even in some cases of severe degree heterogeneity. Therefore, the expected global clustering coefficient $C$ is an order of magnitude higher under the alternative hypothesis than under the null hypothesis for all the null models. While this proposition is an asymptotic result, we present a simulation in the Appendix B to study the finite sample behavior of the observed mean of $C$ in comparison to these bounds.

On the other hand, for average path length (APL), we make the following argument. Let $A$ be a graph generated from the NW-ER $(n,2\delta,\beta)$ model and $A_E$ be the random graph component,  obtained by removing the ring lattice edges from the graph. Clearly, the expected APL of $A$ is smaller than the expected APL of $A_E$ since the additional ring lattice edges can only decrease the path lengths and never increase them. However, note that $A_E$ is generated from the ER model with parameters $\left(n, \frac{2\delta(1-\beta)}{n-1}\right)$. The expected APL of $A_E$ is then given by $O(\frac{\log n}{\log 2\delta(1-\beta)})$. This does not change appreciably from the APL for $\beta=0$ until $\beta$ becomes very close to 1. On the other hand, for $\beta=1$, the value of $L$ is the same as that of the ring lattice, which is deterministically $\frac{n}{4\delta}$. Therefore $L$ for $\beta=1$ is much higher than for other $\beta$ values. These calculations are formalized in Theorems 1 and 2 with probability concentration inequalities to obtain asymptotic level and power of the asymptotic test. 

However, the asymptotic expectation calculations do not give us an idea of the changes in the distribution of $C$ and $L$ as $\beta$ moves away from 0. Therefore as a second piece of evidence, we present a simulation study in Figure \ref{CLdist} in Section 4.3 (Simulation) that shows how the distribution of $C$ and $L$ changes with increasing $\beta$ for different null models.

\subsection{Asymptotic test with ER null model}
We propose an asymptotic test with the test statistic in Equation \ref{test_statistic} and ER as null model. 
We estimate the parameter $p=\frac{2\delta}{n-1}$ in the ER model with the observed network density, $\hat{p} = \frac{2\sum_{i<j}A_{ij}}{n(n-1)}$.
We first define the following two ``population centered" sub-graph statistics 
\begin{align*}
    R_2 & = \frac{1}{\dbinom{n}{3} }\sum_{1 \leq i <j <k \leq n} [(E_{ij}-p)(E_{ik}-p) + (E_{ik}-p) (E_{jk}-p) + (E_{ij}-p) (E_{jk}-p)] \\
    R_3 & = \sum_{1 \leq i <j <k \leq n} (E_{ij}-p)( E_{ik}-p)( E_{kj}-p) / \dbinom{n}{3} 
\end{align*}

Assume $np \to \infty$ and $(1-p)^{-1} = O(1)$. From Theorem 6.1 of \cite{gao2017testing} we have
\[
\sqrt{\dbinom{n}{3}} \begin{pmatrix}
R_2/\sqrt{3p^2(1-p)^2 } \\
R_3/\sqrt{p^3(1-p)^3}
\end{pmatrix} \,
\overset{D}{\rightarrow}
\, \mathcal{MVN}_2 \left(\begin{pmatrix}
    0 \\
    0
\end{pmatrix},
\begin{pmatrix}
    1 & 0 \\
    0 & 1
\end{pmatrix}\right).
\]

We will compute the representations of $T-E[T]$ and $S-E[S]$  in terms of $R_2$ and $R_3$. First, note that $E[T]=p^3$, and $E[S]=3p^2$. Then we have,
\begin{align*}
    R_2 & = \sum_{1 \leq i <j <k \leq n}
    \{(E_{ij}E_{ik} + E_{ik}E_{jk} + E_{ij}E_{jk}) - 2p (E_{ij} + E_{ik} + E_{jk}) + 3p^2\}/ \dbinom{n}{3}\\
    & = S - 6p\hat{p} + 3p^2.
\end{align*}
Rearranging we have
\begin{equation}
    S-3p^2 = R_2 + 6p(\hat{p}-p).
\end{equation}
Moreover, we have
\begin{align*}
    R_3 & = \frac{1}{\dbinom{n}{3}}\sum_{1 \leq i <j <k \leq n}
     \{(E_{ij}E_{ik} E_{jk} - p(E_{ij}E_{ik} + E_{ik}E_{jk} + E_{ij}E_{jk})) \\
    &  \quad \quad + p^2 (E_{ij} + E_{ik} + E_{jk}) - p^3\} \\
    & = T -pR_2 - 3p^2 \hat{p} + 2p^3.
\end{align*}
Then 
\begin{equation}
T-p^3 = R_3 + pR_2 + 3p^2 (\hat{p}- p).
\end{equation}
Note, we always assume $p>>1/n$. However, we note that for different regimes of growth rates on $p$, different terms in $T-p^3$ dominates. For this purpose, we define the two stochastic-o notations as follows. The notation $X_n = o_p(a_n)$, or equivalently $X_n/a_n = o_p(1)$, is used to mean that the sequence $X_n/a_n \to 0$ in probability as $n\to \infty$. The notation $X_n = O_p(a_n)$ is used to mean that the sequence $X_n/a_n$ is bounded by a finite $M$ with high probability as $n \to \infty$.  In particular, from the result  on the joint asymptotic distribution of $R_2$ and $R_3$ stated above, we note
\[
R_3 = O_p\left(\frac{p^{1.5}}{n^{1.5}}\right), \quad pR_2 = O_p\left(\frac{p^{2}}{n^{1.5}}\right),
\]
and from the central limit theorem,
\[
3p^2 (\hat{p}-p) = O_p\left(\frac{p^{2.5}}{n}\right).
\]
When $1/n <<p<<1/\sqrt{n}$, then $R_3 + pR_2$ dominates, while when $p >> 1/\sqrt{n}$, the other term $3p^2 (\hat{p}- p)$ dominates.

On the other hand for $S-3p^2$, we note that 
\[
R_2 = O_p\left(\frac{p}{n^{1.5}}\right), \quad 6p(\hat{p}-p) = O_p\left(\frac{p^{1.5}}{n}\right),
\]
and consequently, the term $6p(\hat{p}-p)$ always dominates for any $p >>1/n$.
With these, the following theorem characterizes the asymptotic normality of the clustering coefficient for various regimes of growth rates on $p$.

\begin{theorem}
Assume $p>>\frac{1}{n}$ and $(1-p)^{-1}=O(1)$. Under the hypothesis of $\beta=0$ in NW-ER model (i.e., under pure ER model), we have the following asymptotic distribution for $C$.
\begin{enumerate}
    \item For $\frac{1}{n} <<p<< \frac{1}{\sqrt{n}}$, we have
    \[
\sqrt{\frac{n^3p}{6(1+2p)(1-p)^2}}(C -p) \overset{D}{\to} N (0,1).
\]
\item For $\frac{1}{\sqrt{n}} <<p $ and for constant $p$,
\begin{align*}
& \sqrt{\frac{n(n-1)}{2p(1-p)}}\left( C - p\right)  \overset{D}{\to} N \left(0,1\right).
\end{align*}
\end{enumerate}
\label{C-dist}
\end{theorem}

The proof of this and the following theorems can be found in the Appendix A. Therefore, in the regime, $\frac{1}{n} <<p<< \frac{1}{\sqrt{n}}$, the clustering coefficient $C$ is $O_p(\frac{1}{\sqrt{n^3p}}),$ which is between $O_p(\frac{1}{\sqrt{n^{(2+\epsilon)}}})$ and $O_p(\frac{1}{\sqrt{n^{(2.5-\epsilon)}}})$, for some small $\epsilon>0$, depending upon the growth rate on $p$. On the other hand, for $p>>\frac{1}{\sqrt{n}} $, the clustering coefficient is  $O_p(\frac{\sqrt{p}}{n}),$ which is between $O_p(\frac{1}{\sqrt{n^{(2.5-\epsilon)}}})$ and $O_p(\frac{1}{n})$ depending upon the growth rate of $p$.

For fixed p, the following covariance matrix for the joint asymptotic distribution of $T,S$ that also contains terms of $O(\frac{1}{n})$ may provide better approximation \cite{reinert2010random}:
\begin{equation}
\Sigma_{T,S}  = \begin{pmatrix}
1 + \frac{1+p-2p^2}{3p^2(n-2)}
 & 1 + \frac{1-p}{2p(n-2)} \\
1 + \frac{1-p}{2p(n-2)} & 
1 + \frac{1-p}{4p(n-2)}
\end{pmatrix}.
\label{fullSigma}
\end{equation}
Let $\Sigma_{ij}$ denote the $(i,j)$th element of the $\Sigma_{T,S}$ matrix. Define,
\[
\Sigma_C= 9\Sigma_{11} - 12 \Sigma_{12} + 4 \Sigma_{22}.
\]
Then
\begin{align*}
& \sqrt{\frac{n(n-1)}{2p(1-p)\Sigma_C}}\left( C - p\right)  \overset{D}{\to} N \left(0,1\right).
\end{align*}

We find in our empirical simulations that in finite samples, the above scaling on $C$ acts almost as an interpolation between the two scaling on $C$ that are valid for different regimes of growth on $p$, and provides a better approximation for most situations. 

We propose to use the following rejection cutoffs for the test with level at most $\alpha$.
\begin{enumerate}
\item For a pre-specified desired asymptotic level $\alpha$, the cutoff $K_{1,\alpha}(\hat{p})$ is given as
\begin{equation}
K_{1,\alpha}(\hat{p}) = \begin{cases}
     \left(\hat{p} + Z_{\alpha} \sqrt{\frac{2\hat{p}(1-\hat{p})}{n(n-1)}}\right) & p>>\frac{1}{\sqrt{n}},\\
\left(\hat{p} + Z_{\alpha} \sqrt{\frac{6(1+2\hat{p})(1-\hat{p}^2)}{n^3\hat{p}}}\right)& \frac{1}{n} <<p<< \frac{1}{\sqrt{n}} .    
\end{cases}
\label{K_alpha}
\end{equation}
where $Z_{\alpha}$ is the $100*(1-\alpha)$th upper quantile of the standard normal distribution (a constant as a function of $n$).
\item $K_2$ is $ \frac{(2+\epsilon) \log(n)}{\log(n\hat{p})}$ for any $\epsilon>0$. We use $\epsilon =0.0001$ in our simulations and real data analysis.
\end{enumerate}

The next two results characterize the asymptotic level and power of the test on $C$ and $L$ whose proofs are based on concentration inequalities. We use the notation $NW(n,p,\beta)$ to mean the NW-ER model with $n$ nodes, $\beta$ as mixing proportion, and degree $2\delta =(n-1)p$.
\begin{theorem}
Assume $ \frac{n p}{\log n} \to \infty$, i.e., $p >> \frac{\log n}{n}$, and $p C_n \to 0$, where $C_n$ is a sequence such that $C_n \to \infty$ at any rate as $n \to \infty$. Let $K_{1,\alpha}(\hat{p})$ be the cutoff as defined in Equation \ref{K_alpha}.  Then
\begin{enumerate}
    \item
$P(C>K_{1,\alpha}(\hat{p})) \to \alpha, \text{ when } A \sim ER (n,p)$, and 
\item  
$P(C>K_{1,\alpha}(\hat{p})) \rightarrow 1, \text{ when } A \sim NW (n,p,\beta) 
$,
as $n\rightarrow \infty$.
\end{enumerate}
\label{C-power}
\end{theorem}
The above theorem shows that the asymptotic level of the test is $\alpha$, and the asymptotic power converges to 1 under the alternative model $NW(n,p,\beta)$. Note that from Theorem \ref{C-dist} we already have $P(C_A > K_{1,\alpha}(p_0)) \to \alpha$. In the proof of this theorem, we show that $K_{1,\alpha}(\hat{p})= K_{1,\alpha}(p_0) + o_p(1)$, which then leads to the result on asymptotic level. For the result on asymptotic power, we further show that $K_{1,\alpha}(\hat{p})$ does not deviate much from $p_0$, and in particular, $K_{1,\alpha}(\hat{p}) = p_0 + o_p(p_0)$. Then we show that $C_A$ when $A$ comes from the alternative model  $NW(n,p,\beta)$ is higher than a constant multiple of $p_0C_n$ with high probability where $C_n$ converges to $\infty$ at any rate. 

\begin{theorem}
Suppose 
$
\frac{np}{\log(n)} \rightarrow \infty
$
as $n \rightarrow \infty$,
and 
$
p < 1/4.
$
Then, in this range of $p$, 
using $K_2 = \frac{(2+\epsilon) \log(n)}{\log(n\hat{p})}$, we have
\begin{enumerate}
    \item 
$P[L>K_2] \rightarrow 0$
when $\beta=0$, i.e., $A \sim ER(n,p)$,
\item  $P[L>K_2] \rightarrow 0$
when $A \sim NW(n,p,\beta)$ for some $0 < \beta < 1$, and 
\item  $P[L>K_2] \rightarrow 1$
when $\beta=1$, i.e., $A$ is a ring lattice, as $n \to \infty$.
\end{enumerate}
\label{L-power}
\end{theorem}

While in the above two theorems we have assumed the mixing proportion $\beta$ is constant as a function of $n$, in the next result we study a specific type of alternative hypothesis and associated power by making $\beta$ a function of $n$. In particular we study power for alternatives of the form $\beta = \frac{h}{n^l}$ and $\beta = 1-\frac{h}{n^{l}}$, for constants $h,l>0$. This result studies the ``local" power of the test as $\beta$ is slightly bigger than 0 and slightly smaller than 1.
\begin{theorem}
As $\beta$ approaches 0 and $A \sim NW(n,p,\beta)$, we have $P(C>K_{1,\alpha}(\hat{p})) \rightarrow 1$, as long as $\beta \geq \frac{1}{n^{1/2-\epsilon}}$.
As $\beta$ approaches 1 and $A \sim NW(n,p,\beta)$, we have $P[L>K_2] \rightarrow 0$ as long as 
 $\beta \leq 1- \frac{1}{n^l}$ where $l = \frac{\epsilon}{12} \frac{\log (np)}{\log n}$.
    \label{betapower}
\end{theorem}

\noindent \textbf{Remark:}
We conclude this subsection by synthesizing the theoretical results in the context of the intersection test.
Recall, from \eqref{test_statistic}, that the null hypothesis is rejected if the intersection criterion
$\{C> K_{1,\alpha} \} \cap  \{L < K_2 \}$
is satisfied.
Note that the probability of the intersection event is bounded above by the the probability of either event,
i.e., 
$$P[\{C> K_{1,\alpha} \} \cap  \{L < K_2 \}] \le \min \left(P[C> K_{1,\alpha}], P[L < K_2] \right).$$ 
Now consider the following three cases (also see Table \ref{tab:intuition} for a heuristic version):
\begin{enumerate}
    \item When $\beta = 0$, the network is generated purely from the ER model, and the correct decision is to not reject.
    Theorems \ref{C-dist} and \ref{C-power} ensure that $P[C> K_{1,\alpha}]$ converges to $\alpha$ asymptotically, which means that the probability of rejection (type-I error) is asymptotically bounded above by $\alpha$, the nominal significance level.
    \item When $\beta = 1$, the network is a pure ring lattice, and once again the correct decision is to not reject.
    Theorem \ref{L-power} ensures that $P[L < K_2]$ converges to zero asymptotically, which means that the probability of rejection (type-I error) goes to zero.
    \item When $0 < \beta < 1$, the network is generated from the small-world model, and the correct decision is to reject.
    Theorem \ref{C-power} ensures that $P[C < K_{1,\alpha}]$ converges to zero asymptotically and Theorem \ref{L-power} ensures that $P[L > K_2]$ converges to zero asymptotically.
    Therefore, the probability of the union event, $\{C> K_{1,\alpha} \} \cup  \{L < K_2 \}$, goes to zero, which means that the probability of rejection (power) goes to one.

    Furthermore, when $\beta$ approaches 0 from above, 
    Theorem \ref{L-power} ensures that $P[L > K_2]$ converges to zero
    and Theorem \ref{betapower} ensures that $P[C < K_{1,\alpha}]$ converges to zero as long as $\beta \geq \frac{1}{n^{1/2-\epsilon}}$.
    Therefore, the power of the test goes to one as long as $\beta \geq \frac{1}{n^{1/2-\epsilon}}$.

    Finally, when $\beta$ approaches 1 from below, 
    Theorem  \ref{C-power} ensures that $P[C < K_{1,\alpha}]$ converges to zero asymptotically, and
    Theorem \ref{betapower} ensures that $P[L > K_2]$ converges to zero
     s long as 
 $\beta \leq 1- \frac{1}{n^l}$ where $l = \frac{\epsilon}{12} \frac{\log (np)}{\log n}$.
    Therefore, the power of the test goes to one as long as $\beta \leq 1- \frac{1}{n^l}$.

\end{enumerate}

\subsection{Bootstrap test}
In the Bootstrap version of the test, we determine the cutoffs through a parametric bootstrap procedure which involves fitting the respective null models to the observed data to estimate the parameters of the null models. An adequate number ($B$) of graphs are sampled from the fitted null distribution parameters. We define the test by letting $K_1$ be the 95th percentile of the distribution of $C$ and $K_2$ to be a very high percentile (e.g.,  the 99th percentile) of the distribution of $L$. Therefore a network will be called ``small-world" if the observed $C$ is \textit{higher} than the $95$th percentile of the reference bootstrap distribution for $C$ and the observed $L$ is \textit{lower} than $99$th percentile of the reference bootstrap distribution for $L$. For fitting SBM and DCSBM to the observed networks, we estimate the community assignments with a spectral clustering procedure that involves the following steps: (i) eigendecomposition of the adjacency matrix and creation of the matrix of $k$ eigenvectors $U_{n \times k}$ that correspond to the $k$ largest eigenvalues in absolute value, (ii) projection of rows of $U$ to the unit circle and (iii) k-means clustering of the rows of the matrix $U$. Finally, the model parameters are estimated using method of moments. The spectral clustering method is known to be consistent for the problem of estimating community structure from SBM and DCSBM when the number of communities are known \cite{gao2017achieving,lei2015consistency}. In practice, for our real data results we pick the number of communities $k$ through a heuristic method called Louvain modularity optimization method \cite{blondel08}. One can also use alternative methods for detecting communities from DCSBM graphs, e.g., using the SCORE method in \cite{jin2021improvements,jin2022mixedmembershipestimationsocial}.
While we have used the spectral method described above for our empirical results, we find that the results on the real data do not differ much if we use the SCORE method instead (Figure \ref{fig:comm_det_comp} in the Appendix).

\subsection{Remark: Test for $C$ in inhomogeneous random graphs}
Our asymptotic test and the associated theoretical results on asymptotic distribution of $C$, and power of the test are focused on homogeneous random graphs (ER random graphs). In this section, we consider an inhomogeneous random graph where edges $A_{ij}$ between pairs of vertices $(i,j)$ are generated independently from Bernoulli distributions with probabilities $\theta_{ij}$. The CL, SBM and DCSBM with non-random community labels and degree parameters are submodels of this model. We describe the asymptotic limiting distribution of $C$ under this model and prove the asymptotic level and power of a testing procedure to detect $\beta>0$. The average path length $L$ is a difficult quantity to probabilistically characterize under the inhomogeneous random graph model and hence we do not study it here.  Let $\theta_{\max}=\max_{ij} \theta_{ij}$, $\theta_{\min}=\min_{ij} \theta_{ij}$, and $\tilde{\theta}=\sum_{1 \leq i<j\leq n}\theta_{ij}/\dbinom{n}{2}$. The statistics $S,T,C$ are all defined as before since those are model-agnostic definitions. Define the following quantities including expectations of $S$ and $T$ under this model.
\begin{align*}
    s(\theta) &= E[S]  = \frac{1}{\dbinom{n}{3}} \sum_{1\leq i<j<k \leq n} (\theta_{ij}\theta_{ik} + \theta_{ij}\theta_{jk} + \theta_{jk}\theta_{ik}) \\
    t(\theta) &= E[T]  = \frac{1}{\dbinom{n}{3}} \sum_{1\leq i<j<k \leq n} (\theta_{ij}\theta_{ik}\theta_{jk})\\
    \Theta_T & = \frac{1}{\dbinom{n}{3}} \sum_{1\leq i<j<k \leq n} (\theta_{ij}\theta_{ik}\theta_{jk})(1-\theta_{ij})(1-\theta_{ik})(1-\theta_{jk}).
\end{align*}
These quantities can be efficiently computed from the matrix of probabilities $\Theta$ \cite{gao2017testing2}. We have the following result on the asymptotic limiting distribution of $C$.
\begin{theorem}
    Assume $\frac{1}{n}<<\theta_{\min} \asymp \tilde{\theta} \asymp \theta_{\max} << \frac{1}{\sqrt{n}}$. Then for the inhomogeneous random graph model described above the limiting distribution of clustering coefficient as $n \to \infty$ is
    \[
    \sqrt{\frac{\dbinom{n}{3}s(\theta)^2}{9 \Theta_T}}  \left(C- \frac{3t(\theta)}{s(\theta)}\right) \overset{D}{\to} N(0,1).
    \]
    \label{C-inhomo}
\end{theorem}

Therefore we propose an asymptotic test for $C$ with the following cutoff. Let $K_{1,\alpha}(\theta)$ be the $100*(1-\alpha)$\% quantile of this asymptotic distribution, i.e.,
\[
K_{1,\alpha}(\theta) = \frac{3t(\theta)}{s(\theta)} + Z_{\alpha} \sqrt{\frac{9\Theta_T}{\dbinom{n}{3}s(\theta)^2}}.
\]

It is easy to see that $\frac{t(\theta)}{s(\theta)}= O(\tilde{\theta})$, and  $\sqrt{\frac{9\Theta_T}{\dbinom{n}{3}s(\theta)^2}} = O(\sqrt{\frac{\tilde{\theta}}{n^3}})$. Therefore $K_{1,\alpha}(\theta) = O(\tilde{\theta})$.

We assume that we have a good enough estimator for the probabilities $\hat{\theta}$, such that the following two assumptions hold.
\begin{enumerate}
\item[A.1] $\frac{t(\hat{\theta})}{s(\hat{\theta})} = \frac{t(\theta)}{s(\theta)} + o_p(\tilde{\theta})$
    \item[A.2] $K_{1,\alpha}(\hat{\theta}) = K_{1,\alpha}(\theta) + o_p(1)$.
\end{enumerate}
The above two assumptions enable us to obtain $K_{1,\alpha}(\hat{\theta}) = 3\frac{t(\theta)}{s(\theta)} + o_p(1)$. Consequently, we have the following result.

\begin{theorem}
    Consider the NW superimposed inhomogeneous random graph model with $\beta \in [0,1]$. Assume $\frac{\log n}{n}<<\theta_{\min} \asymp \tilde{\theta} \asymp \theta_{\max} << \frac{1}{\sqrt{n}}$, and $\frac{t(\theta)}{s(\theta)}C_n \to 0$, for a sequence $C_n \to \infty$ at any rate. Further assume we have an estimator of the parameters $\theta$ that satisfies the properties in assumptions A.1 and A.2. Then as $n\rightarrow \infty$,
\begin{enumerate}
    \item
$P(C>K_{\alpha}(\hat{\theta})) \to \alpha, \text{ when } \beta=0$, and 
\item  
$P(C>K_{\alpha}(\hat{\theta})) \rightarrow 1, \text{ when } \beta>0 
$.
\end{enumerate}
\label{C-power-inhomo}
\end{theorem}

\section{Simulation}

\begin{figure}
\begin{subfigure}{0.3 \linewidth}
\includegraphics[width=\linewidth]{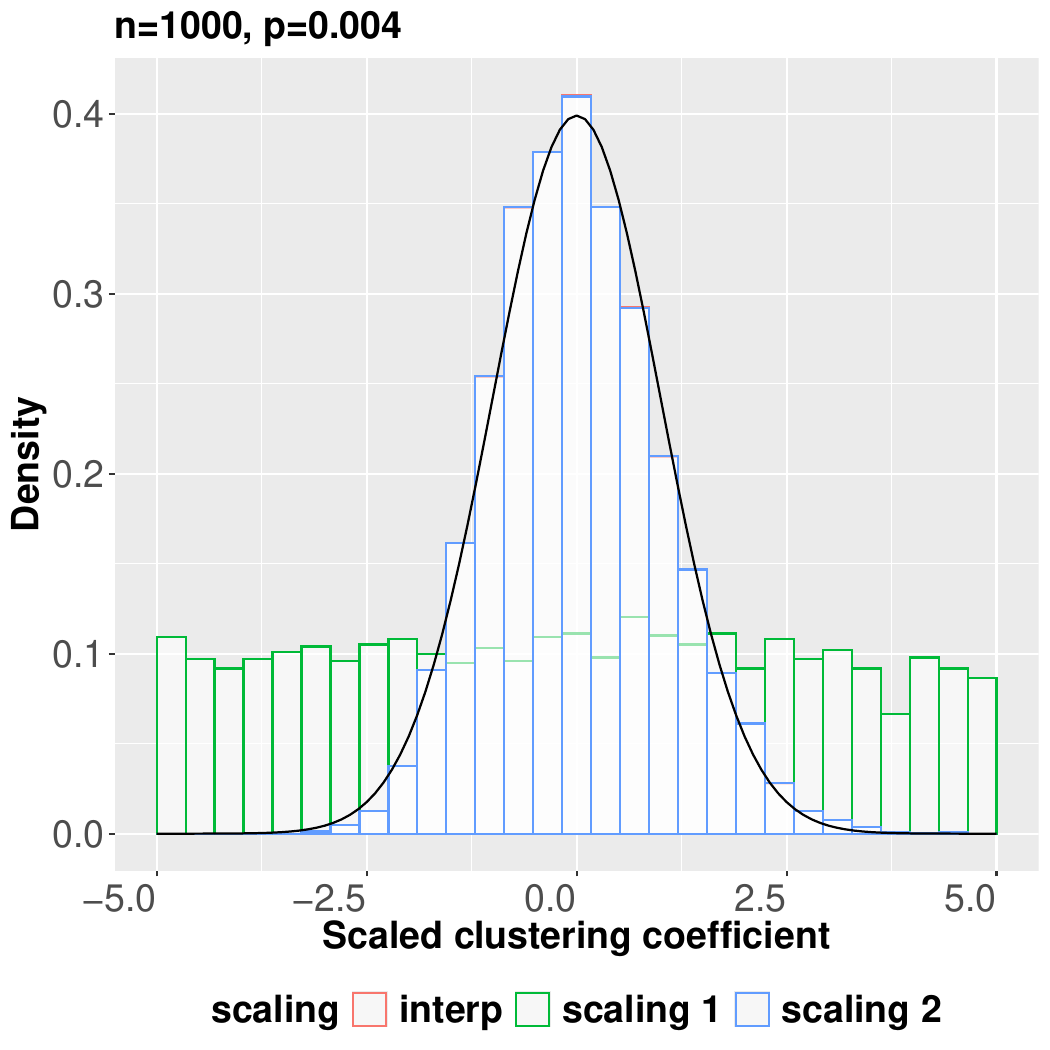}
\end{subfigure}%
\begin{subfigure}{0.3 \linewidth}
\includegraphics[width=\linewidth]{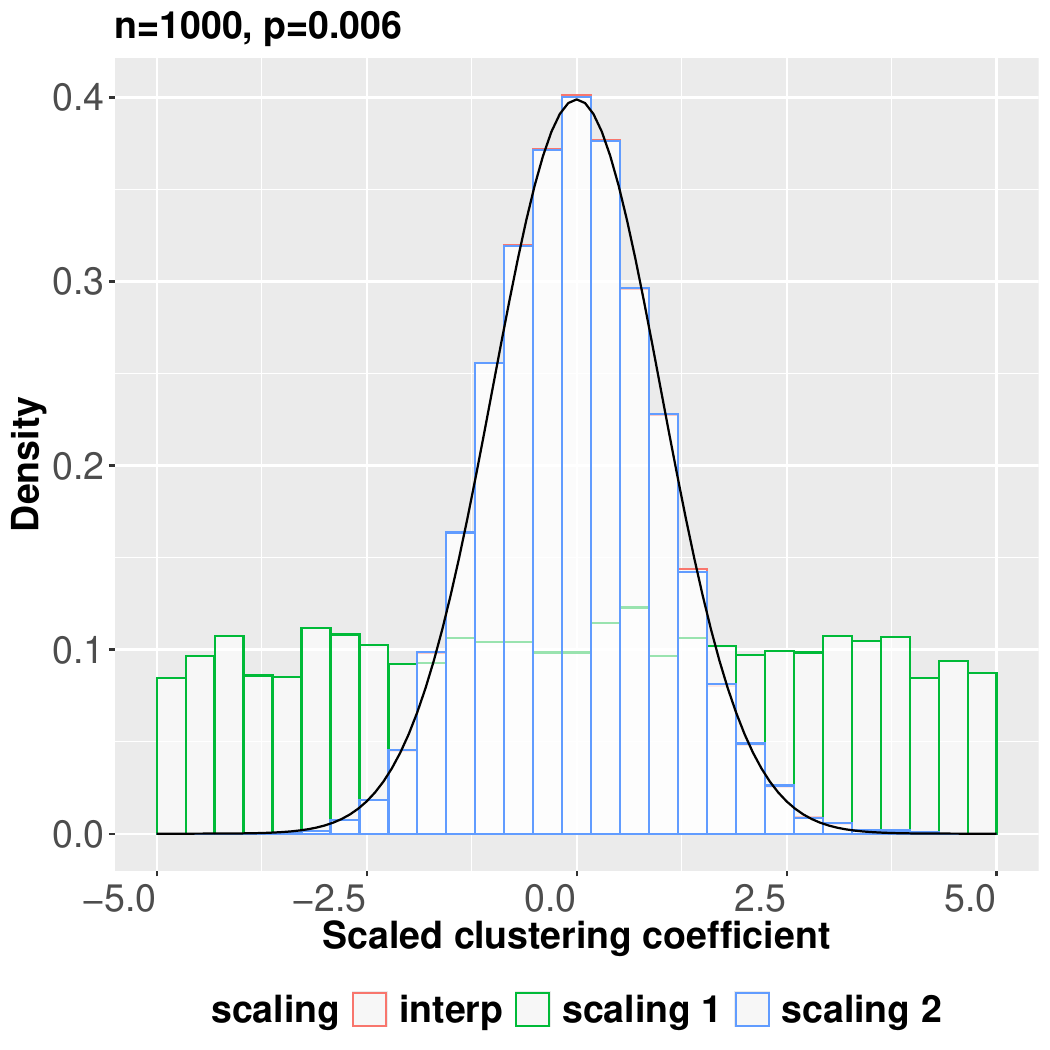}
\end{subfigure}%
\begin{subfigure}{0.3 \linewidth}
\includegraphics[width=\linewidth]{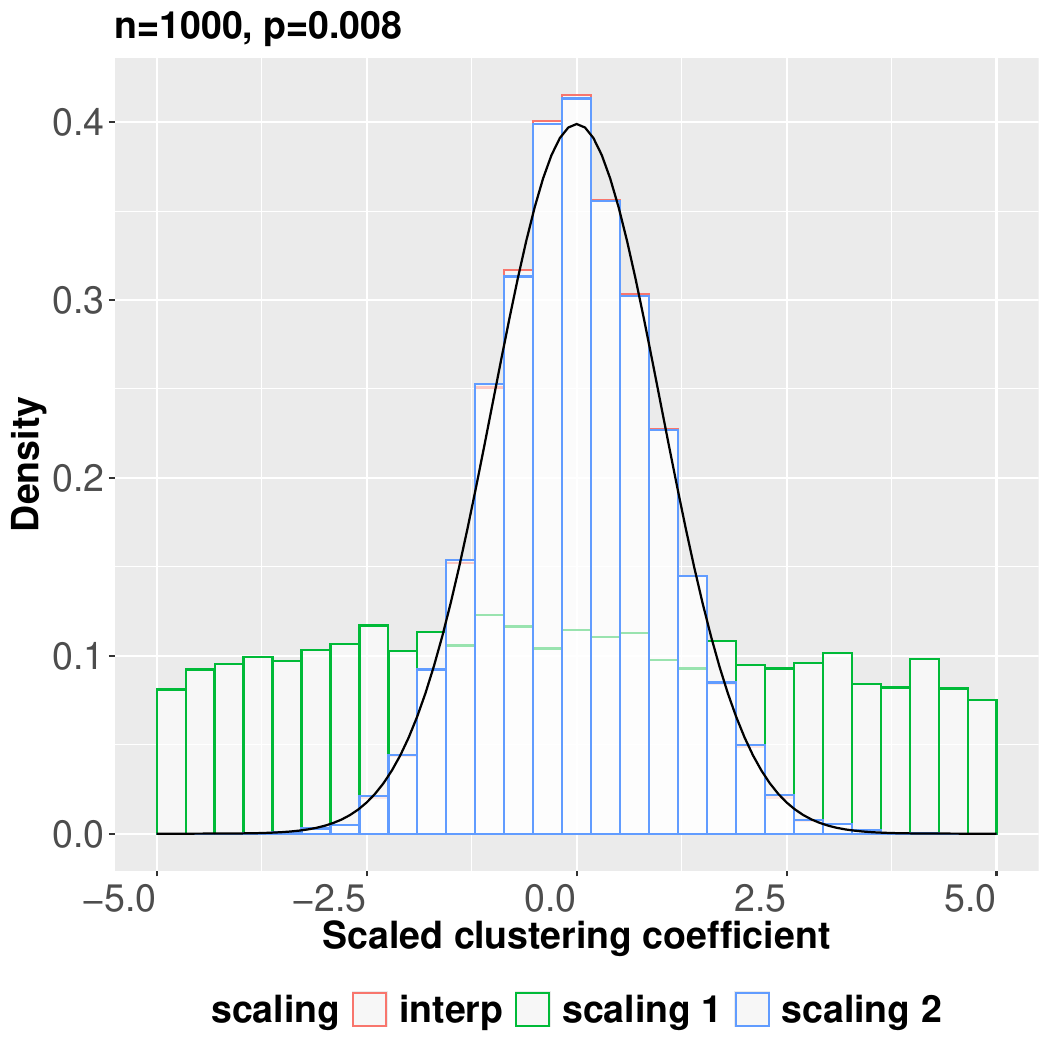}
\end{subfigure}
\begin{subfigure}{0.3 \linewidth}
\includegraphics[width=\linewidth]{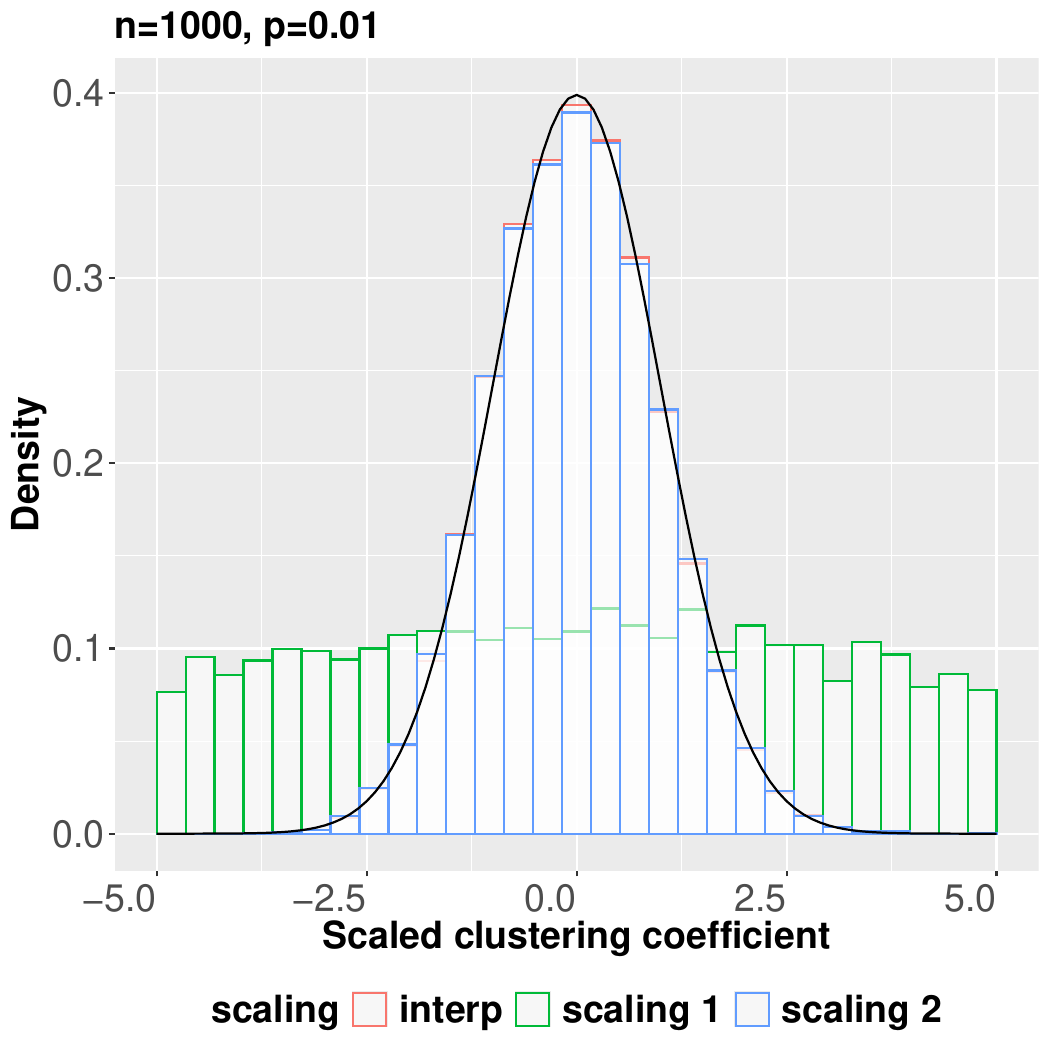}
\end{subfigure}%
\begin{subfigure}{0.3 \linewidth}
\includegraphics[width=\linewidth]{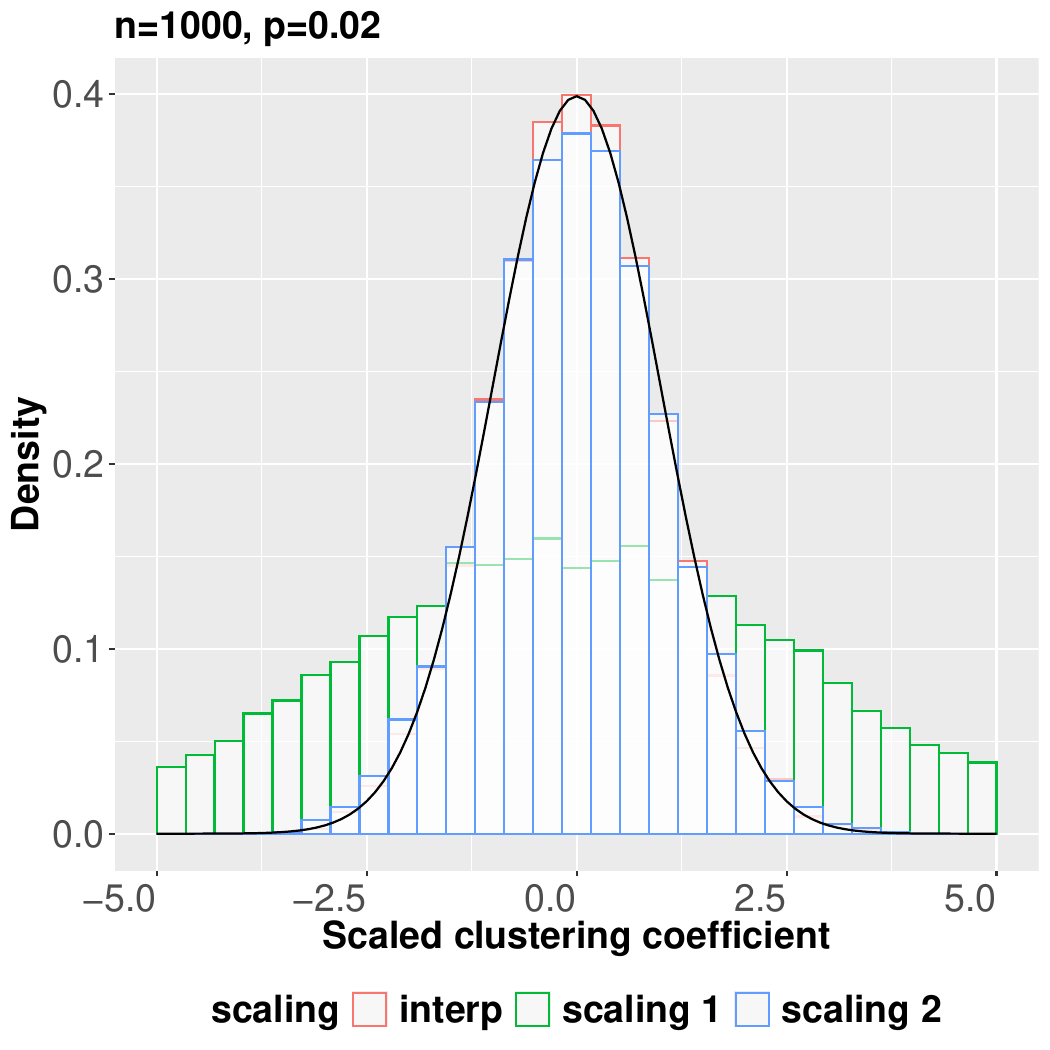}
\end{subfigure}%
\begin{subfigure}{0.3 \linewidth}
\includegraphics[width=\linewidth]{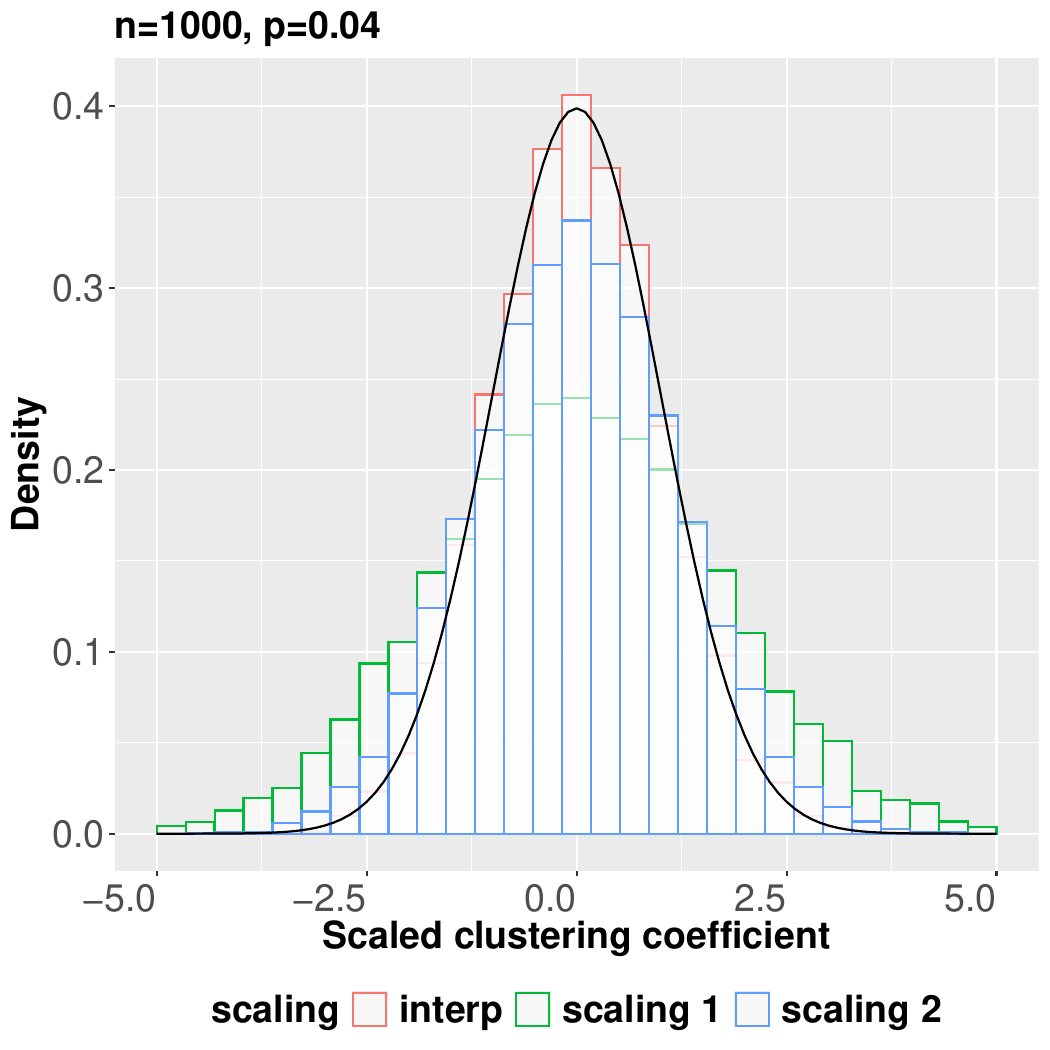}
\end{subfigure}
\begin{subfigure}{0.3 \linewidth}
\includegraphics[width=\linewidth]{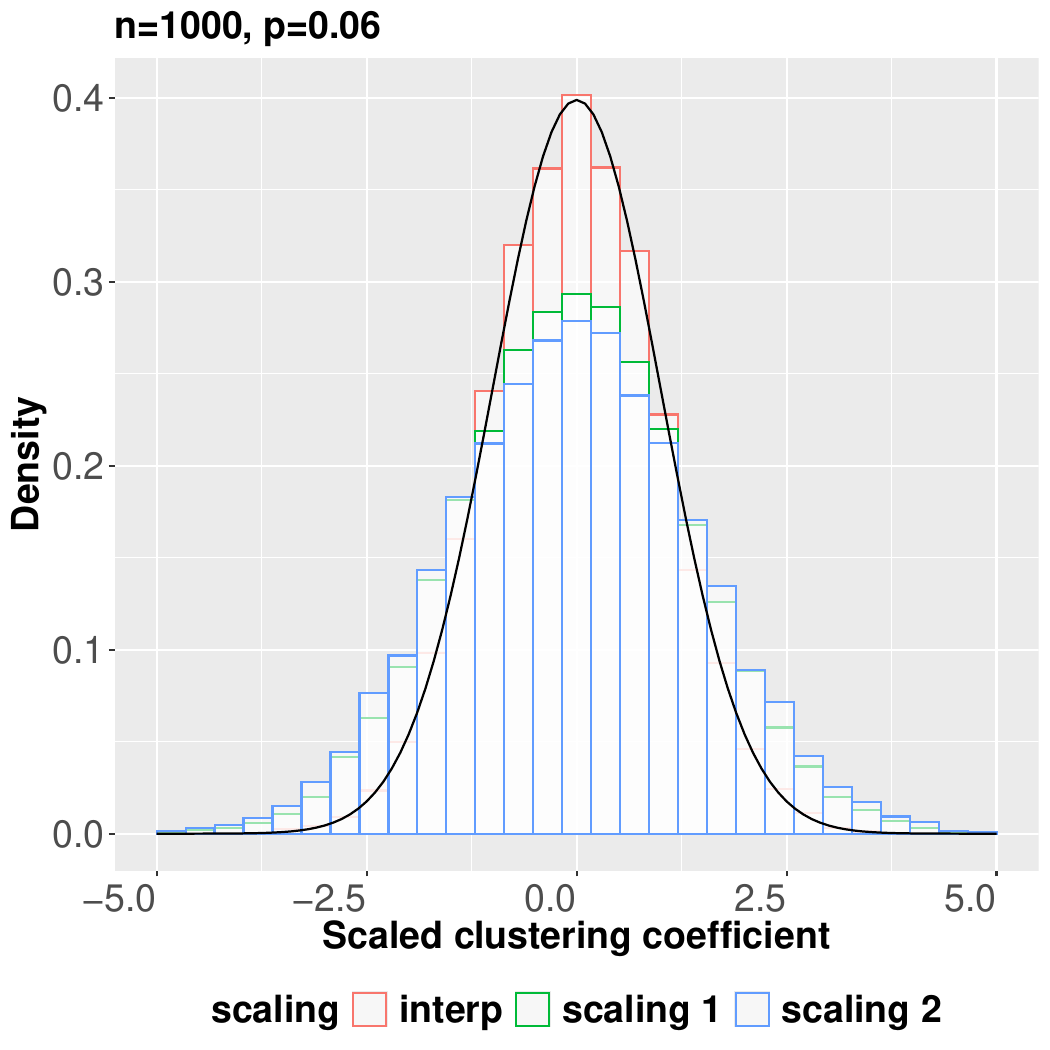}
\end{subfigure}%
\begin{subfigure}{0.3 \linewidth}
\includegraphics[width=\linewidth]{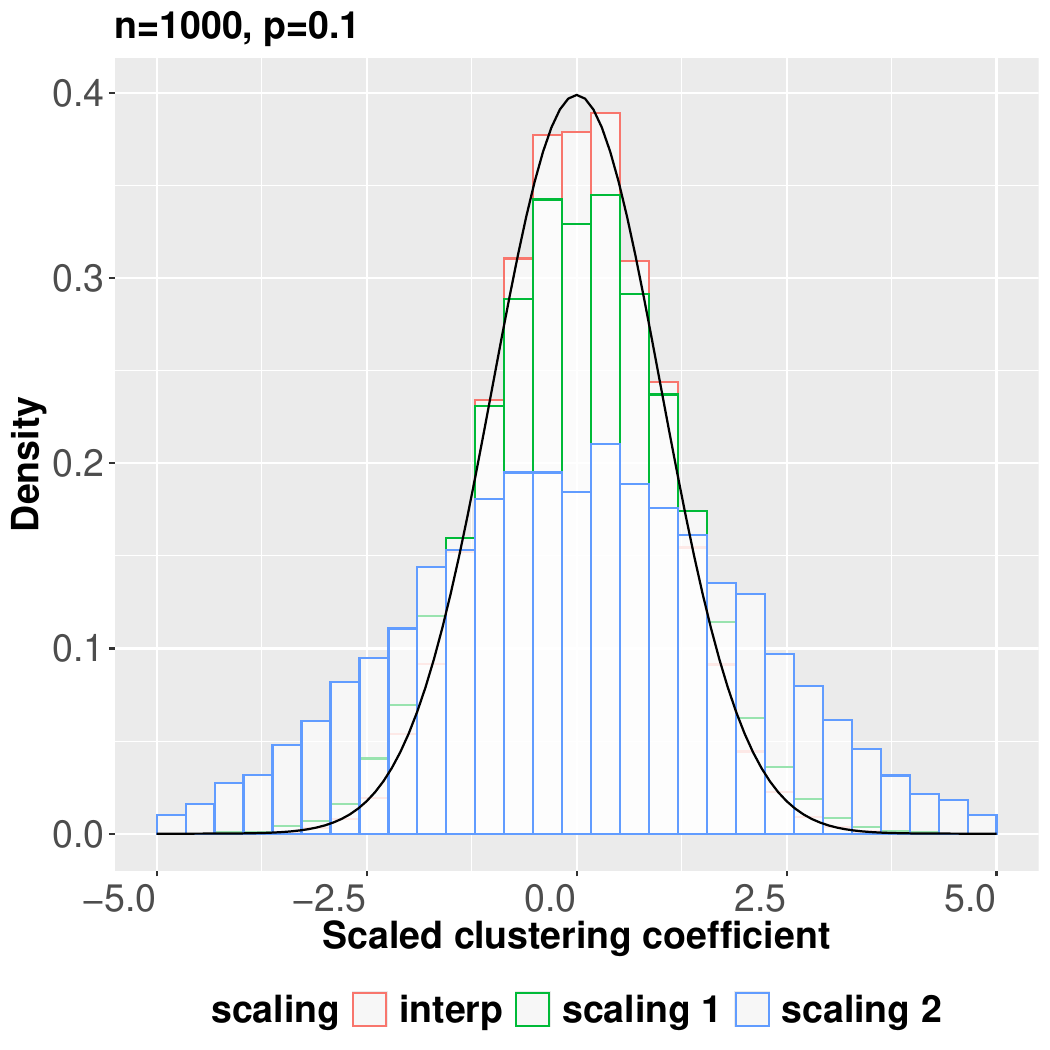}
\end{subfigure}%
\begin{subfigure}{0.3 \linewidth}
\includegraphics[width=\linewidth]{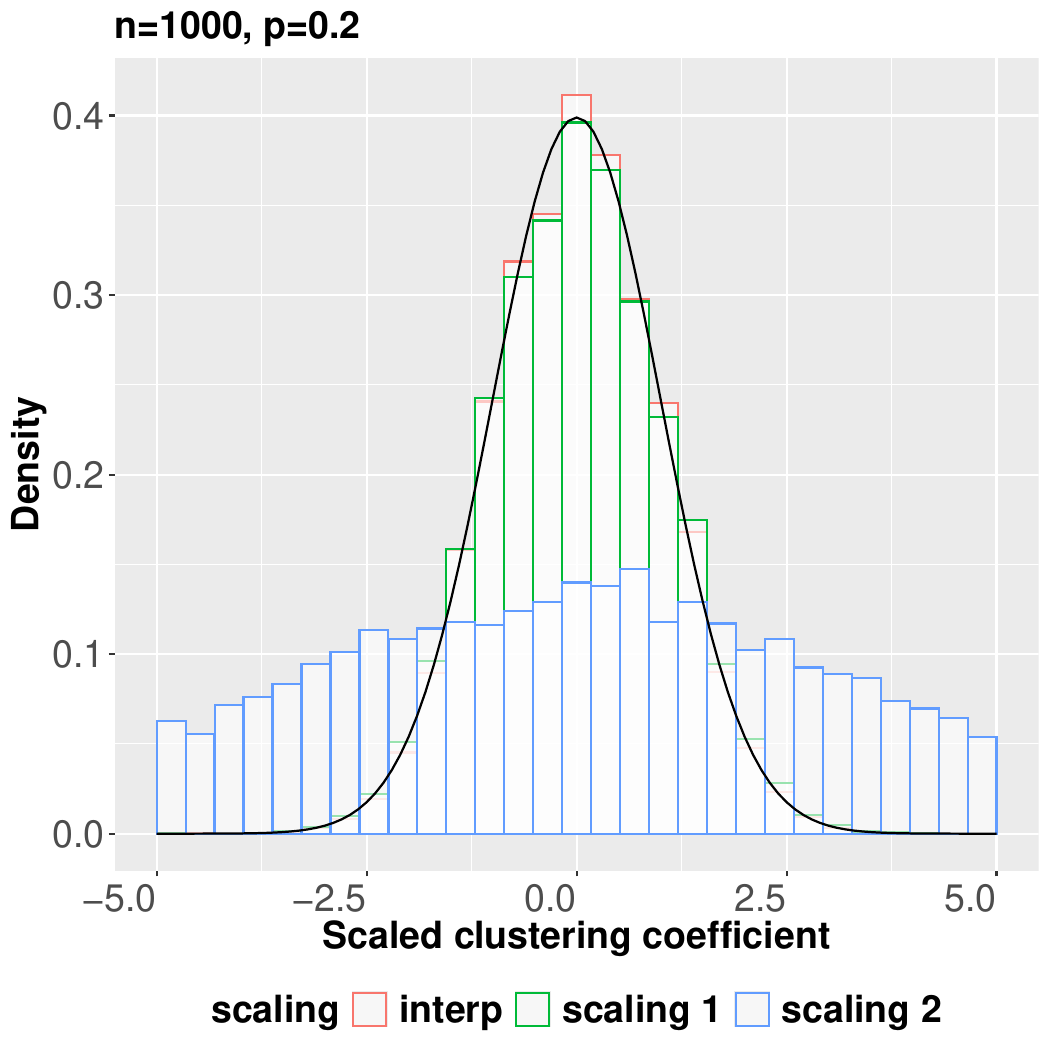}
\end{subfigure}
    \caption{Comparison of various scaling factors in Theorem \ref{C-dist} with 10000 simulations from the ER model compared with the density function of standard normal distribution. We fix $n=1000$ and vary $p$ from 0.004 to 0.20.}
    \label{fig:Cdist-comp}
\end{figure}

\begin{figure}
\begin{subfigure}{0.3 \linewidth}
\includegraphics[width=\linewidth]{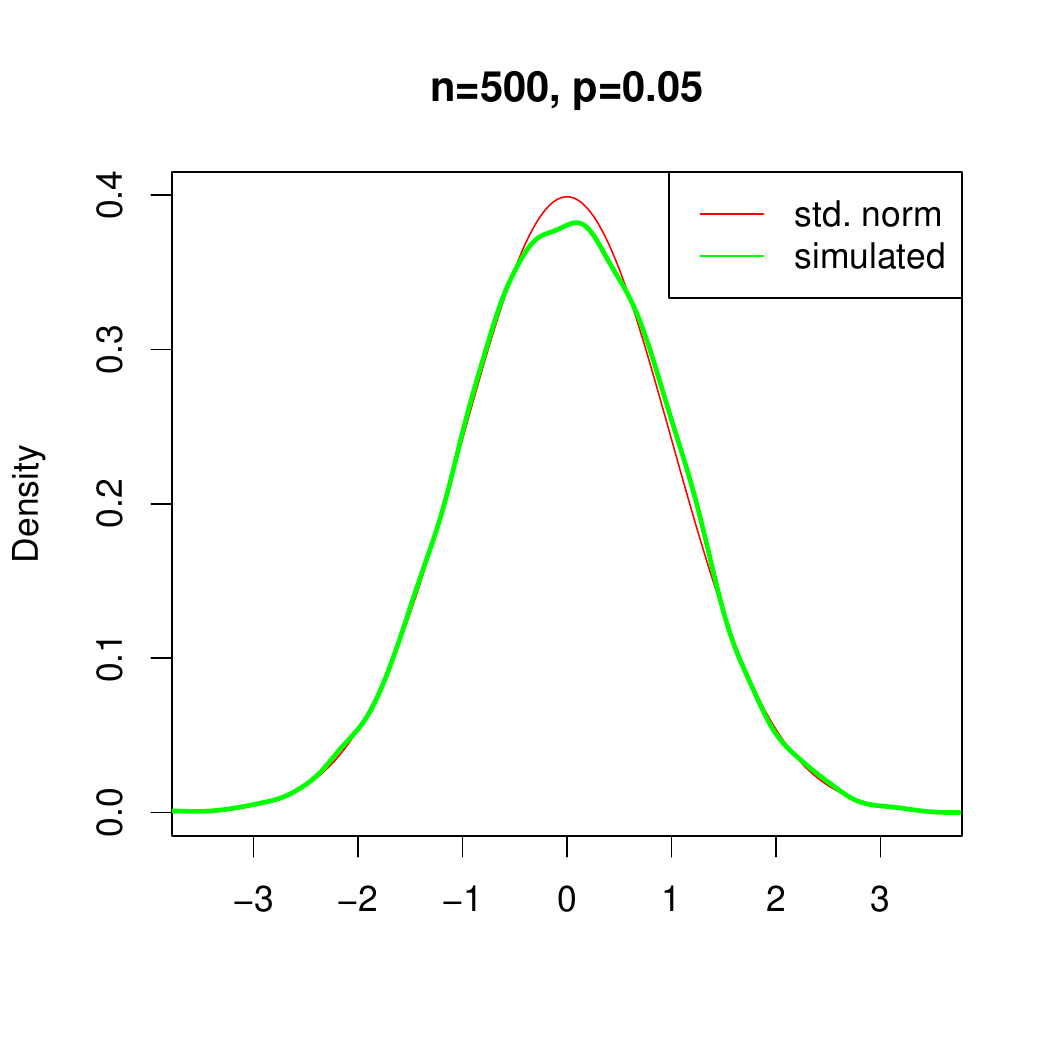}
\end{subfigure}%
\begin{subfigure}{0.3 \linewidth}
\includegraphics[width=\linewidth]{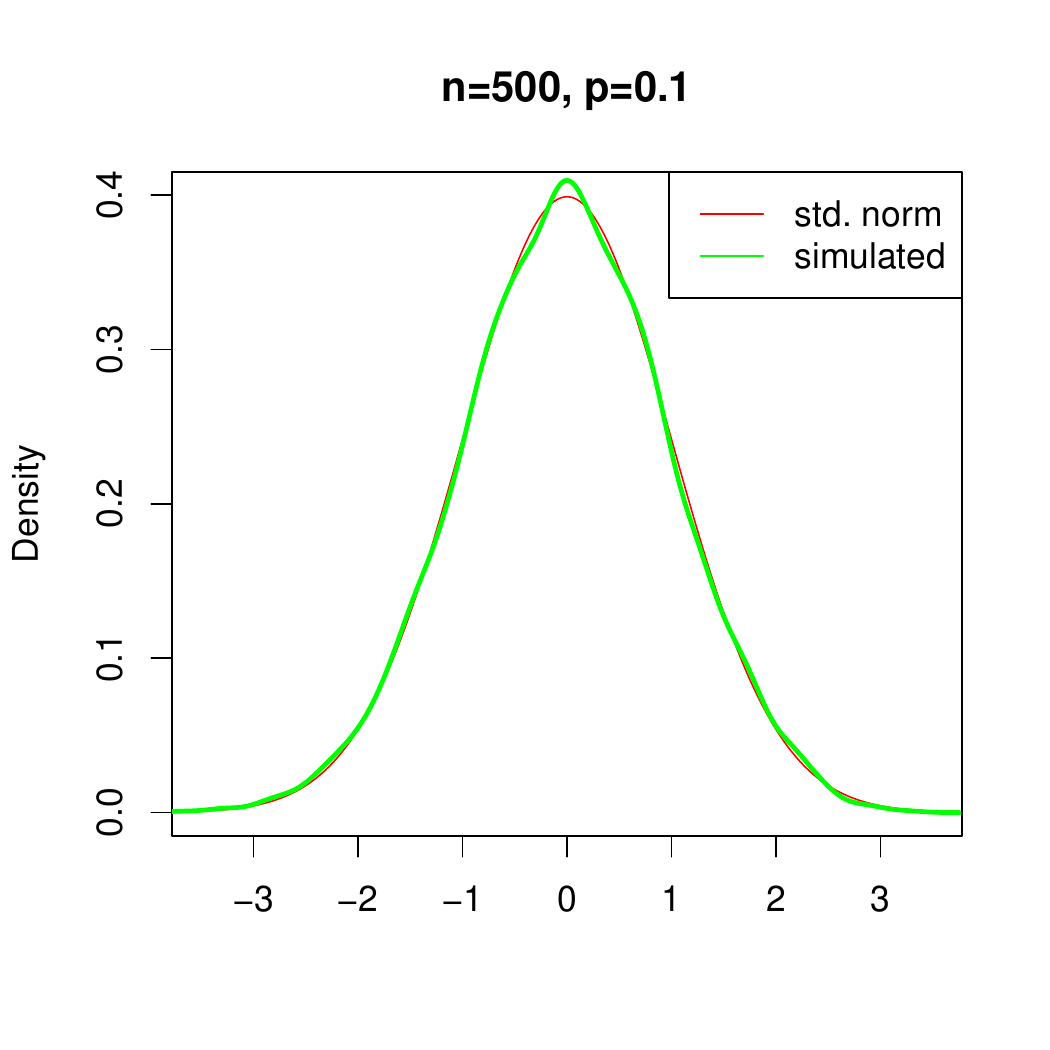}
\end{subfigure}%
\begin{subfigure}{0.3 \linewidth}
\includegraphics[width=\linewidth]{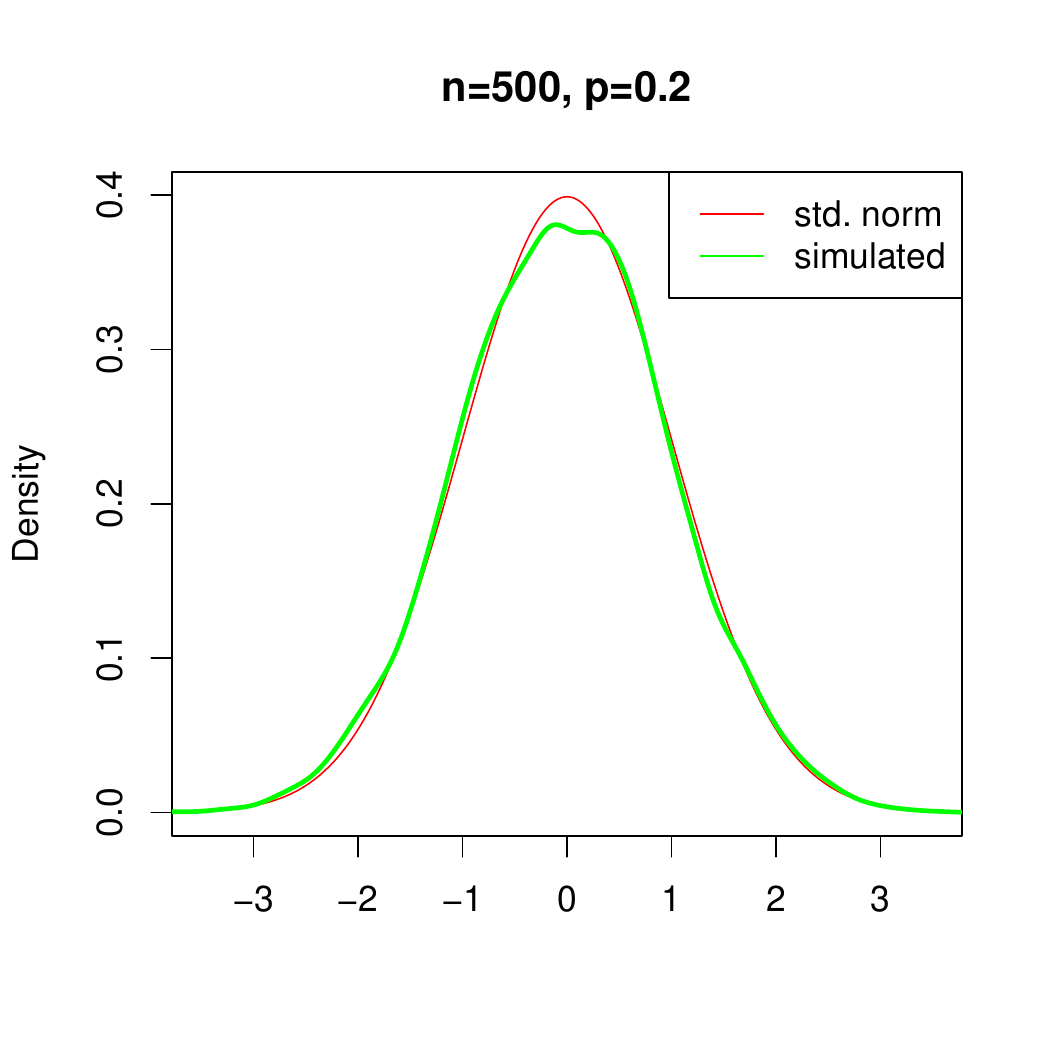}
\end{subfigure}
\begin{subfigure}{0.3 \linewidth}
\includegraphics[width=\linewidth]{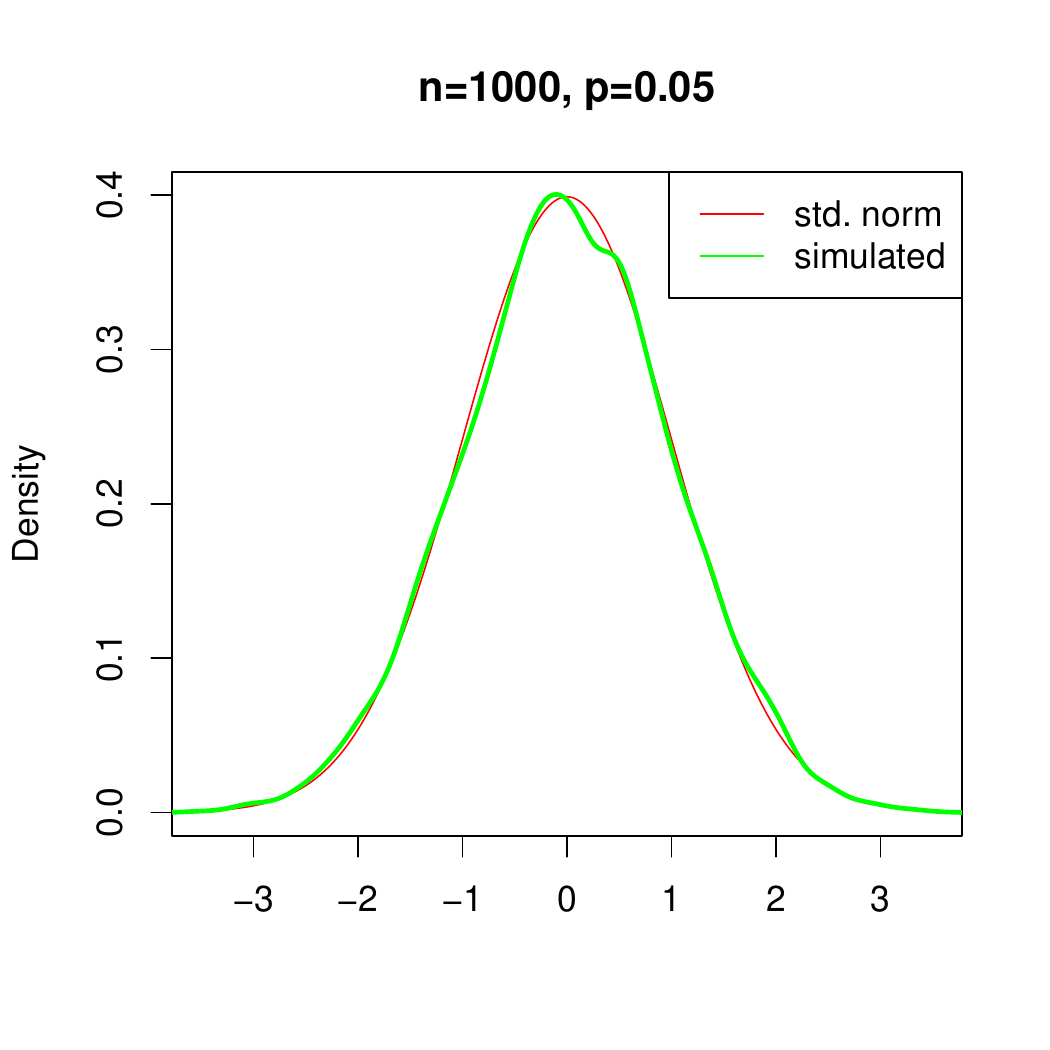}
\end{subfigure}%
\begin{subfigure}{0.3 \linewidth}
\includegraphics[width=\linewidth]{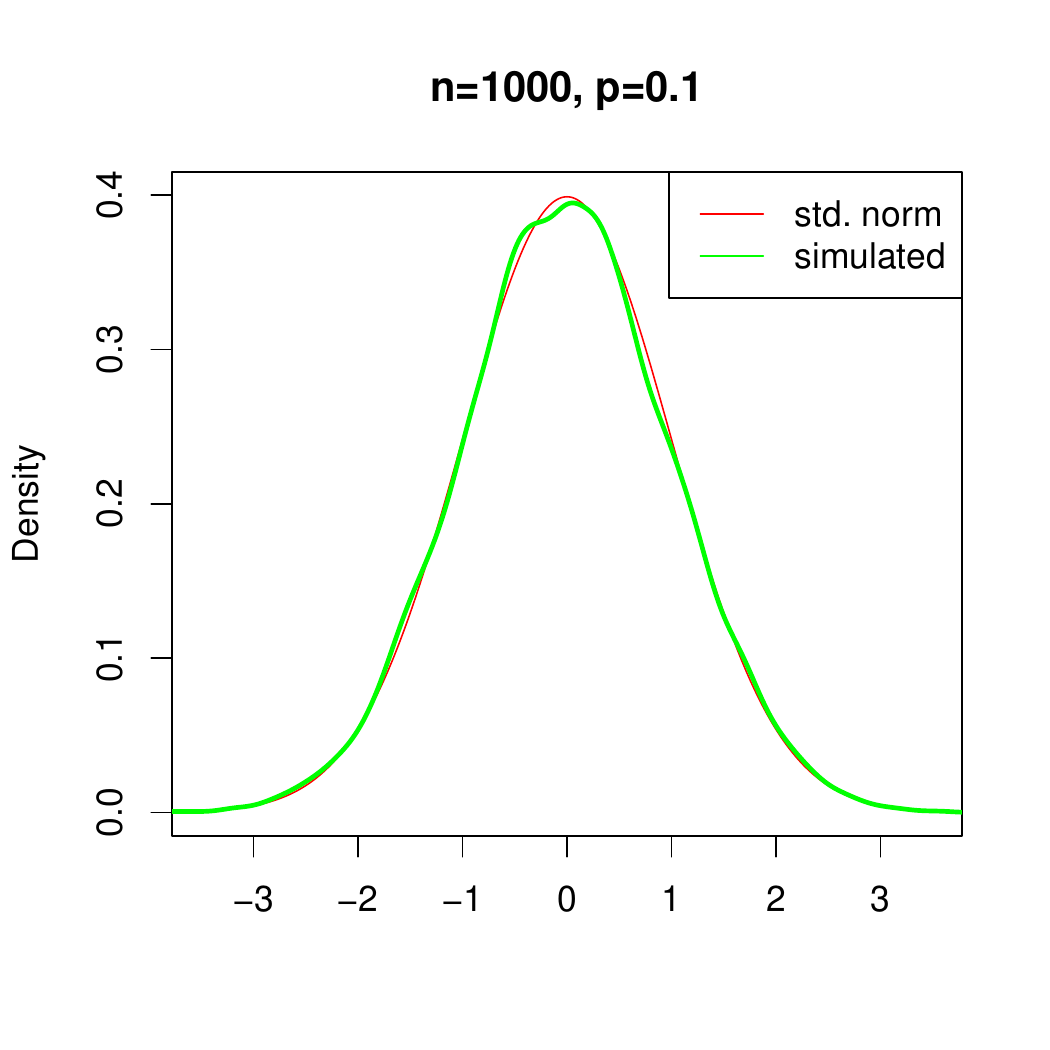}
\end{subfigure}%
\begin{subfigure}{0.3 \linewidth}
\includegraphics[width=\linewidth]{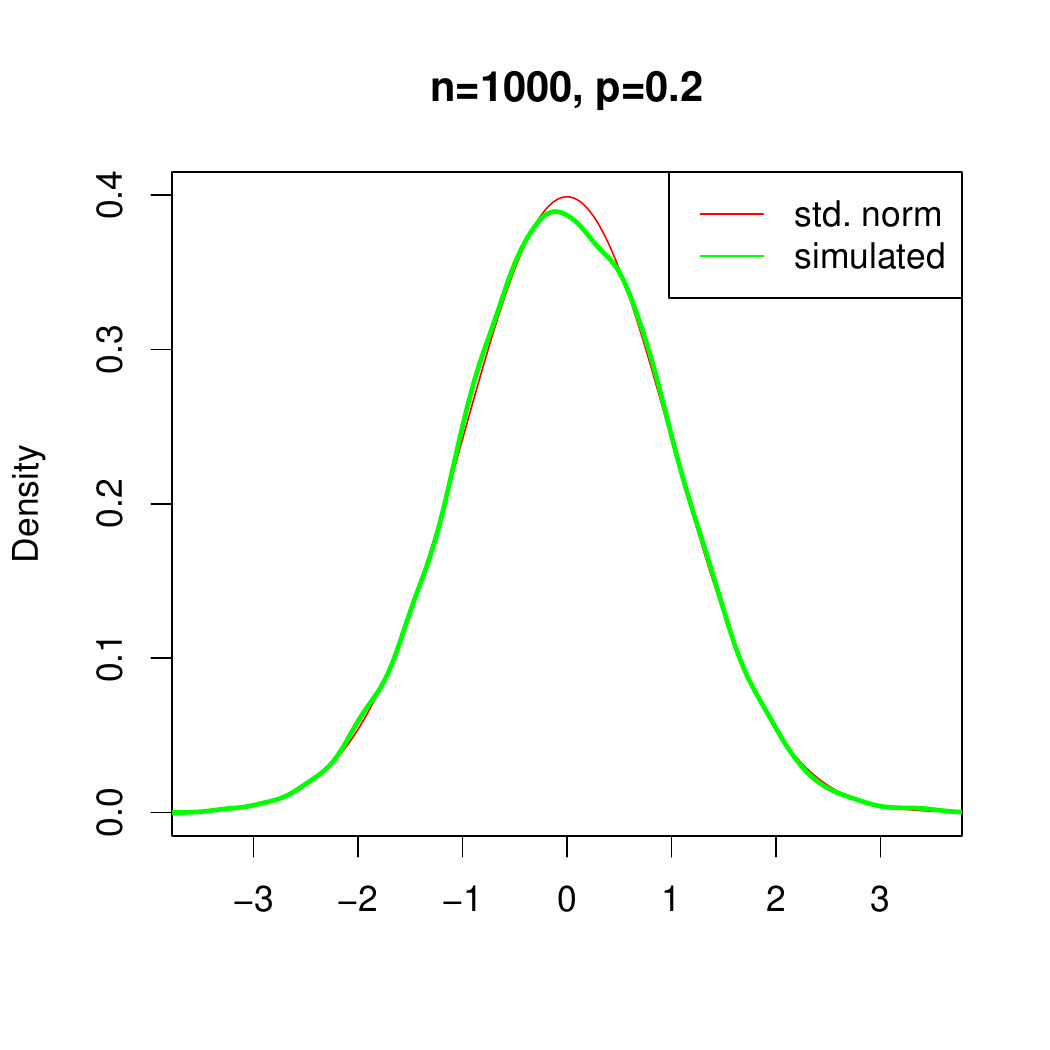}
\end{subfigure}
\begin{subfigure}{0.3 \linewidth}
\includegraphics[width=\linewidth]{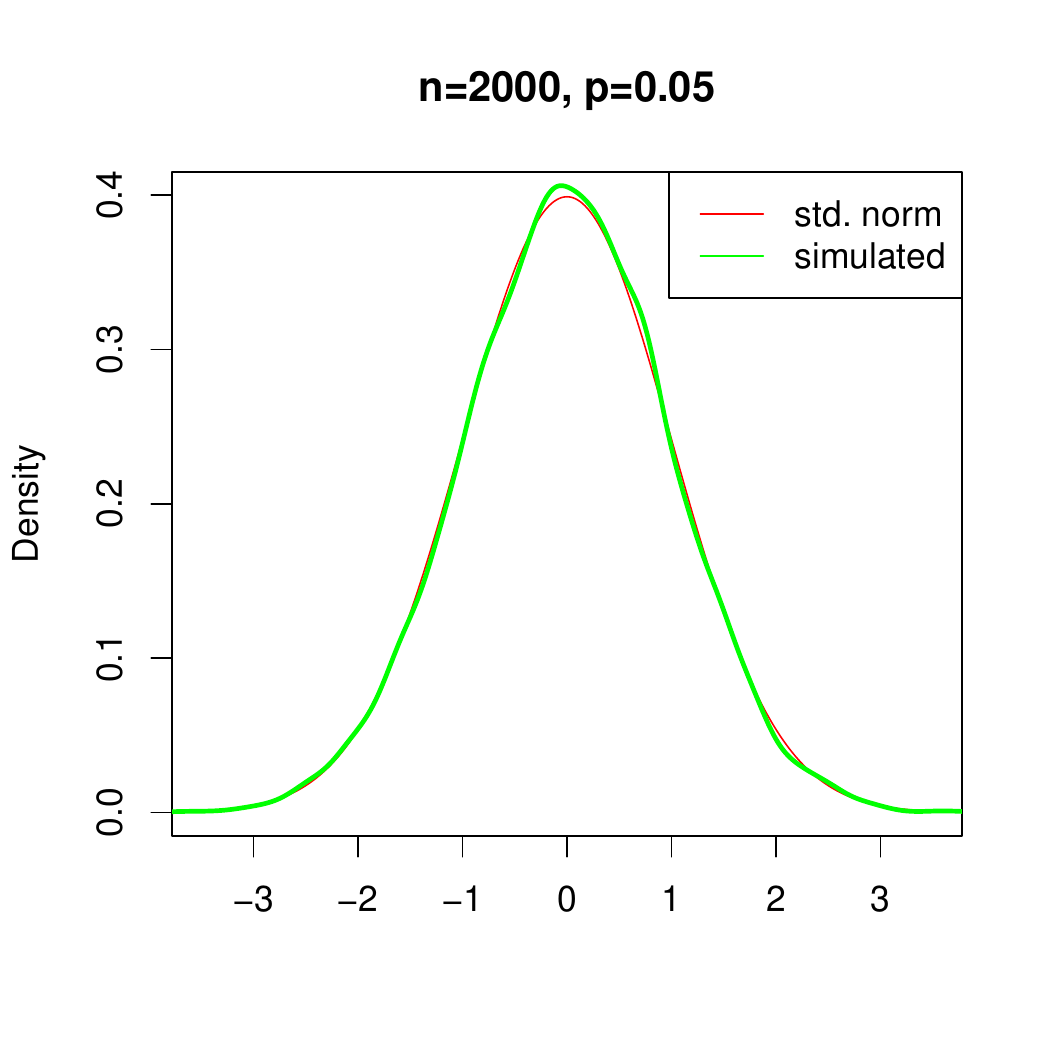}
\end{subfigure}%
\begin{subfigure}{0.3 \linewidth}
\includegraphics[width=\linewidth]{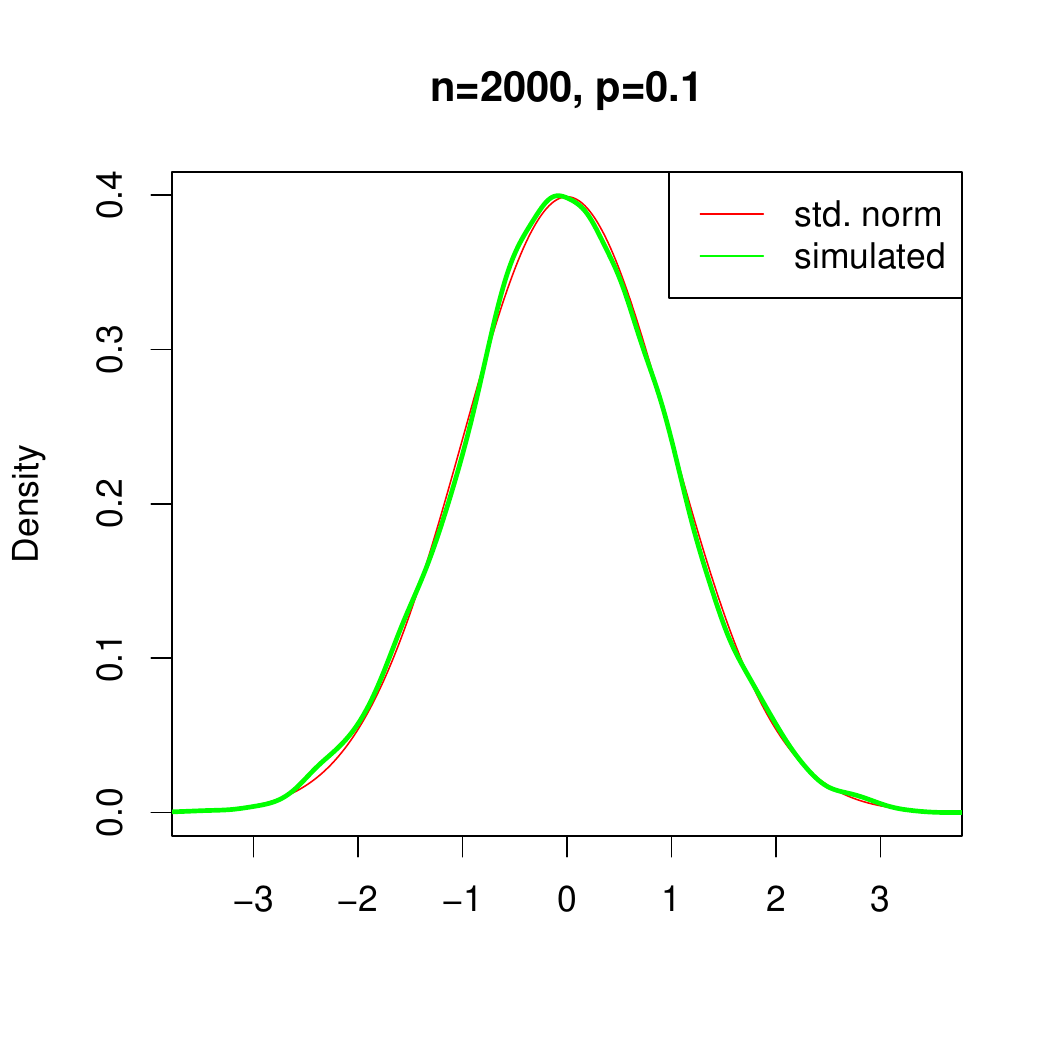}
\end{subfigure}%
\begin{subfigure}{0.3 \linewidth}
\includegraphics[width=\linewidth]{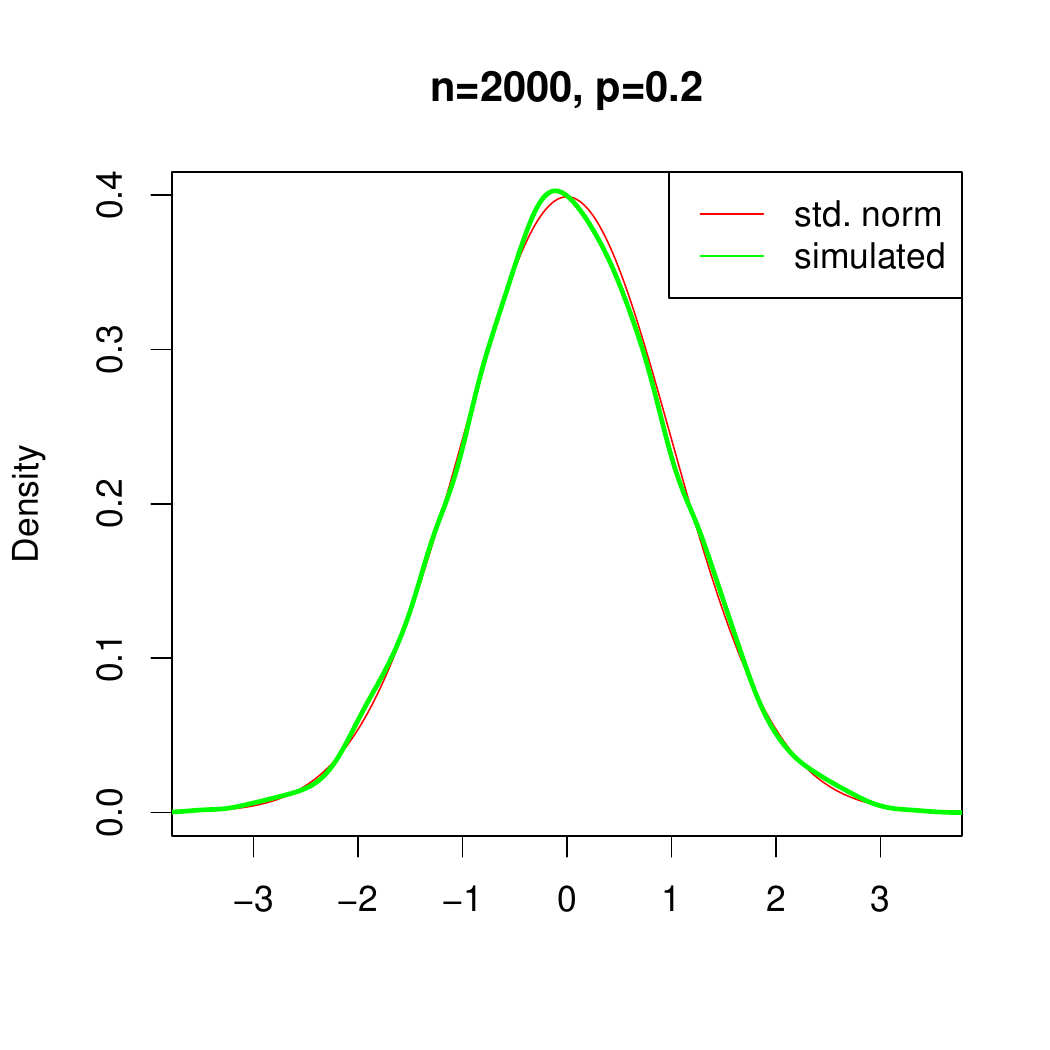}
\end{subfigure}
    \caption{The observed distribution of $C$ appropriately scaled as per the interpolating scaling with 10000 simulations from the ER model compared with the density function of standard normal distribution .}
    \label{fig:Cdist}
\end{figure}

\subsection{Finite sample behavior of the distribution of $C$ in Theorem \ref{C-dist}}
We study the finite sample approximation provided by the asymptotic distribution of $C$ under the ER model derived in Theorem \ref{C-dist} through two simulations. 

In the first simulation, we fix $n=1000$ and vary $p$ from $0.004$ to $0.20$ in order to investigate the transition in the asymptotic normality result observed in Theorem \ref{C-dist}. We generate 10000 graphs from these ER models and compute the clustering coefficient $C$ in each case. Then we transform these empirical values of $C$ using the two scalings predicted by Theorem \ref{C-dist}.  We then plot the observed histograms of scaled values of $(C-p)$ of these values and compare it with the density function of the standard normal distribution evaluated between $(-3.5, 3.5)$.  In Figure \ref{fig:Cdist-comp} we compare 3 scaling factors on $C-p$, namely, the one given in part (1) of Theorem \ref{C-dist} (scaling 2), the one given in part (2) of Theorem \ref{C-dist} (scaling 1), and the one which include $\Sigma_C$ with the scaling in part (2), that we call ``interpolating" scaling.  We see that as Theorem \ref{C-dist} predicted, for small values of $p$, the scaling given in part (1) of the theorem provides accurate approximation, and as $p$ increases, this approximation starts performing worse and the approximation with scaling in part (2) starts getting better, until for larger values of $p$ when the scaling in part (2) provides an accurate approximation. Perhaps somewhat surprisingly, we see that the interpolation scaling performs very well for all ranges of values of $p$. Overall, we can see the transition in the asymptotic normality result from Theorem \ref{C-dist} for different growth rates on $p$ in our finite sample simulation as well.

In order investigate the interpolating scaling $\sqrt{\frac{n(n-1)}{2p(1-p)\Sigma_C}}(C-p)$ further empirically we conduct another simulation.  We generate 10000 random graphs for each of the 9 ER models resulting from varying $n=(500, 1000, 2000)$ and $p=(0.05, 0.1, 0.2)$ and compute the empirical clustering coefficient ($C$) in each case. Then we transform these empirical values of $C$ according to the specified scaling, plot the empirical density of these values, and compare it with the density function of the standard normal distribution evaluated between $(-3.5, 3.5)$ in Figure \ref{fig:Cdist}. The theoretical asymptotic distribution matches the simulated one closely for all values of $n$ and $p$ indicating that the distribution is a good fit.

\begin{figure*}
\centering
\begin{subfigure}{0.31 \linewidth}
\includegraphics[width=\linewidth]{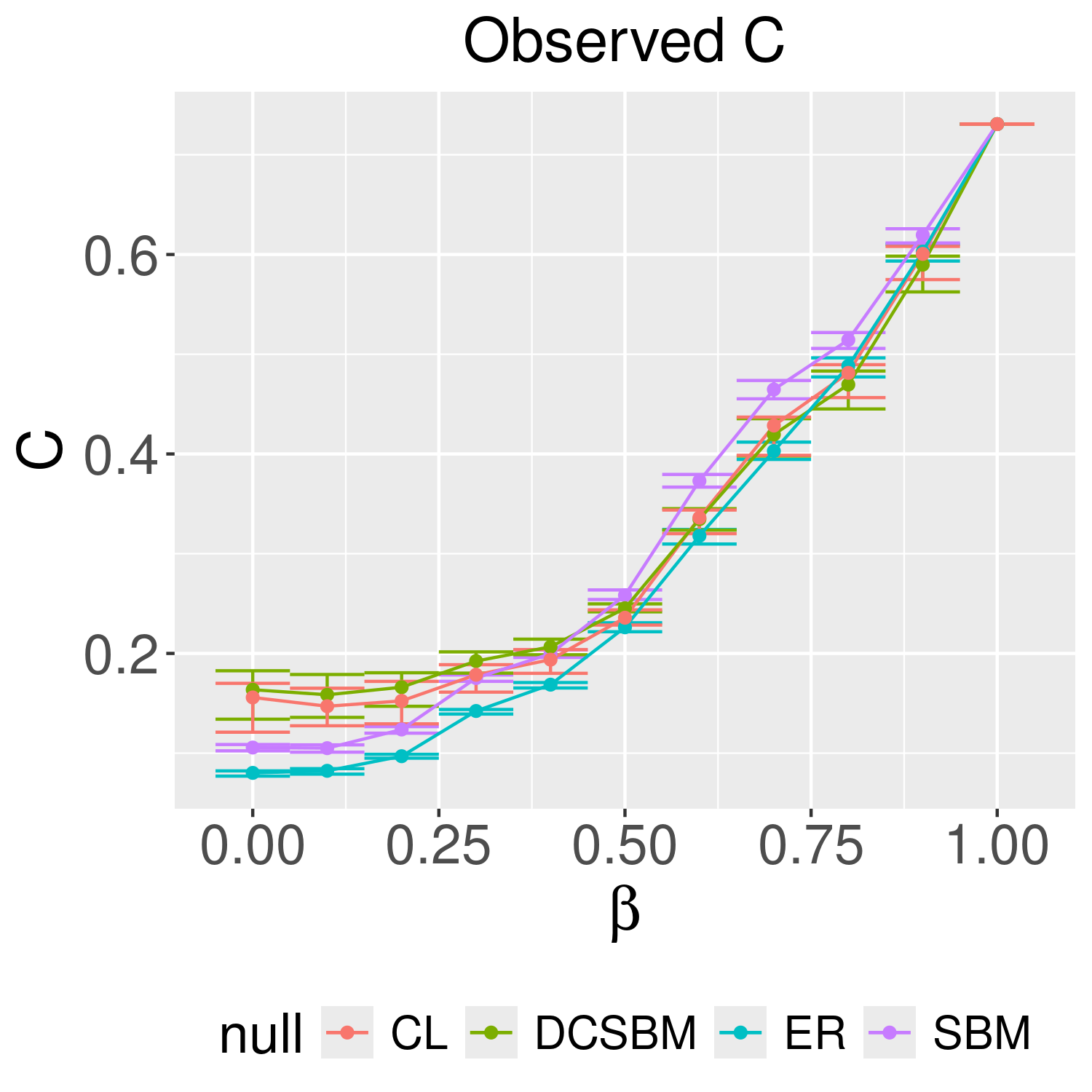}
\end{subfigure}
\begin{subfigure}{0.31 \linewidth}
\includegraphics[width=\linewidth]{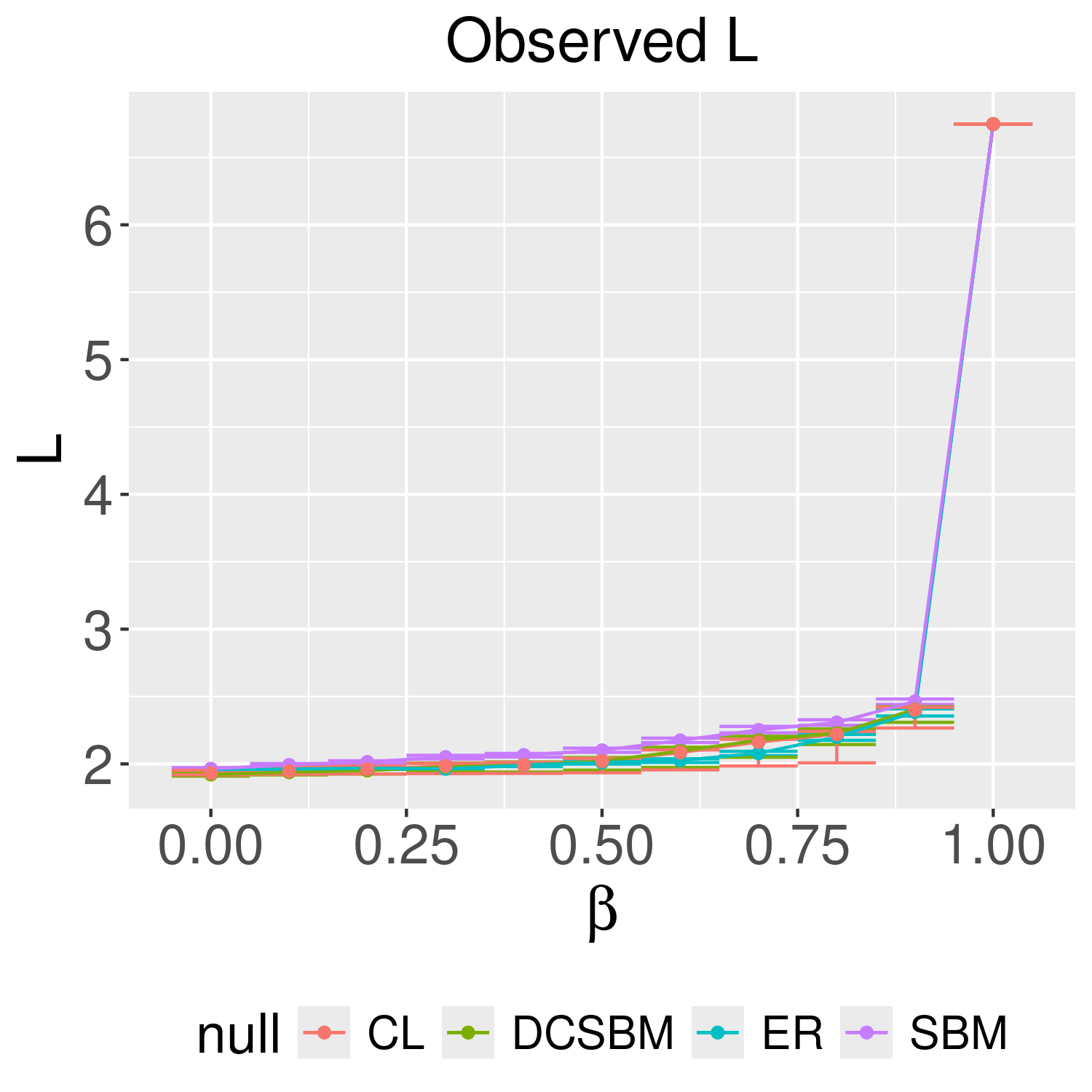}
\end{subfigure}%
\begin{subfigure}{0.31 \linewidth}
\includegraphics[width=\linewidth]{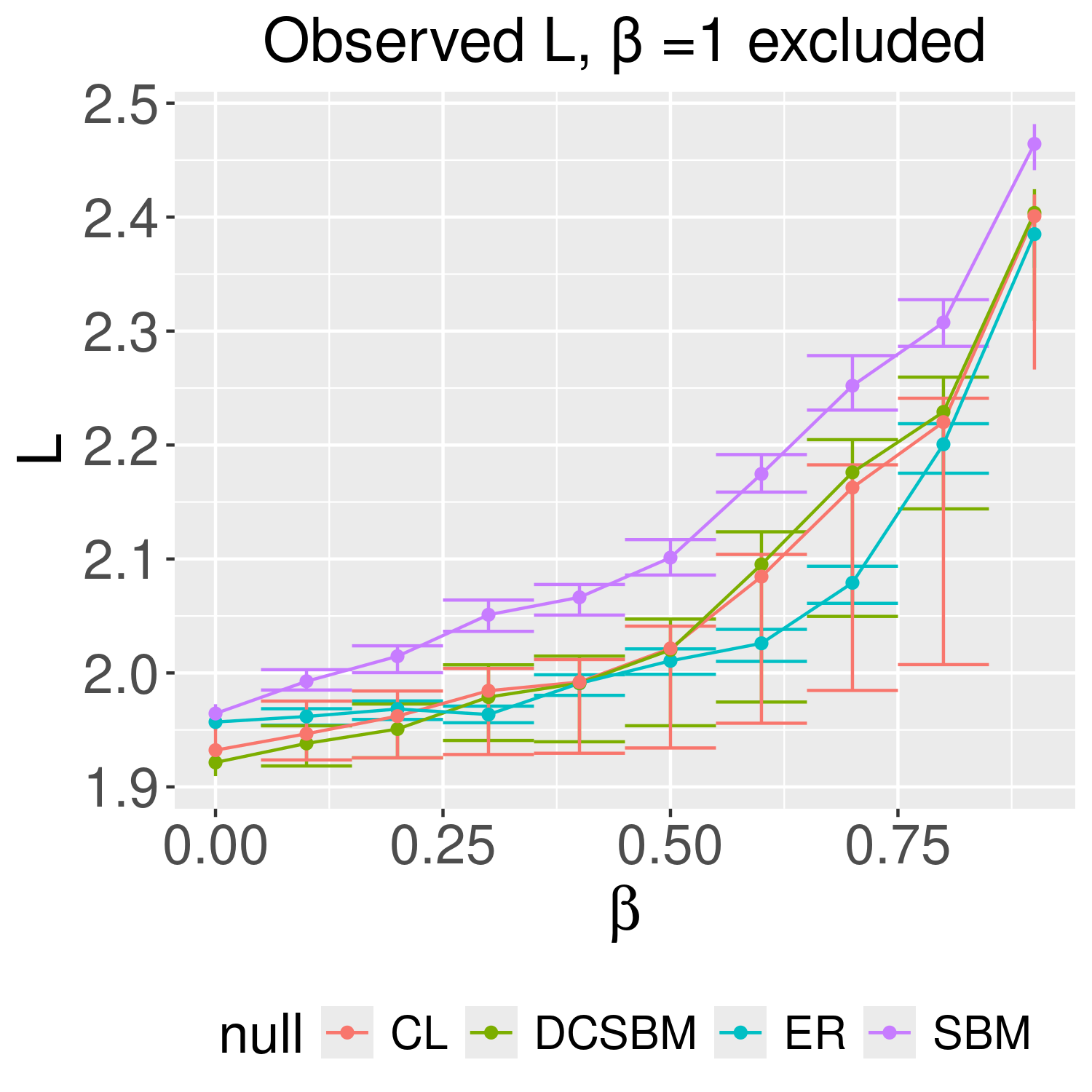}
\end{subfigure}
\caption{Median along with 1\% and 99\% quantiles for observed distribution of $C$ and $L$ with increasing $\beta$ for ER, CL, SBM, and DCSBM.}
\label{CLdist}
\end{figure*}

\begin{figure*}
\centering
\begin{subfigure}{0.31 \linewidth}
\includegraphics[width=\linewidth]{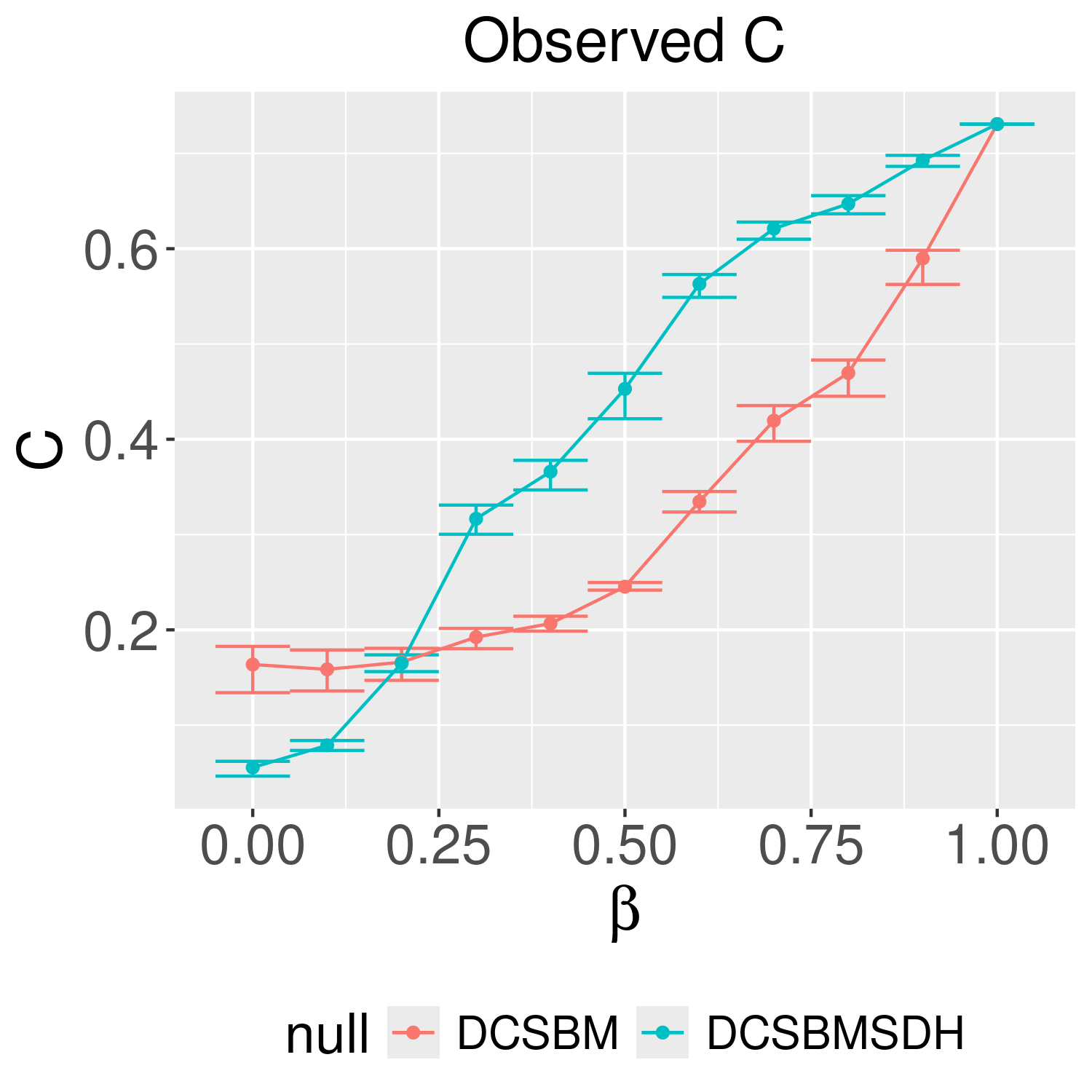}
\end{subfigure}
\begin{subfigure}{0.31 \linewidth}
\includegraphics[width=\linewidth]{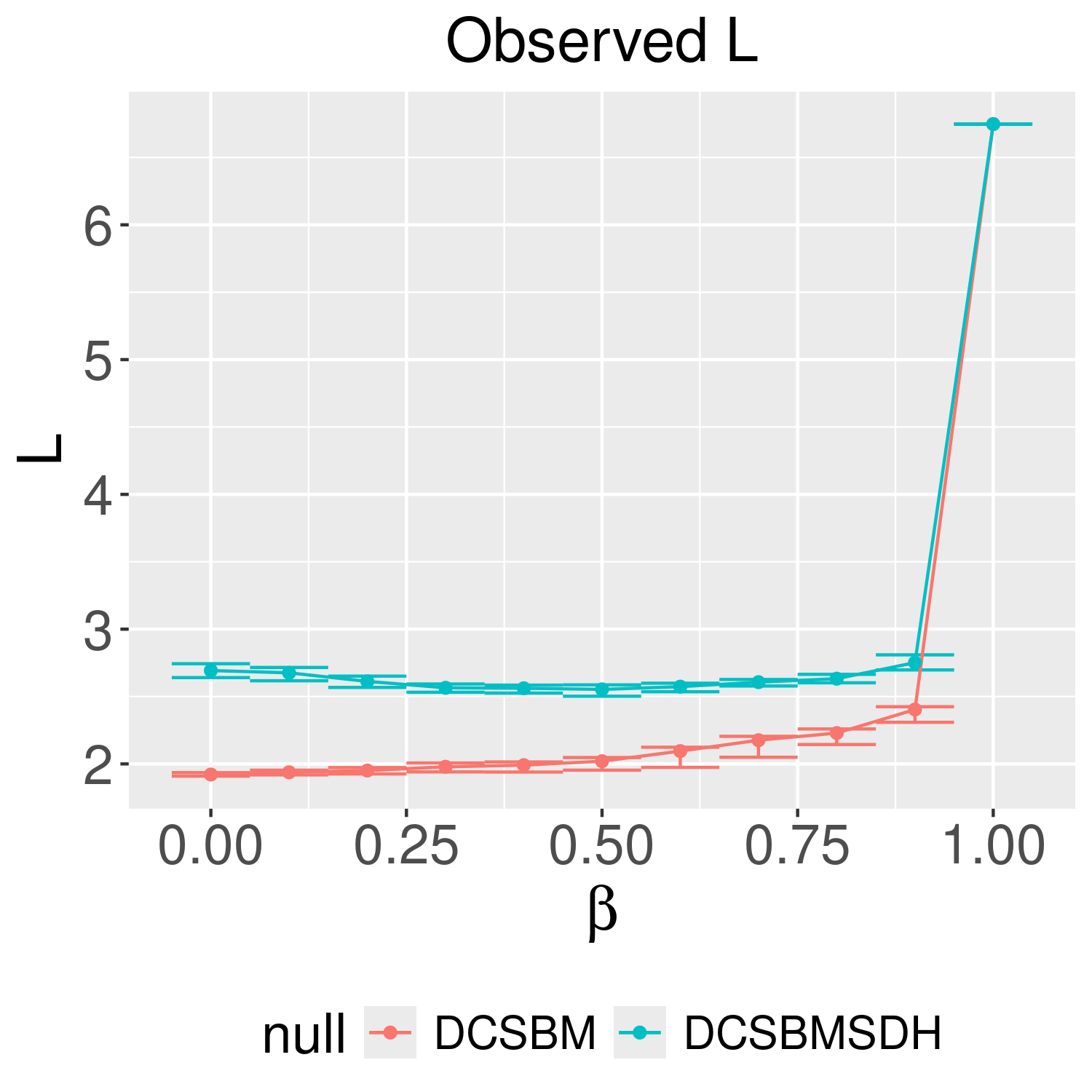}
\end{subfigure}%
\begin{subfigure}{0.31 \linewidth}
\includegraphics[width=\linewidth]{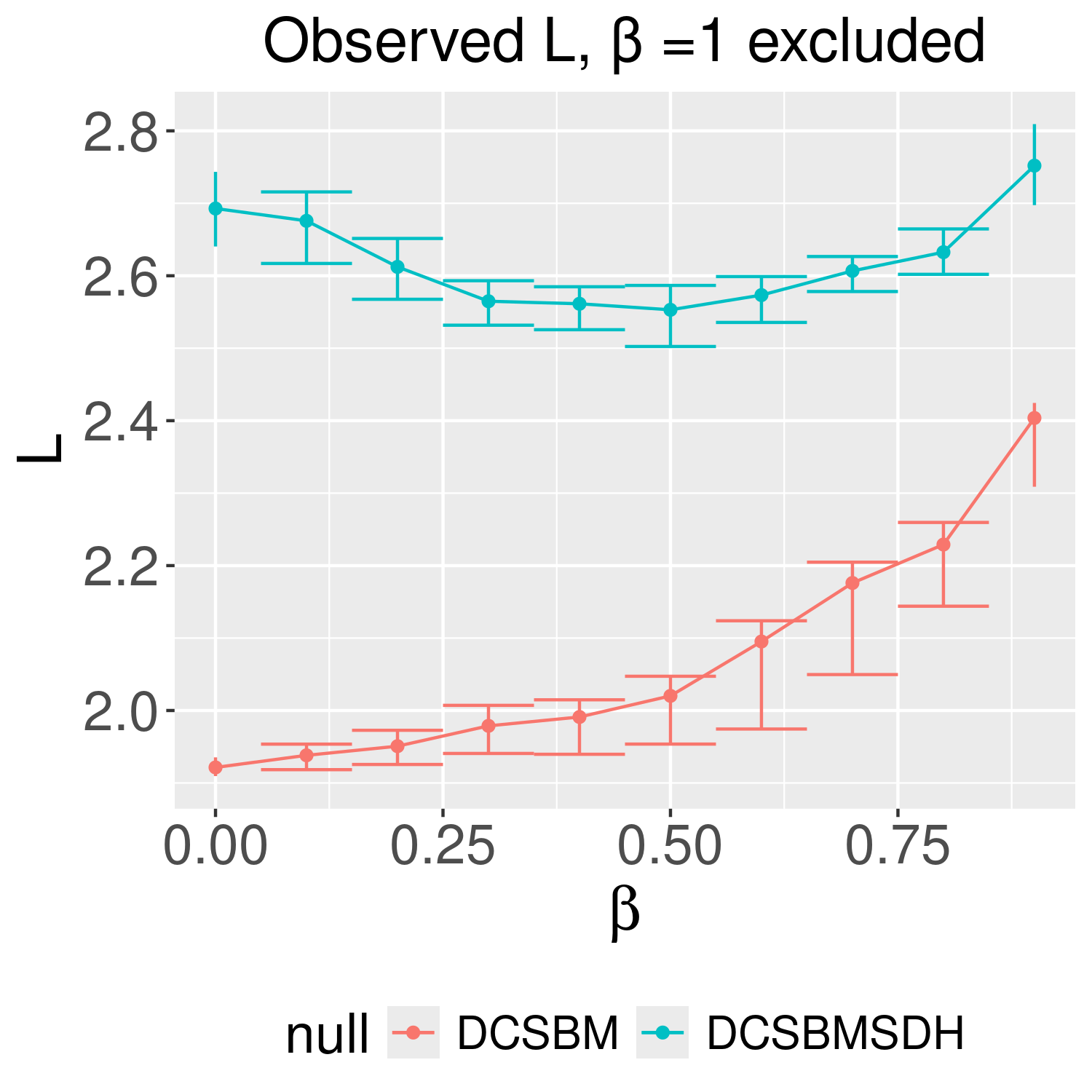}
\end{subfigure}
\caption{Median along with 1\% and 99\% quantiles for observed distribution of $C$ and $L$ with increasing $\beta$ for DCSBM with low (orange) and high (blue) degree heterogeneity.}
\label{CLdcbm}
\end{figure*}

\subsection{Distribution of $C$ and $L$ under the superimposed model in simulation}

To understand the changes in the distribution of $C$ and $L$ as $\beta$ is varied we perform a simulation study. We generate data from the superimposed Newman Watts models with ER, CL, SBM and DCSBM null models by varying $\beta$. We keep $n=500$ and $2\delta = 40$ and generate 150 networks for each $\beta$ value. Figure \ref{CLdist}(a) and (b) presents the median along with 1\% and 99\% quantiles of the observed values of $C$ and $L$ over these 150 networks respectively. Figure \ref{CLdist}(c) is the same figure as Figure \ref{CLdist}(b), but without $\beta=1$ to better observe the differences for smaller $\beta$ values. We make a number of observations from these figures. First, as $\beta$ increases both the 1\% and 99\% quantiles of $C$ values steadily increases for all null models and the 1\% quantiles for $\beta >0$ quickly become larger than the 99\% quantiles of $\beta=0$. On the other hand for $L$ the increase in the 1\% and 99\% quantiles is slower with increasing $\beta$, and in fact the 1\% quantiles for $\beta>0$ remain smaller than the 99\% quantiles for $\beta=0$ for many values of $\beta$ until eventually at $\beta=1$, the values increase rapidly. This gives credence to the fact that there is a range of $\beta$ values where $C$ is large compared to $\beta=0$ while $L$ is comparable to $\beta=0$. Second, we find differences in behavior of the different null models. Both $C$ and $L$ are highly concentrated around their median for the NW-ER model for all values of $\beta$. However, for NW models which also account for degree heterogeneity, i.e., NW-CL and NW-DCSBM, the intervals between 1\% and 99\% quantiles are quite large. This is especially the case for $L$. Therefore for NW-CL model, we note that there is a large range of $\beta$ values for which 1\% quantile of $C$ is larger than 99\% quantile of $\beta=0$, while the 1\% quantile of $L$ is smaller than the 99\% quantile of $\beta=0$. Finally, comparing among the null models, it appears that NW-SBM generally produces higher median $C$ and $L$ values compared to NW-ER for almost all $\beta$ values. The NW-CL and NW-DCBM models also produce higher median $C$ and $L$ values compared to NW-ER, however, the range of values between 1\% and 99\% quantiles is very wide.

We perform another simulation to assess the impact of severe degree heterogeneity.  To generate the DCSBM part of the superimposed networks with severe degree heterogeneity, we use the setup in \cite{jin2022mixedmembershipestimationsocial} and generate $1/\theta_i \sim U(1, 8)$, where $\theta_i$ is the degree heterogeneity parameter for node $i$. We obtain quantiles by generating 150 networks with $n=500$ and $2\delta = 40$ over different values of $\beta$. The results are presented in Figure \ref{CLdcbm}, where the blue curve represents the case with severe degree heterogeneity. We note that as $\beta$ increases, both 1\% and 99\% quantiles of $C$ for networks with severe degree heterogeneity increase at a faster rate compared to networks with low degree heterogeneity. On the other hand, both quantiles for $L$ first reduce up to $\beta \approx 0.5$ before increasing again in the severe degree heterogeneity case. 

\begin{figure*}
\begin{subfigure}{0.4 \linewidth}
\includegraphics[width=\linewidth]{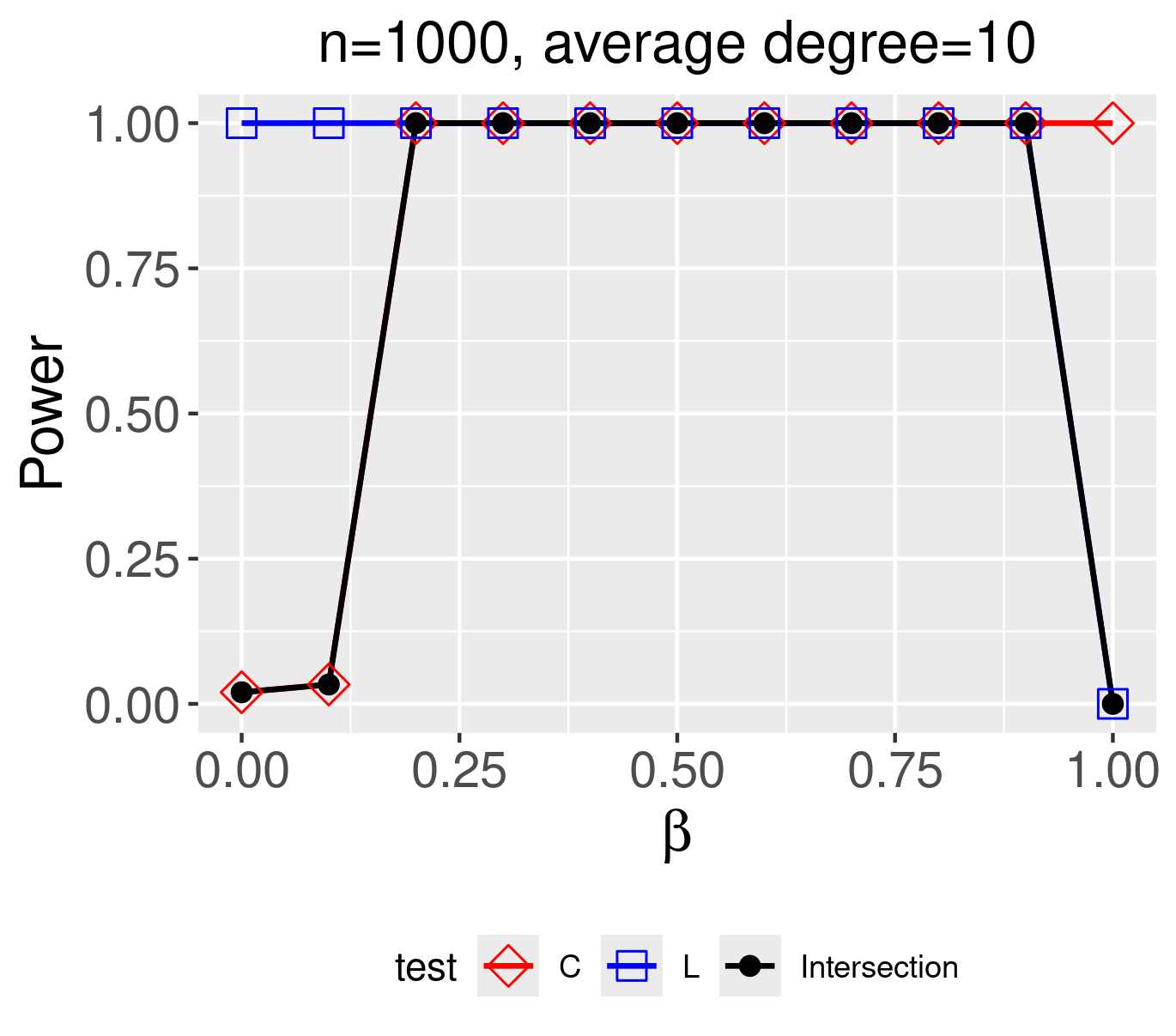}
\end{subfigure}%
\begin{subfigure}{0.4 \linewidth}
\includegraphics[width=\linewidth]{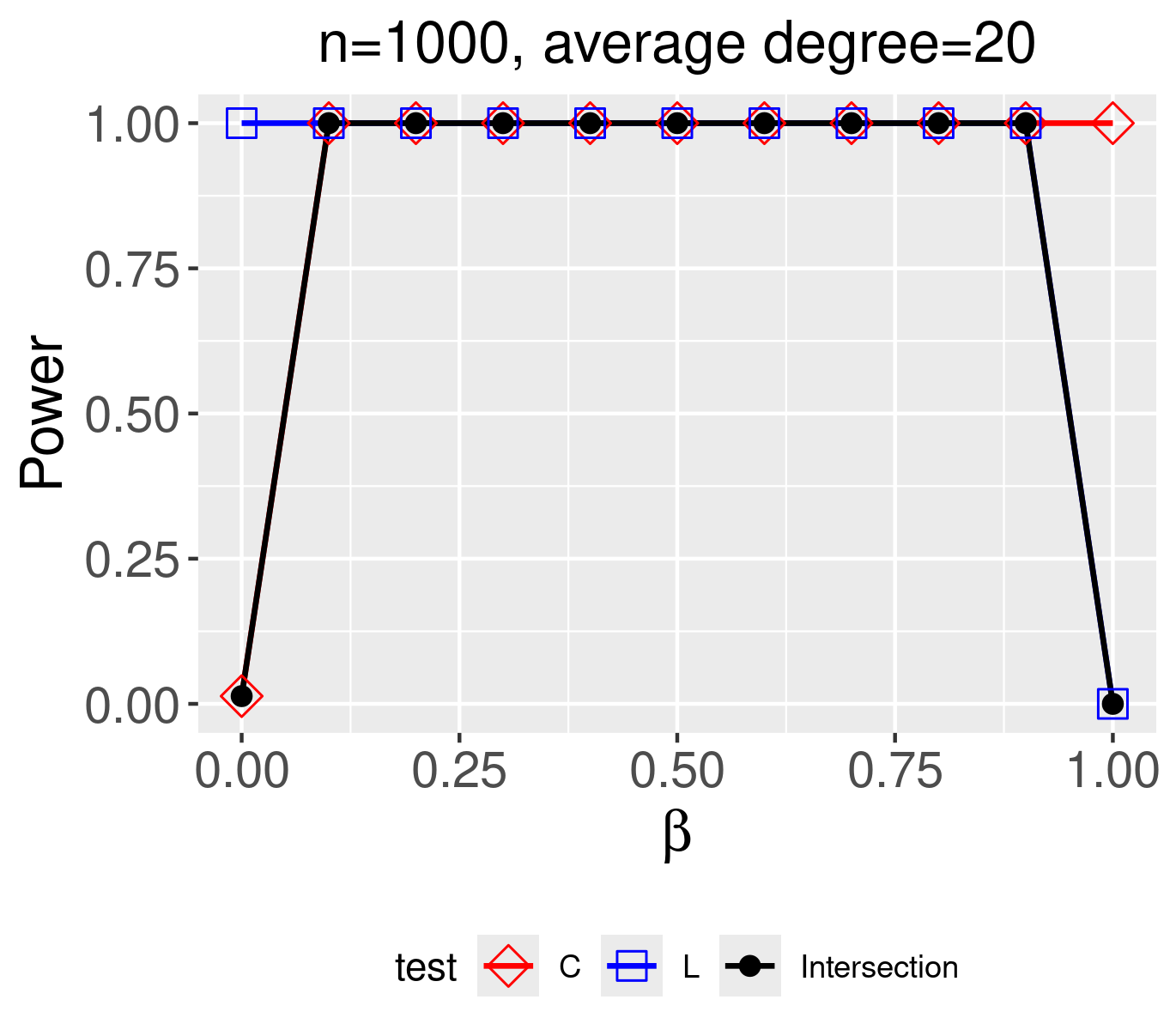}
\end{subfigure}
\begin{subfigure}{0.4 \linewidth}
\includegraphics[width=\linewidth]{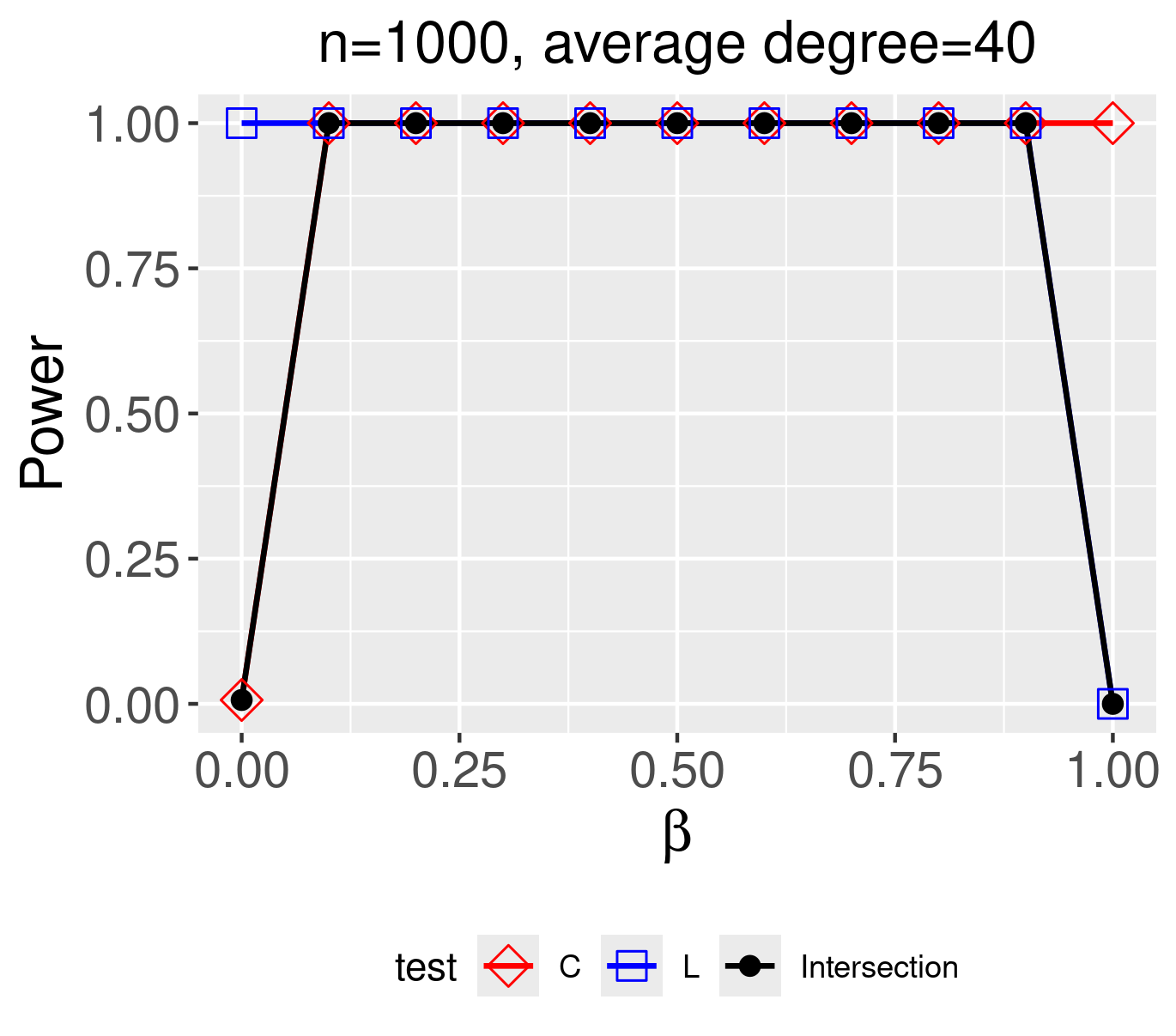}
\end{subfigure}%
\begin{subfigure}{0.4 \linewidth}
\includegraphics[width=\linewidth]{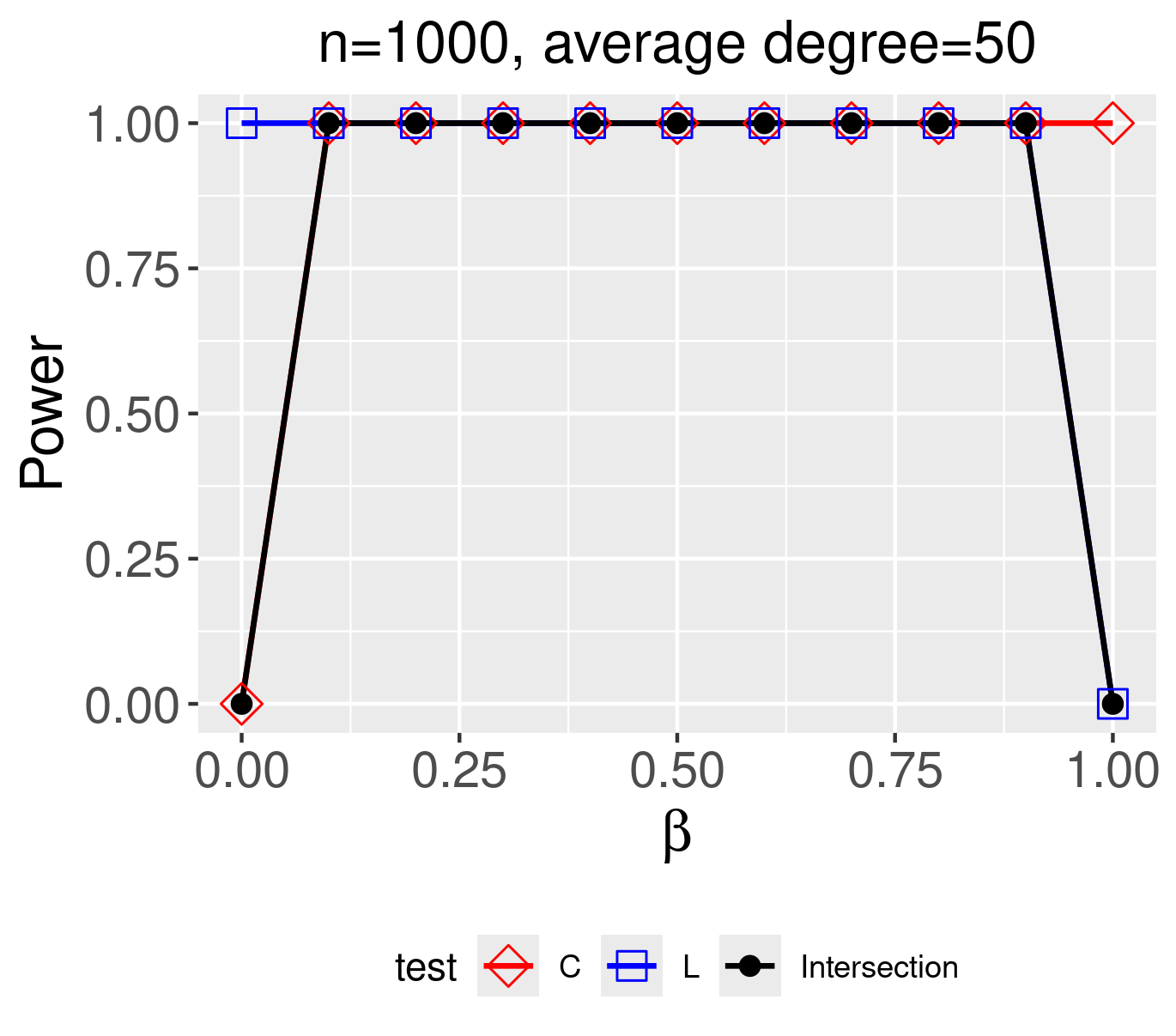}
\end{subfigure}

\begin{subfigure}{0.4 \linewidth}
\includegraphics[width=\linewidth]{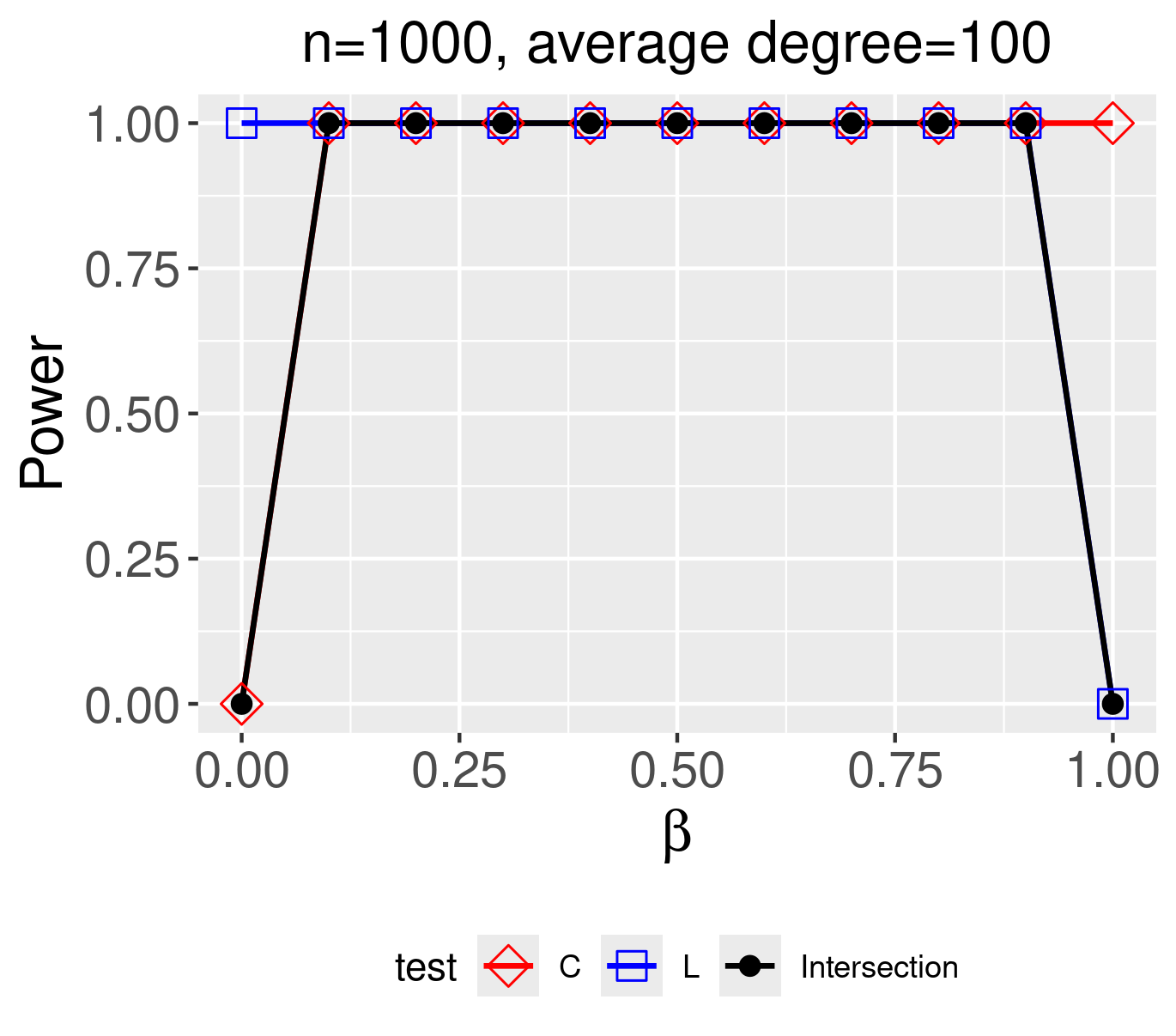}
\end{subfigure}%
\begin{subfigure}{0.4 \linewidth}
\includegraphics[width=\linewidth]{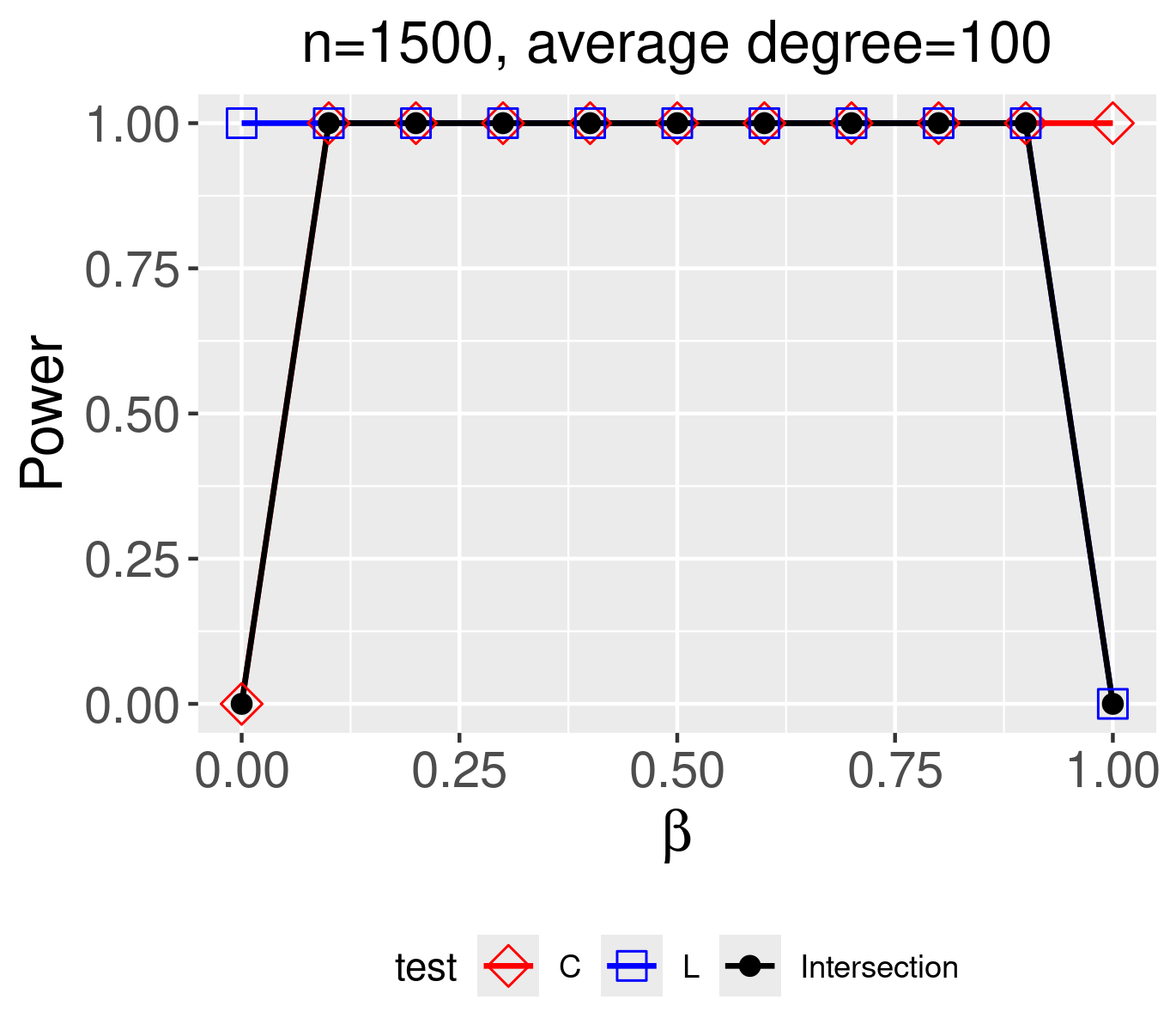}
\end{subfigure}
\begin{subfigure}{0.4 \linewidth}
\includegraphics[width=\linewidth]{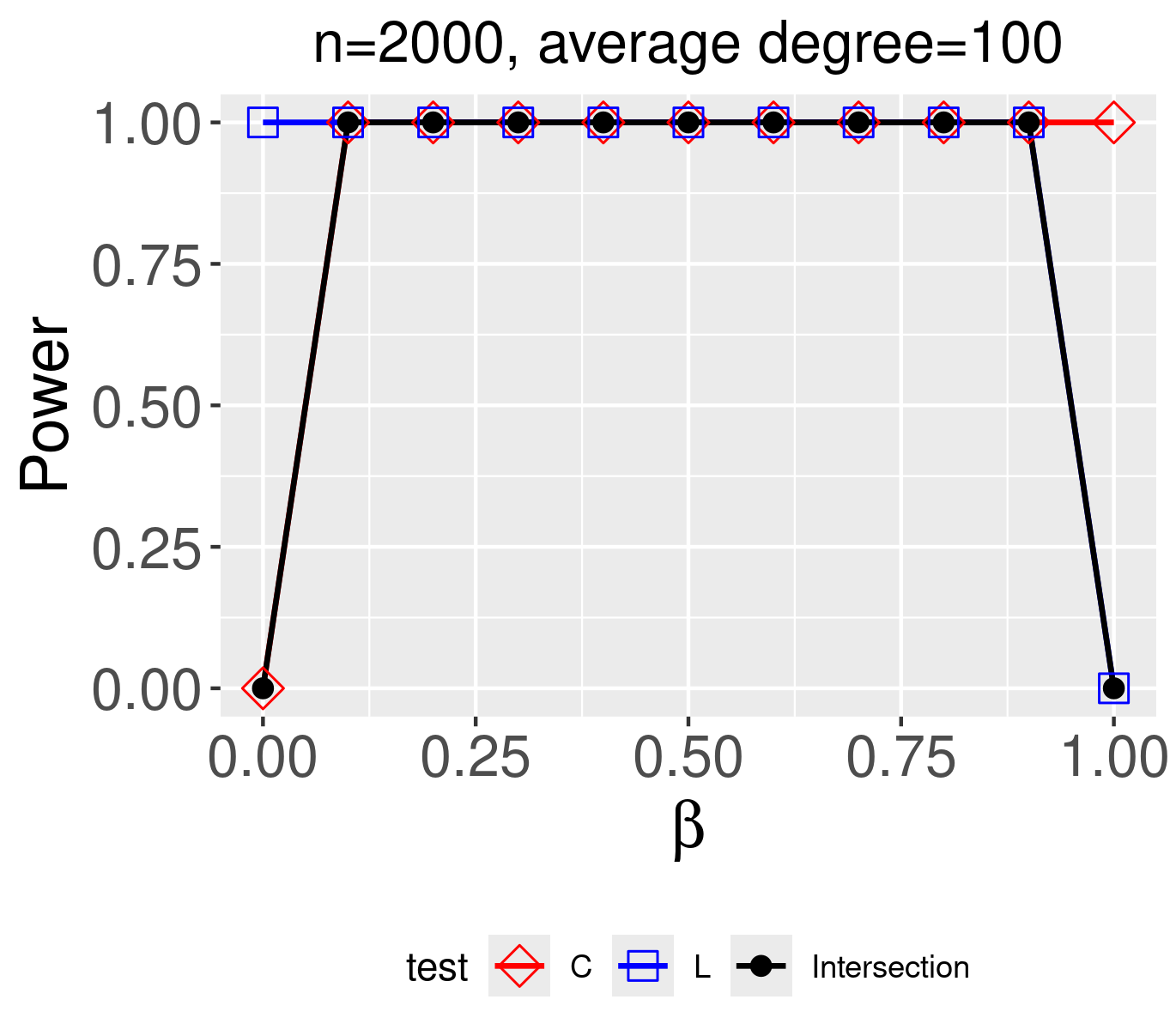}
\end{subfigure}%
\begin{subfigure}{0.4 \linewidth}
\includegraphics[width=\linewidth]{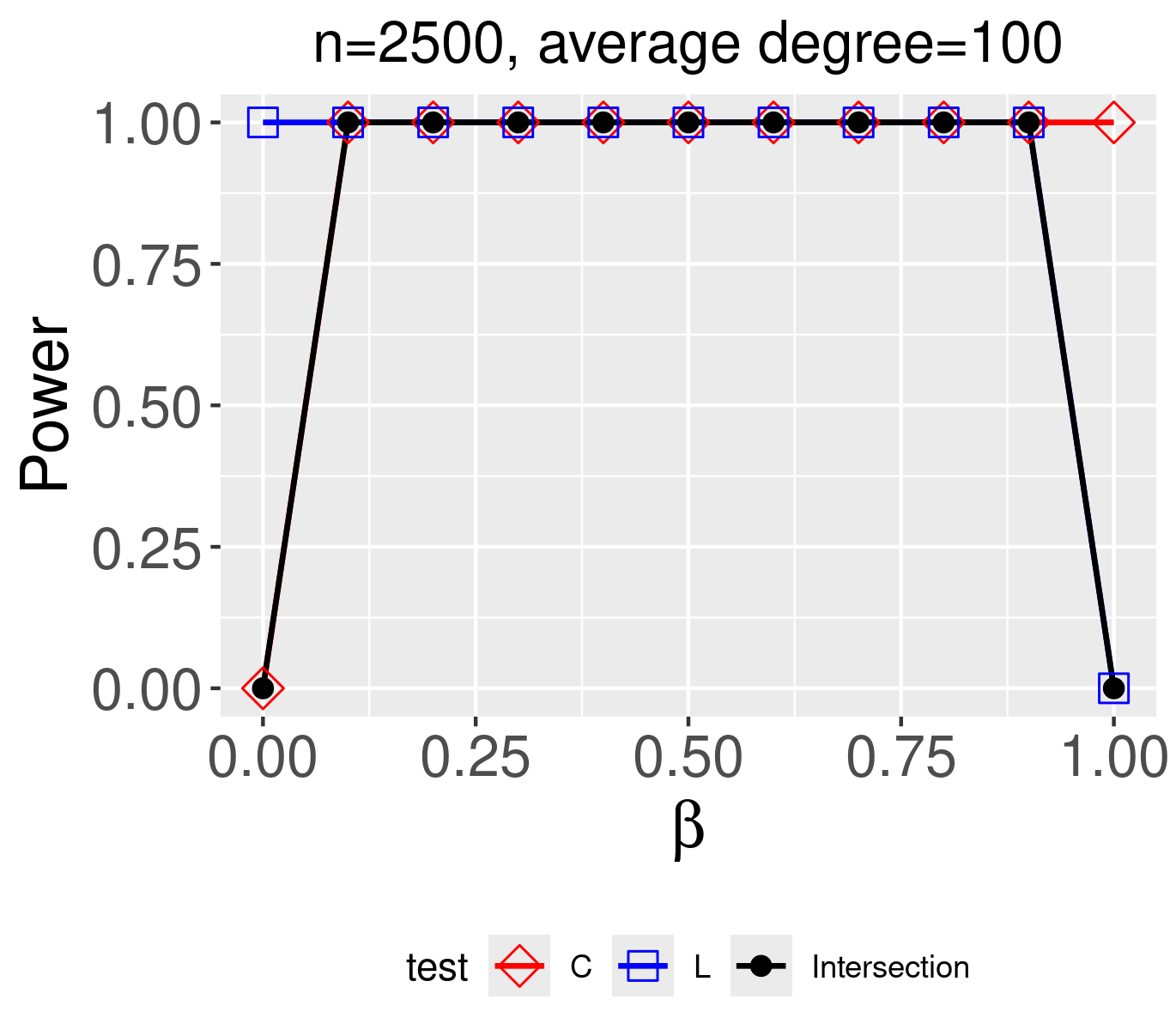}
\end{subfigure}%
    \caption{Power curves for the asymptotic test. The X-axis represents values of $\beta$. In the first 4 figures, we fix $n$ at 1000, and vary average degree from $10$ to $50$, and in the last 4 figures, we fix average degree at 100 and vary $n$ from 1000 to 2500.  The red curve represents the observed rate of rejection of the clustering rule, i.e., fraction of simulated networks with $C>K_{\alpha}$, the blue curve represents the fraction of simulated networks with $L<K_2$, while the black curve represents the empirical power of the intersection test, i.e., $[C>K_{\alpha}, L<K_2]$.}
    \label{fig:A-power}
\end{figure*}

\begin{figure*}
\begin{subfigure}{0.4 \linewidth}
\includegraphics[width=\linewidth]{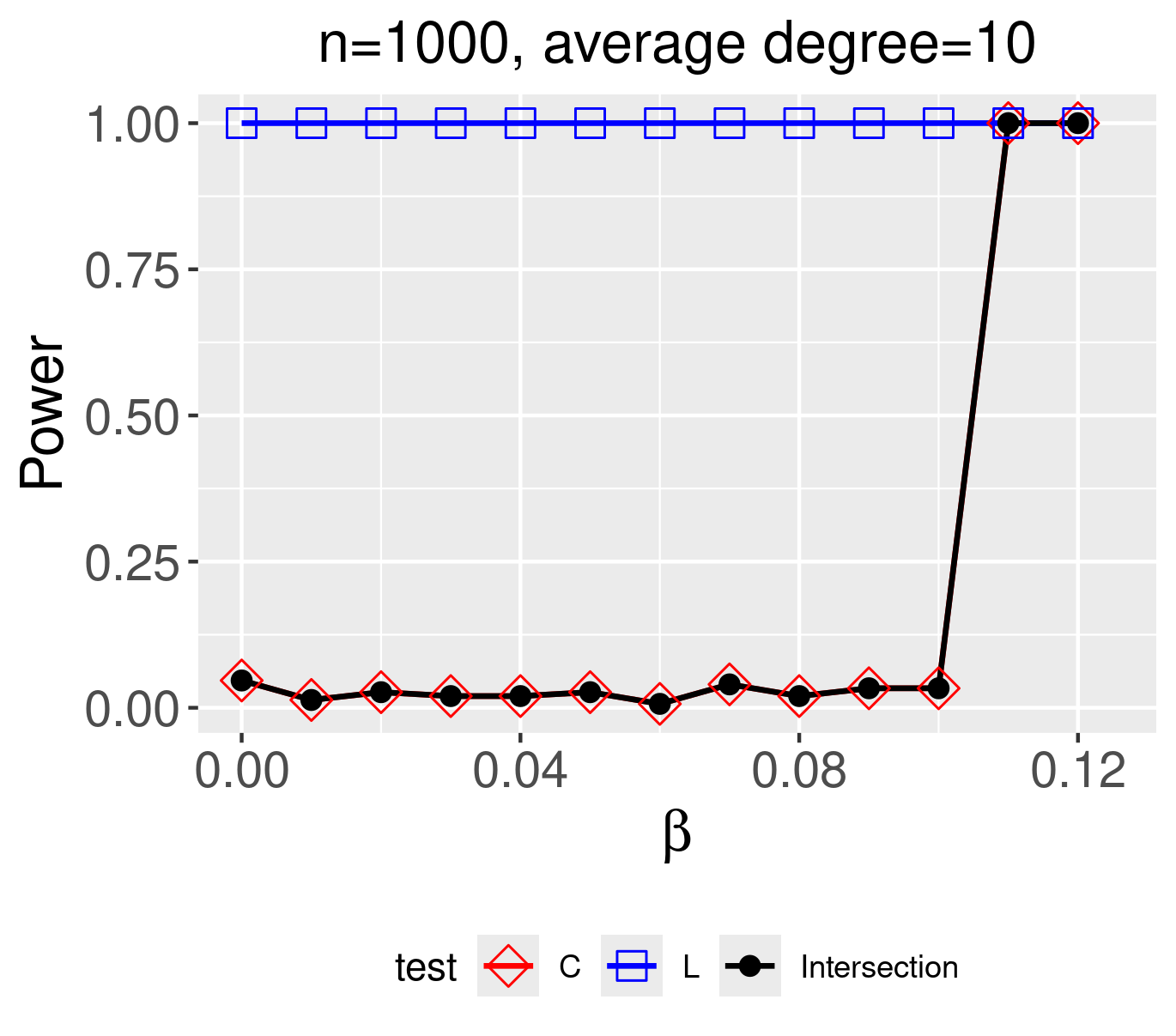}
\end{subfigure}%
\begin{subfigure}{0.4 \linewidth}
\includegraphics[width=\linewidth]{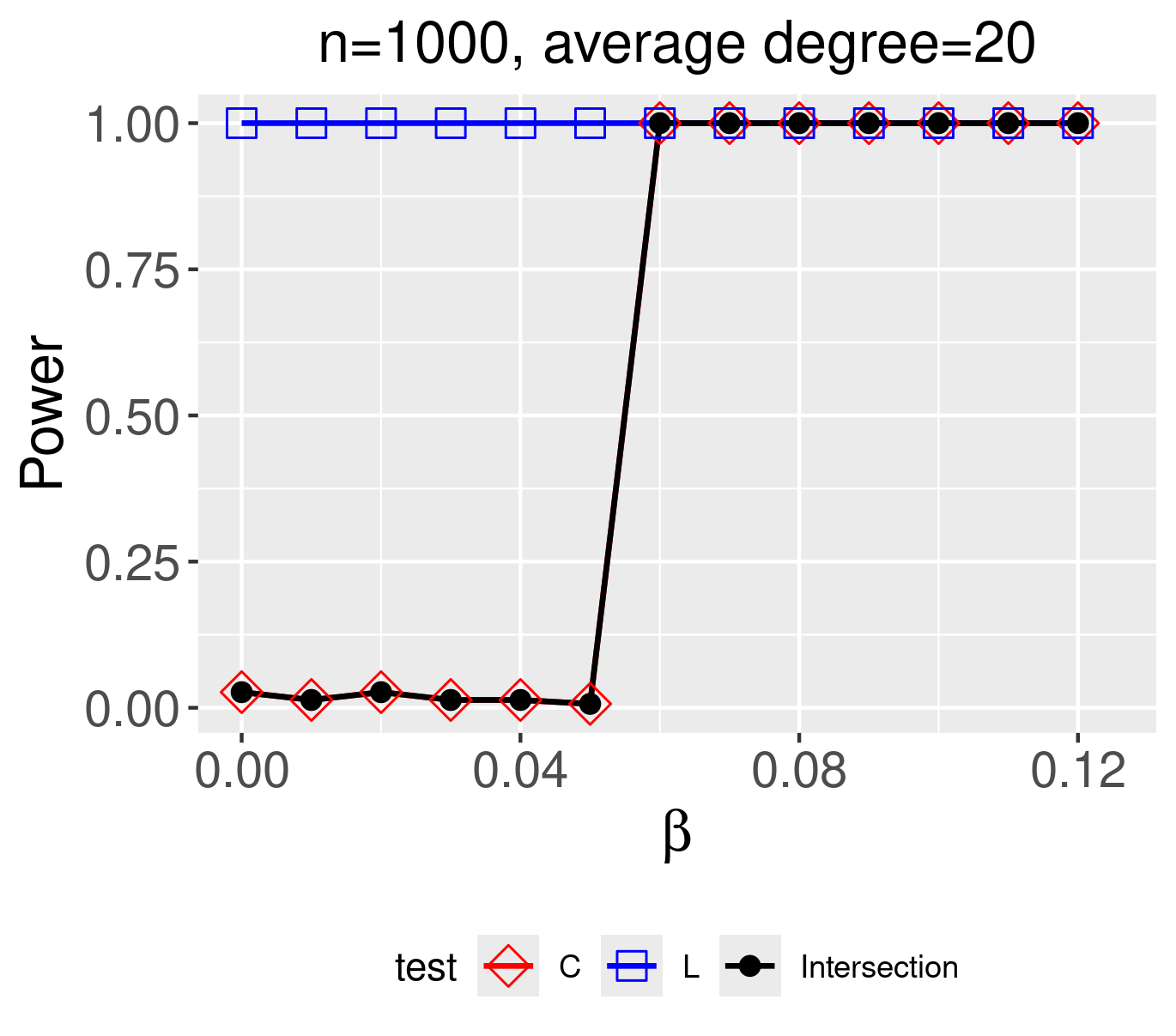}
\end{subfigure}
\begin{subfigure}{0.4 \linewidth}
\includegraphics[width=\linewidth]{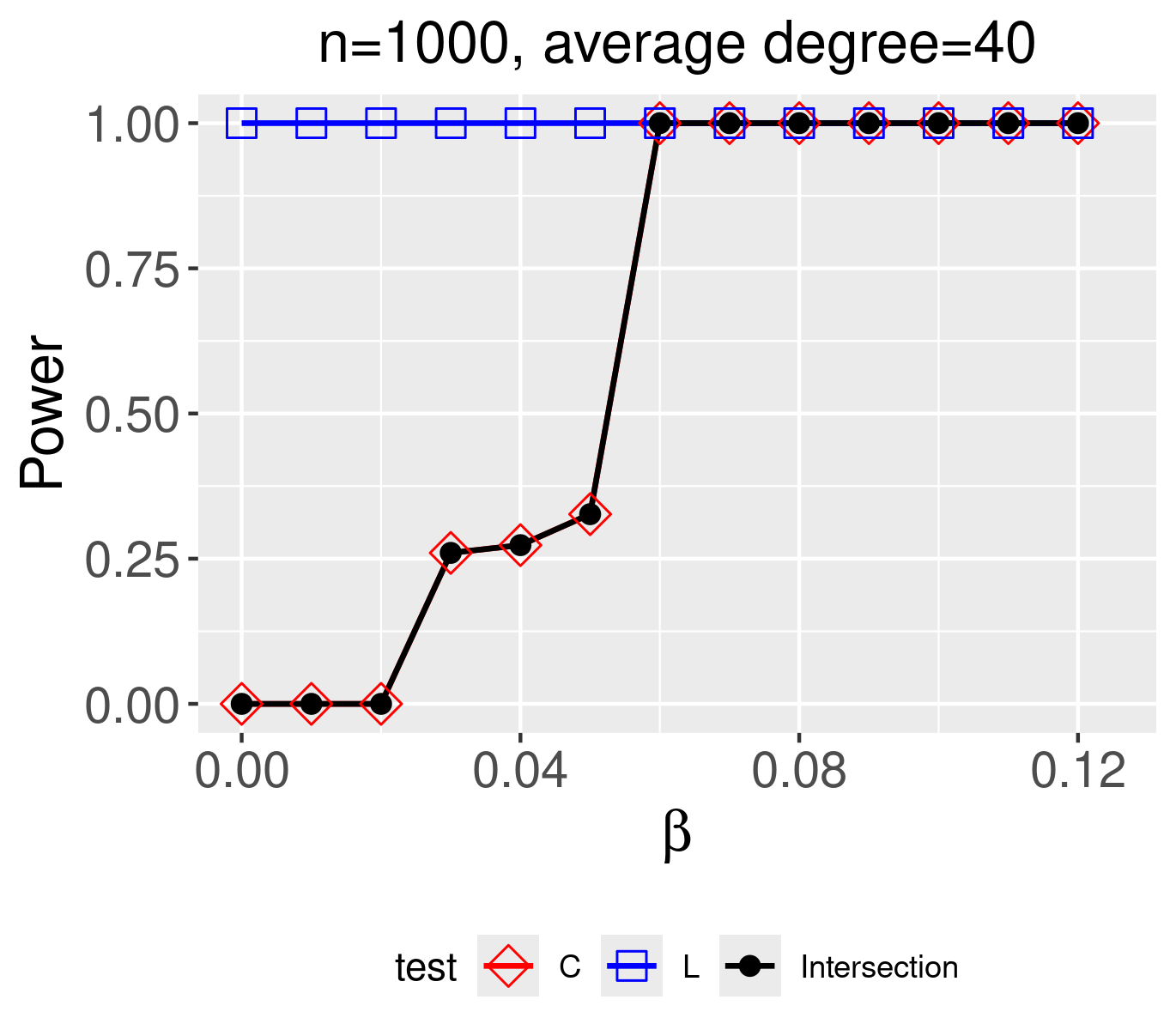}
\end{subfigure}%
\begin{subfigure}{0.4 \linewidth}
\includegraphics[width=\linewidth]{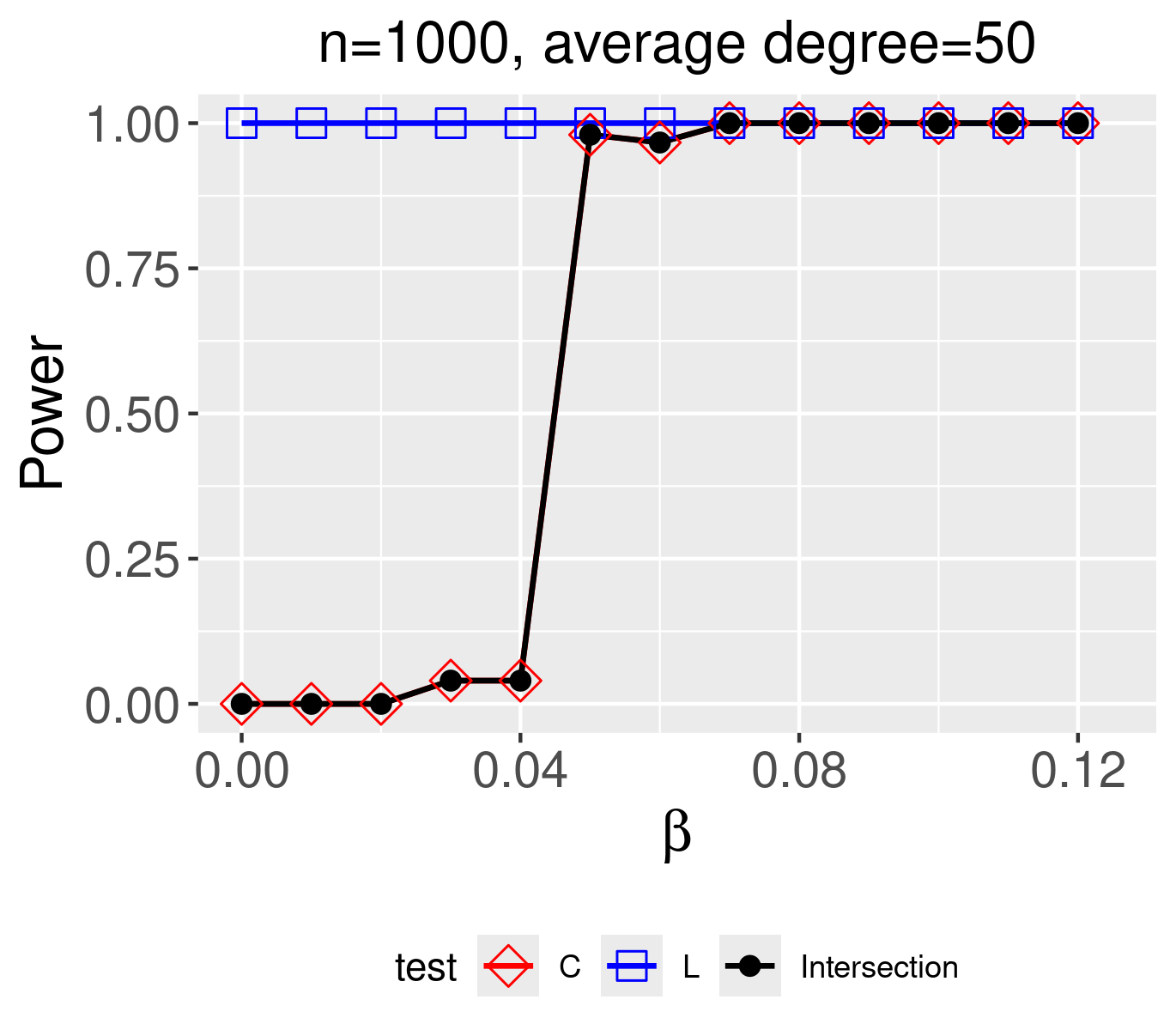}
\end{subfigure}

\begin{subfigure}{0.4 \linewidth}
\includegraphics[width=\linewidth]{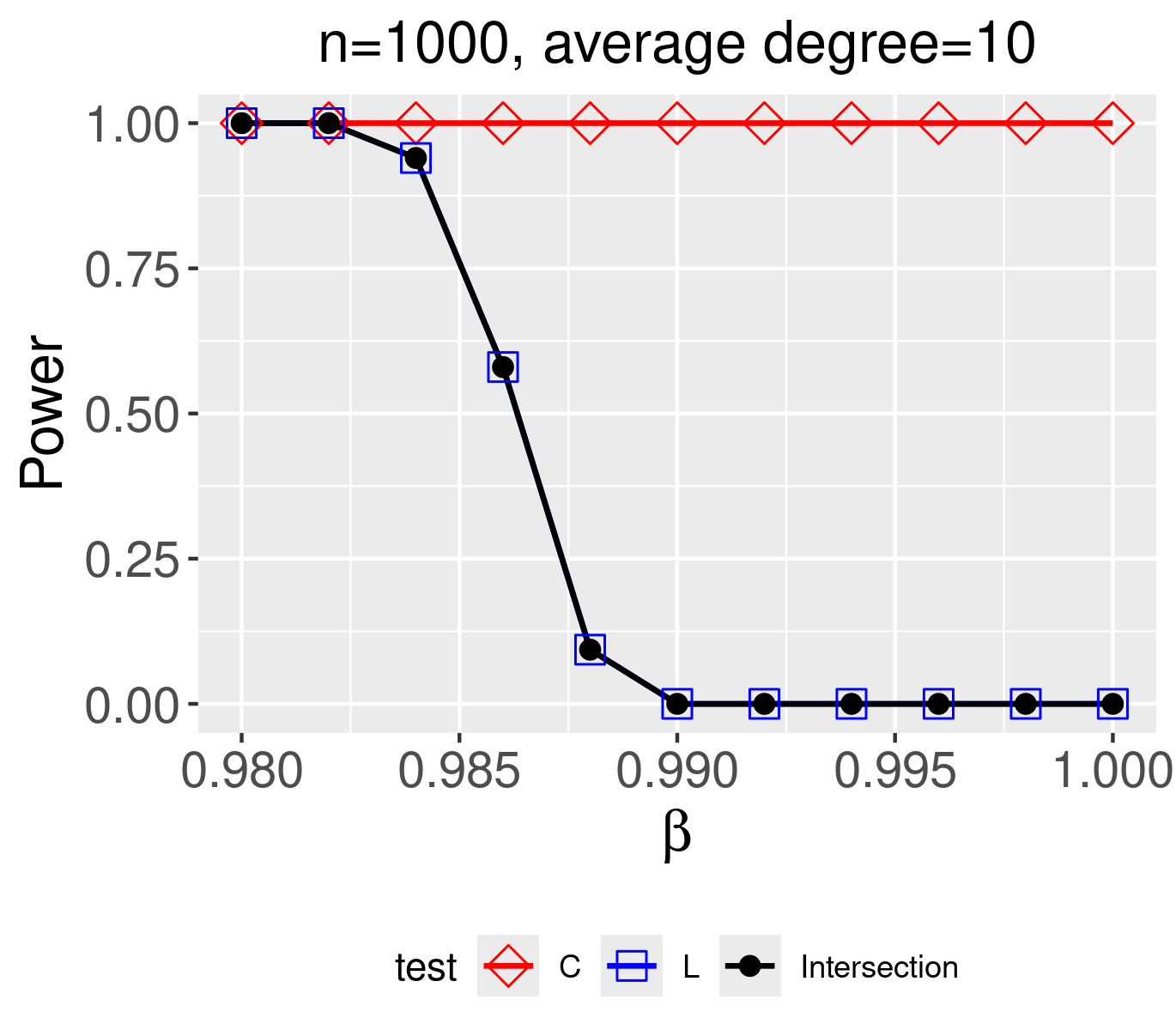}
\end{subfigure}%
\begin{subfigure}{0.4 \linewidth}
\includegraphics[width=\linewidth]{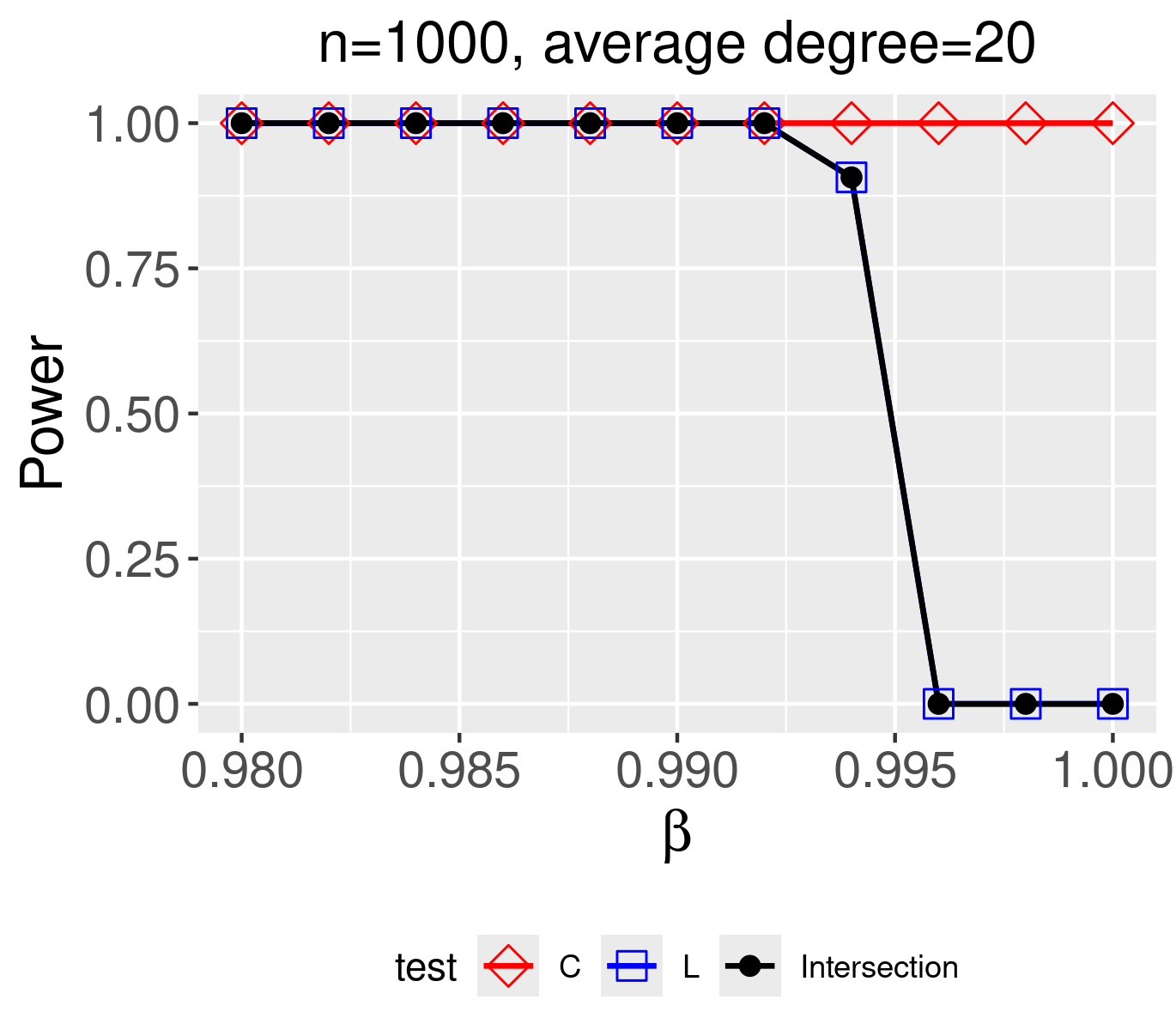}
\end{subfigure}
\begin{subfigure}{0.4 \linewidth}
\includegraphics[width=\linewidth]{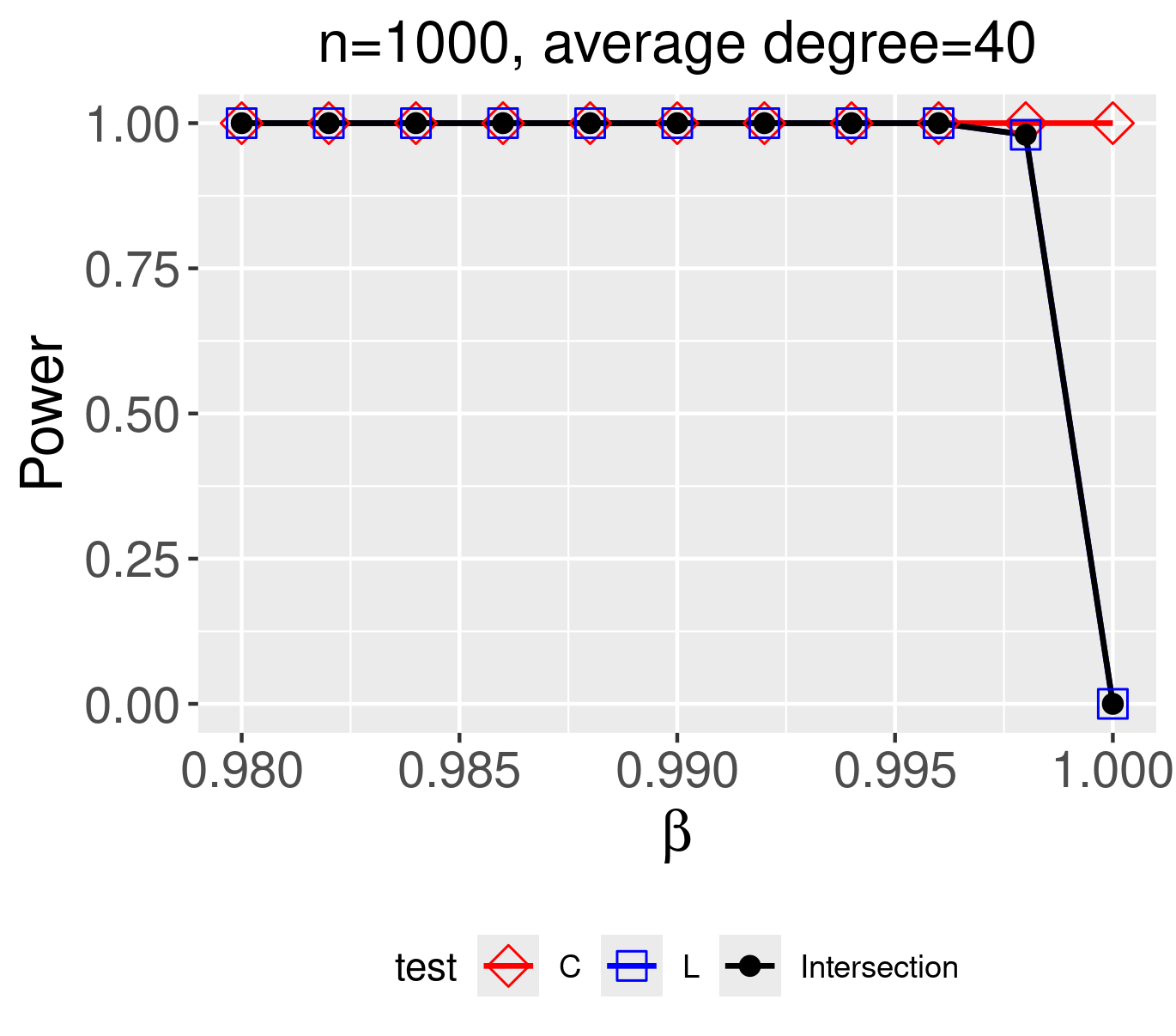}
\end{subfigure}%
\begin{subfigure}{0.4 \linewidth}
\includegraphics[width=\linewidth]{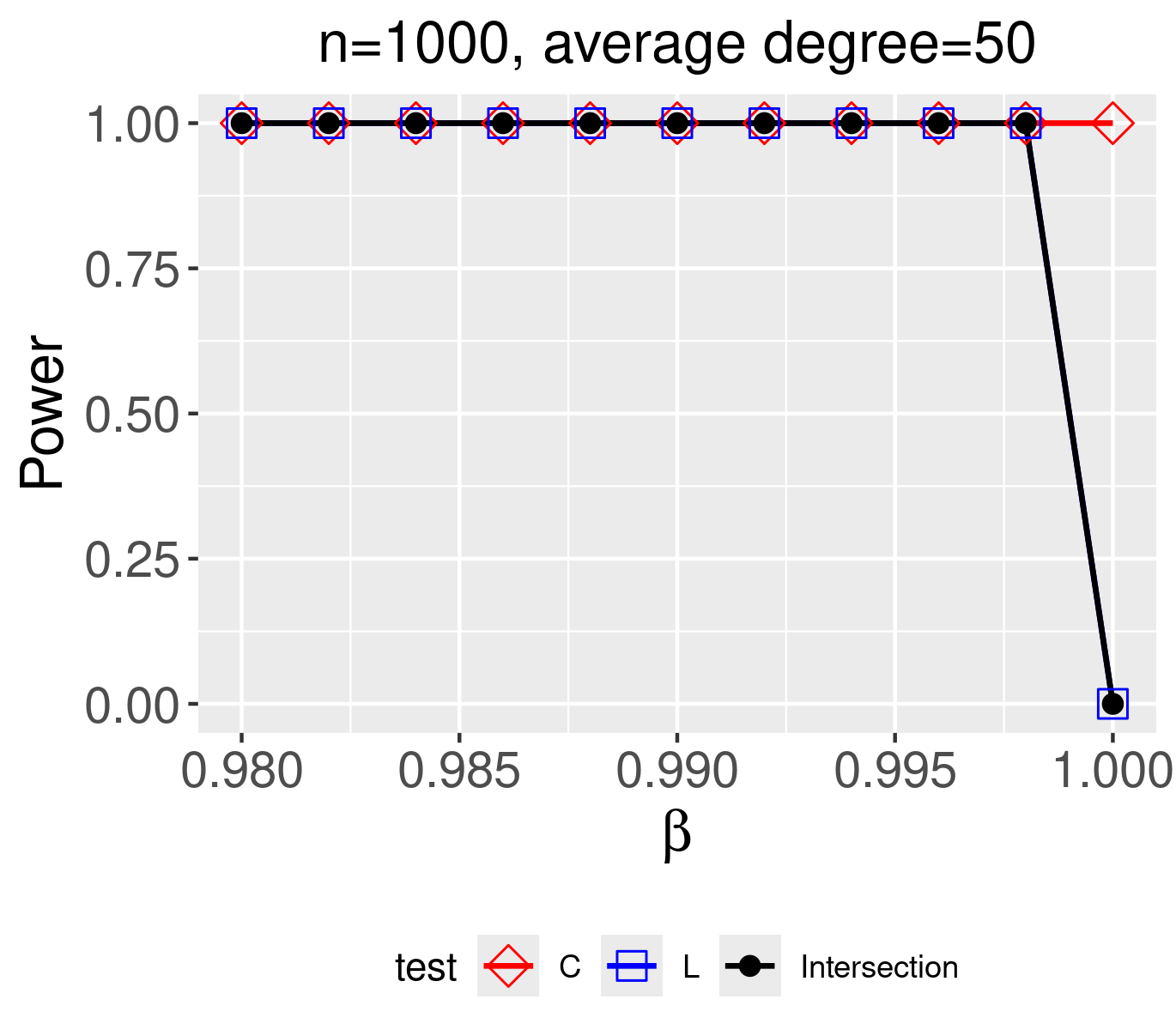}
\end{subfigure}%

    \caption{Power curves for the asymptotic test with $n=1000$ and varying average degree $10, 20, 40, 50$ for values of $\beta$ close to either 0 (the first 4 figures) or 1 (the last 4 figures. The red curve represents the observed rate of rejection of the clustering rule, i.e., fraction of simulated networks with $C>K_{\alpha}$, the blue curve represents the fraction of simulated networks with $L<K_2$, while the black curve represents the empirical power of the intersection test, i.e., $[C>K_{\alpha}, L<K_2]$.}
    \label{fig:A-power-zoom}
\end{figure*}

\subsection{Power of the asymptotic test}
Next, we verify the power of the asymptotic test described through Theorems \ref{C-power} and \ref{L-power} in two sets of simulations. First, we fix $n=1000$ and vary the average degree (which is $2\delta$) as $10,20,30,40,50$. Then we fix average degree at $100$ and vary $n$ as $1000, 1500, 2000, 2500$. The power curves against changing $\beta$ using the asymptotic test are shown in Figure \ref{fig:A-power}. From the figure we note that both at $\beta =0$ and $\beta=1$, the rejection rate of the test is close to 0. The rejection rate curve (power curve) sharply increases to 1 after sufficiently large $\beta$ and stays close to 1 until $\beta=1$. As the average degree increases, the sharp increase in the power curve starts for $\beta$ closer to 0, and at $2\delta = 50$, the power curve is close to 1 for almost all value of $\beta$ in between 0 and 1. Comparing the left and right sides of the power curve, we see that the power goes to 1 slowly as $\beta$ goes away from 0, while it goes to 1 relatively faster as $\beta$ goes away from 1. We need a higher average degree or density of the graph for the power to go to 1 on the left-hand side of the power graph as opposed to on the right-hand side of the power graph.

\begin{figure}[!htbp]
\begin{subfigure}{0.3 \linewidth}
\includegraphics[width=\linewidth]{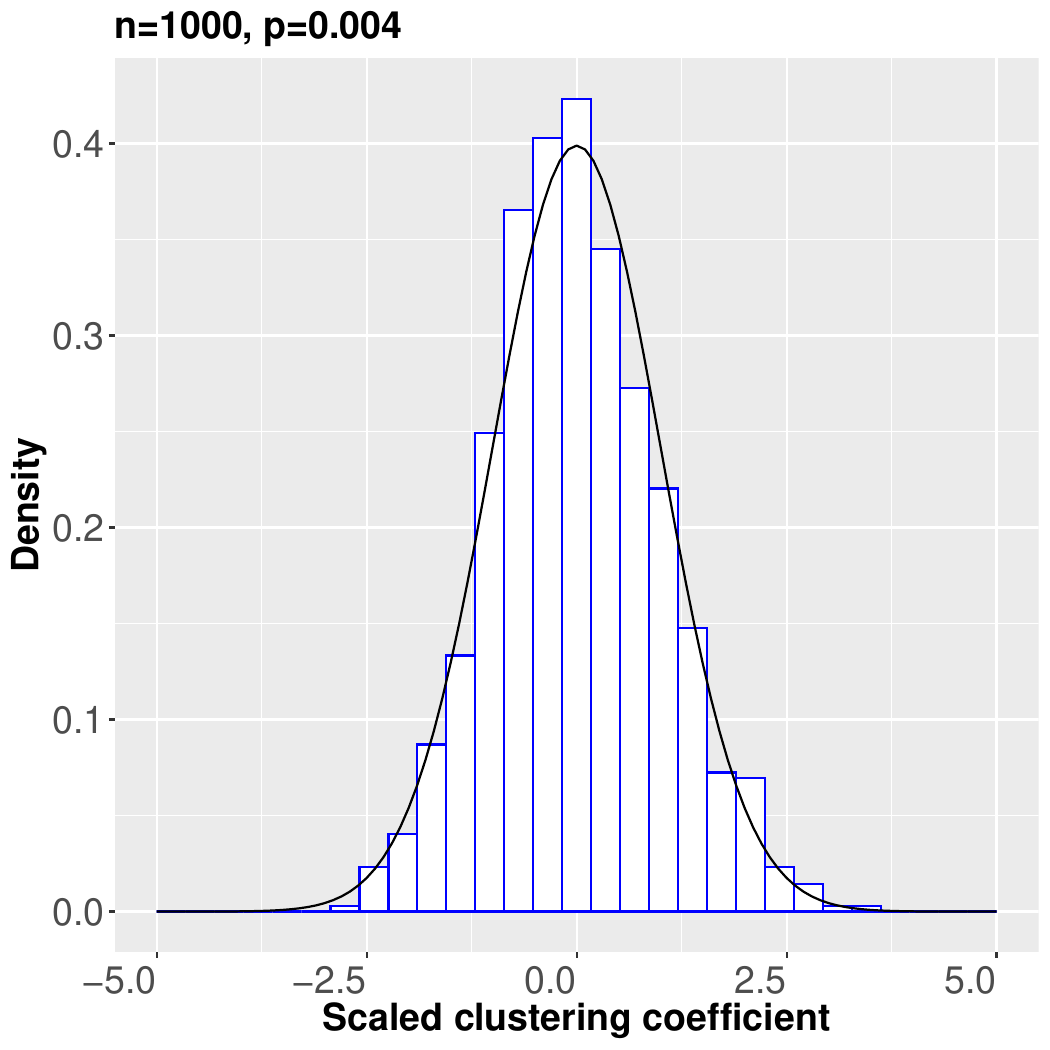}
\end{subfigure}%
\begin{subfigure}{0.3 \linewidth}
\includegraphics[width=\linewidth]{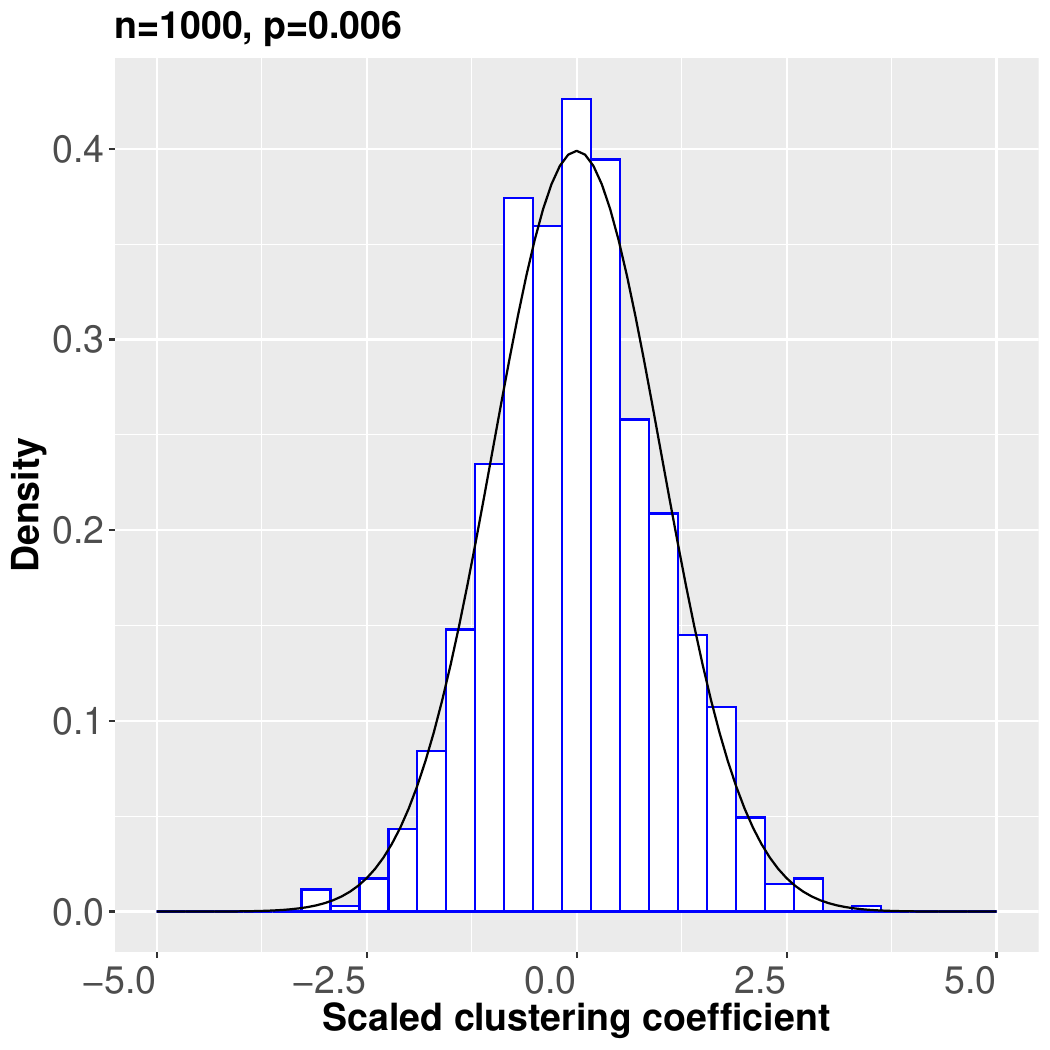}
\end{subfigure}%
\begin{subfigure}{0.3 \linewidth}
\includegraphics[width=\linewidth]{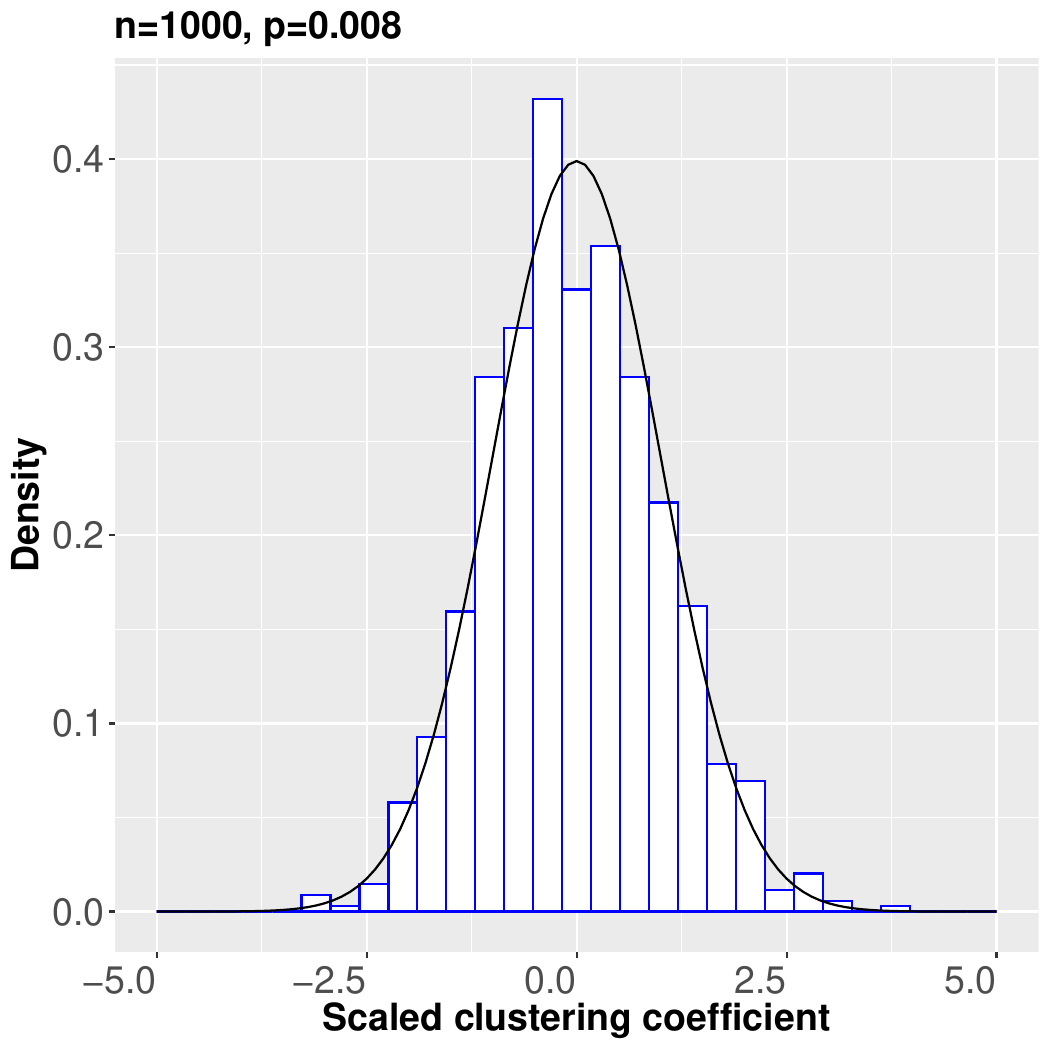}
\end{subfigure}
\begin{subfigure}{0.3 \linewidth}
\includegraphics[width=\linewidth]{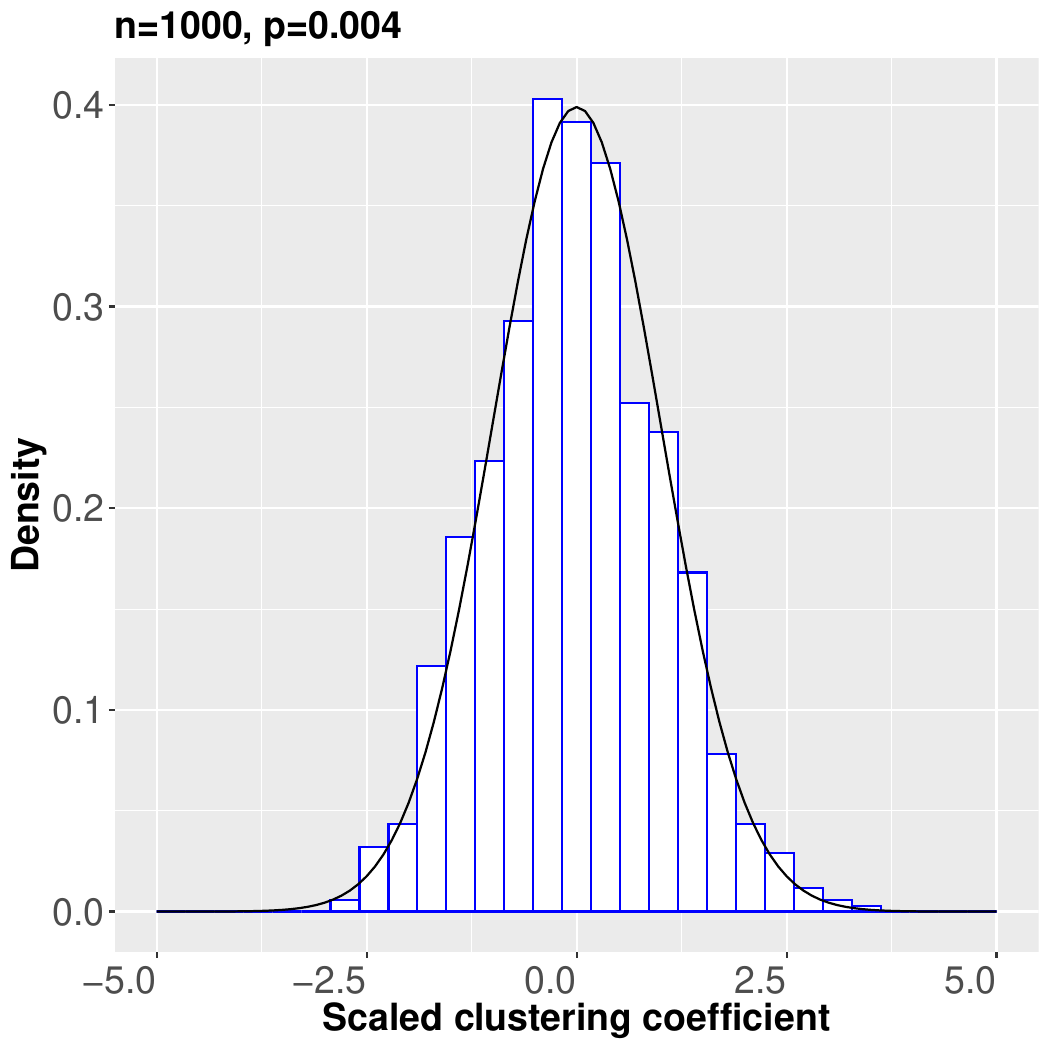}
\end{subfigure}%
\begin{subfigure}{0.3 \linewidth}
\includegraphics[width=\linewidth]{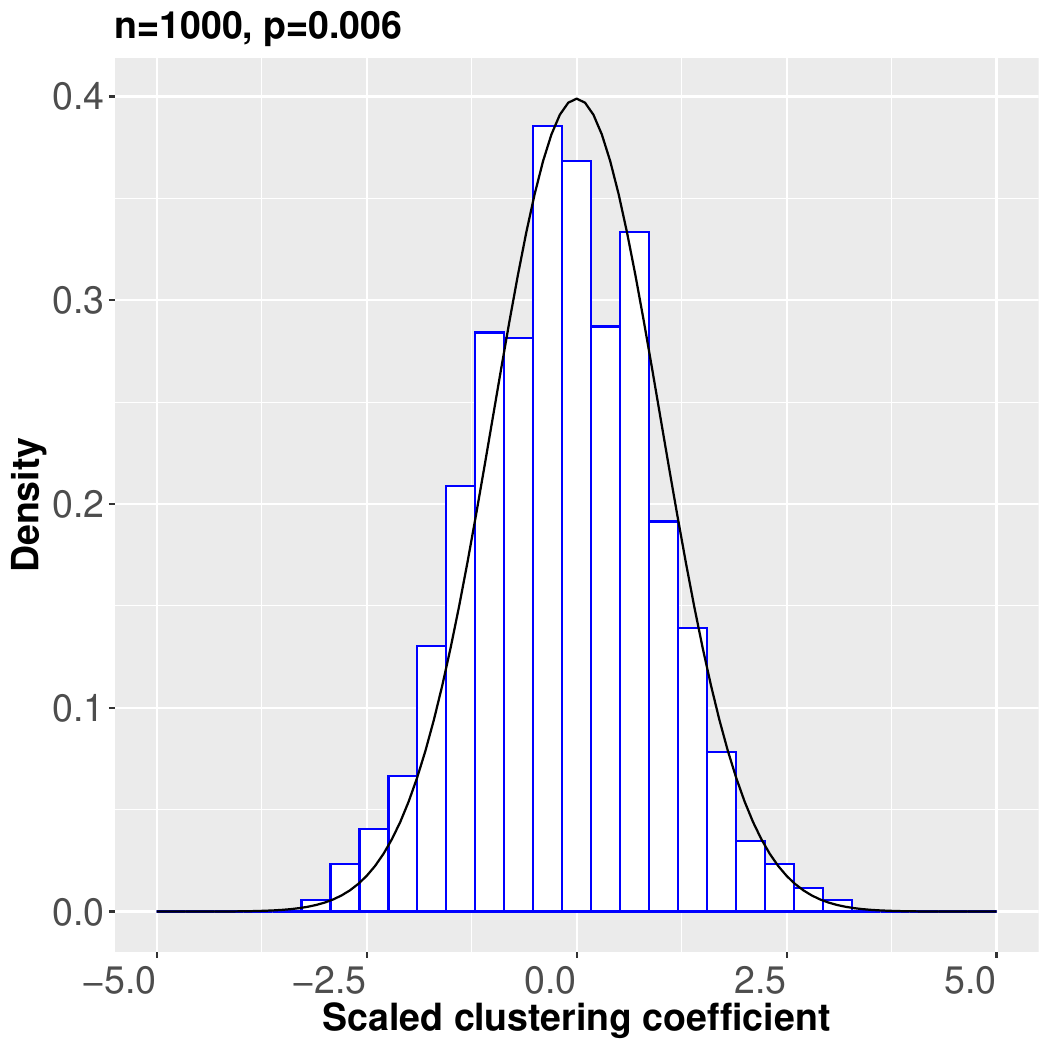}
\end{subfigure}%
\begin{subfigure}{0.3 \linewidth}
\includegraphics[width=\linewidth]{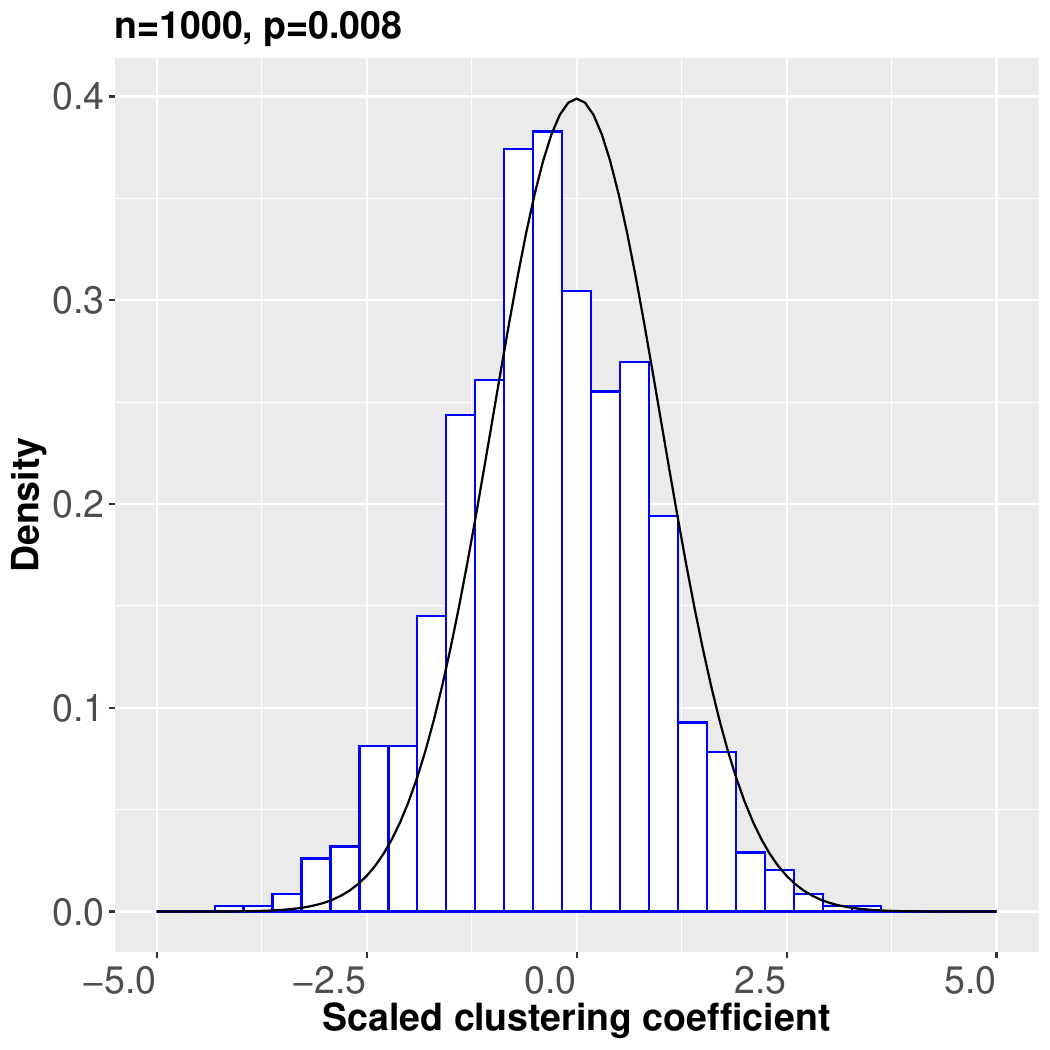}
\end{subfigure}
    \caption{Distribution of scaled clustering coefficient for inhomogeneous random graph generated from (top row) SBM and (bottom row) DCSBM. We fix $n=1000$ and the expected density (indicated as $p$) is varied from, from 0.004 to 0.008. The block matrix parameters are generated randomly from $U(0,1)$ distribution and in the case of DCSBM the degree parameters are generated from $LogNormal(1,0.5)$ distribution.}
    \label{fig:Cdist-SBM}
\end{figure}

An intriguing observation from Figure \ref{fig:A-power} is that there is an asymmetric behavior of the rejection rate on
the left and right ends of the parameter space of $\beta$. To investigate this further we conduct another simulation of the empirical power of the asymptotic test focusing on $\beta$ close to 0, namely, $\beta \in (0, 0.12)$ and $\beta$ close to 1, namely, $\beta \in (0.98, 1)$ in Figure \ref{fig:A-power-zoom}. We see that when the average degree is 10, empirical power is slow to pick up and reaches 1 only when $\beta$ is around 0.10, while the power is already close to 1 when $\beta$ is around 0.98. As the average degree increases, performance of the test in terms of empirical power improves on both sides of the power curve for $\beta$ close to 0 and 1. However, the performance in the side $\beta$ close to 1 is always better than in the side $\beta$ close to 0 highlighting the asymmetry in the power curves. For example when average degree is 50, we see the power reaches 1 around $\beta=0.04$, while it already reaches 1 when $\beta = 0.995$, i.e., only $0.005$ away from $\beta=1$.

When $\beta = 0$, the network is a pure ER model and we know from Theorem \ref{C-dist} that $C \le K_{1,\alpha}$ with probability $(1-\alpha)$.
As $\beta$ increases from zero, ER edges are gradually replaced by ring lattice edges, thereby increasing the clustering in the network.
This drives $C$ higher until $C$ crosses the threshold  $K_{1,\alpha}$ with a high probability and the null hypothesis is rejected.
As we observe from Figure \ref{fig:A-power}, this effect of the ring lattice takes time to build up, especially for smaller values of $n$.
On the other hand, when $\beta = 1$, the network is a pure ring lattice and we know from Theorem \ref{L-power} that $L > K_2$ with probability going to one.
As $\beta$ decreases from 1, ring lattice edges are being progressively replaced by random edges from the ER model.
Even a small number of such random edges can cause a drastic reduction in path lengths, since the addition of an edge can affect path lengths across the network.
This drives the rejection rate to to sharply increase to 1 in the right ends in Figure \ref{fig:A-power}.
This asymmetry also shows up quite prominently in Figure \ref{CLdist}, where we see a gradual increase in $C$ (left panel) but a sharp increase in $L$ (middle panel) as $\beta$ increases.

\subsection{Finite sample behavior of distribution of $C$ for inhomogeneous models}

In this section we assess the finite sample approximation provided by the asymptotic normality result in Theorem \ref{C-inhomo} for SBM and DCSBM graphs. We generate SBM and DCSBM graphs from the fastRG R package \cite{rohe2018note}. The block matrix parameters are generated randomly once from $U(0,1)$ distribution and fixed for the 1000 replications. For DCSBM model, the degree parameters are also generated once from lognormal distribution with logmean 1 and logSD 0.5. Therefore, the probabilities of edges are fixed for the 1000 replications as per the settings of Theorem \ref{C-inhomo}. For both SBM and DCSBM, we consider three expected densities: 0.004, 0.006, and 0.008, and fix $n=1000$. The histogram of observed values of $C$ with appropriate transformation as suggested by Theorem \ref{C-inhomo} is compared with the standard normal density in Figure \ref{fig:Cdist-SBM}. We see the approximation predicted by the theorem works well for this finite sample example for both SBM and DCSBM. However, the approximation deteriorates when conditions differ from the settings of the theorem, i.e., the expected density $\tilde{\theta}$ is higher, or heterogeneity among the probabilities is higher.

\section{Results on real networks}

\begin{table}[h]
\tiny
\captionsetup{justification=centering, width=\linewidth,font=small}
  \begin{subtable}[t]{.33\textwidth}%
    \centering%
    \begin{tabular}{c|c|c|c}
     \hline
        Null & C & L & Decision\\
        \hline
        ER & 0 & 0.99 & Reject \\
        SBM & 0 & 0.016 & Reject \\
        DCSBM & 0 & 0 & Reject \\
        CL & 0 & 0.06 & Reject \\
         \hline
    \end{tabular}
    \caption{C Elegans}\label{table:celeg}
  \end{subtable}%
  \begin{subtable}[t]{.33\textwidth}
    \centering
    \begin{tabular}{c|c|c|c}
    \hline
        Null & C & L & Decision\\
        \hline
        ER & 0 & 1 & Fail to reject \\
        SBM & 0 & 0.932 & Reject \\
        DCSBM & 0 & 0.854 & Reject \\
        CL & 0 & 0.968 & Reject \\
        \hline
    \end{tabular}
    \caption{Dolphins}\label{table:dolph}
  \end{subtable} %
  \begin{subtable}[t]{.33\textwidth}%
    \centering%
    \begin{tabular}{c|c|c|c}
    \hline
        Null & C & L & Decision\\
        \hline
        ER & 0 & 1 & Fail to reject \\
        SBM & 0.002 & 0.668 & Reject \\
        DCSBM & 0 & 0 & Reject \\
        CL & 0 & 1 & Fail to reject \\
        \hline
    \end{tabular}
    \caption{Football}\label{table:fb}
  \end{subtable} \par
  \begin{subtable}[t]{.33\linewidth}
    \centering
  \begin{tabular}{c|c|c|c}
  \hline
        Null & C & L & Decision\\
        \hline
        ER & 0 & 0.536 & Reject \\
        SBM & 0.376 & 0.058 & Fail to reject \\
        DCSBM & 0.012 & 0.018 & Reject \\
        CL & 0.652 & 0.164 & Fail to reject \\
        \hline
    \end{tabular}
    \caption{Karate}\label{table:karate}
  \end{subtable} %
  \hspace{\fill}
  \begin{subtable}[t]{.33\linewidth}%
    \centering%
    \begin{tabular}{c|c|c|c}
    \hline
        Null & C & L & Decision\\
        \hline
        ER & 0 & 0.982 & Reject \\
        SBM & 0 & 0.006 & Reject \\
        DCSBM & 0 & 0.006 & Reject \\
        CL & 0 & 0.028 & Reject \\
        \hline
    \end{tabular}
    \caption{Les Miserables}\label{table:lesmis}
  \end{subtable}%
  \begin{subtable}[t]{.33\linewidth}
    \centering
  \begin{tabular}{c|c|c|c}
  \hline
        Null & C & L & Decision\\
        \hline
        ER & 0 & 1 & Fail to reject \\
        SBM & 0 & 1 & Fail to reject \\
        DCSBM & 0 & 0.9 & Reject \\
        CL & 0 & 1 & Fail to reject \\
        \hline
    \end{tabular}
    \caption{Macaque Cortex}\label{table:maccort}
  \end{subtable} \par 
  \begin{subtable}[t]{.33\linewidth}
    \centering
    \begin{tabular}{c|c|c|c}
    \hline
        Null & C & L & Decision\\
        \hline
        ER & 0 & 1 & Fail to reject \\
        SBM & 0 & 0 & Reject \\
        DCSBM & 0 & 0.036 & Reject \\
        CL & 0 & 0 & Reject \\
        \hline
    \end{tabular}
    \caption{Political Blogs}\label{table:polblogs}
  \end{subtable}%
  \begin{subtable}[t]{.33\linewidth}
    \centering
  \begin{tabular}{c|c|c|c}
  \hline
        Null & C & L & Decision\\
        \hline
        ER & 0 & 1 & Fail to reject \\
        SBM & 0 & 1 & Fail to reject \\
        DCSBM & 0 & 0.99 & Reject \\
        CL & 0 & 1 & Fail to reject \\
        \hline
    \end{tabular}
    \caption{Political Books}\label{label:polbooks}
  \end{subtable}%
  \begin{subtable}[t]{.33\linewidth}
    \centering%
    \begin{tabular}{c|c|c|c}
    \hline
        Null & C & L & Decision\\
        \hline
        ER & 0 & 1 & Fail to reject \\
        SBM & 0 & 0.672 & Reject \\
        DCSBM & 0 & 0.986 & Reject \\
        CL & 0 & 1 & Fail to reject \\
        \hline
    \end{tabular}
    \caption{Power Grid}\label{table:powgrid}
  \end{subtable} \par 
      \centering
  \begin{subtable}[t]{0.33 \linewidth}
  \begin{tabular}{c|c|c|c}
  \hline
        Null & C & L & Decision\\
        \hline
        ER & 0 & 0.508 & Reject \\
        SBM & 0 & 0.22 & Reject \\
        DCSBM & 0 & 0.032 & Reject \\
        CL & 0.972 & 0.012 & Fail to reject \\
        \hline
    \end{tabular}
    \caption{Word Adjacencies}\label{table:wordadj}
  \end{subtable}
  \caption{Results of the bootstrap intersection test on real world networks. }\label{table:realworld}
\end{table}

\begin{figure*}
   \centering
    \begin{subfigure}[b]{\textwidth}
        \centering
\includegraphics[width=\linewidth, height=3.2cm]{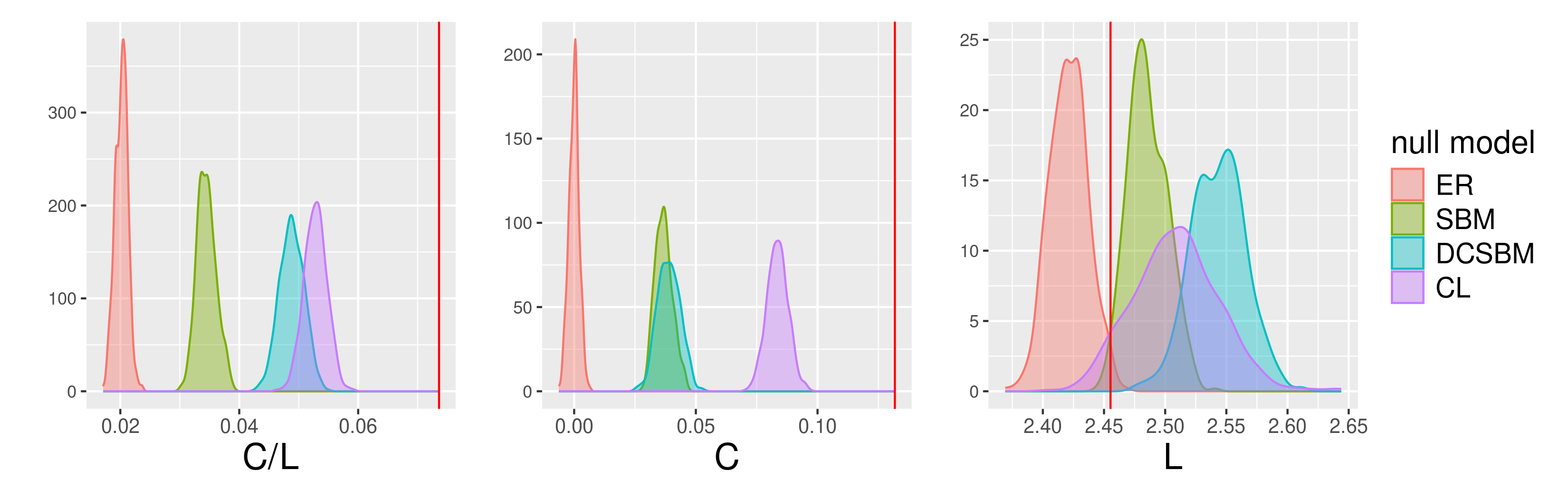}
        \caption{C.Elegans \cite{white86,watts1998collective}}
    \end{subfigure}
    \vskip\baselineskip
    
    \begin{subfigure}[b]{\textwidth}
        \centering
        \includegraphics[width=\linewidth, height=3.2cm]{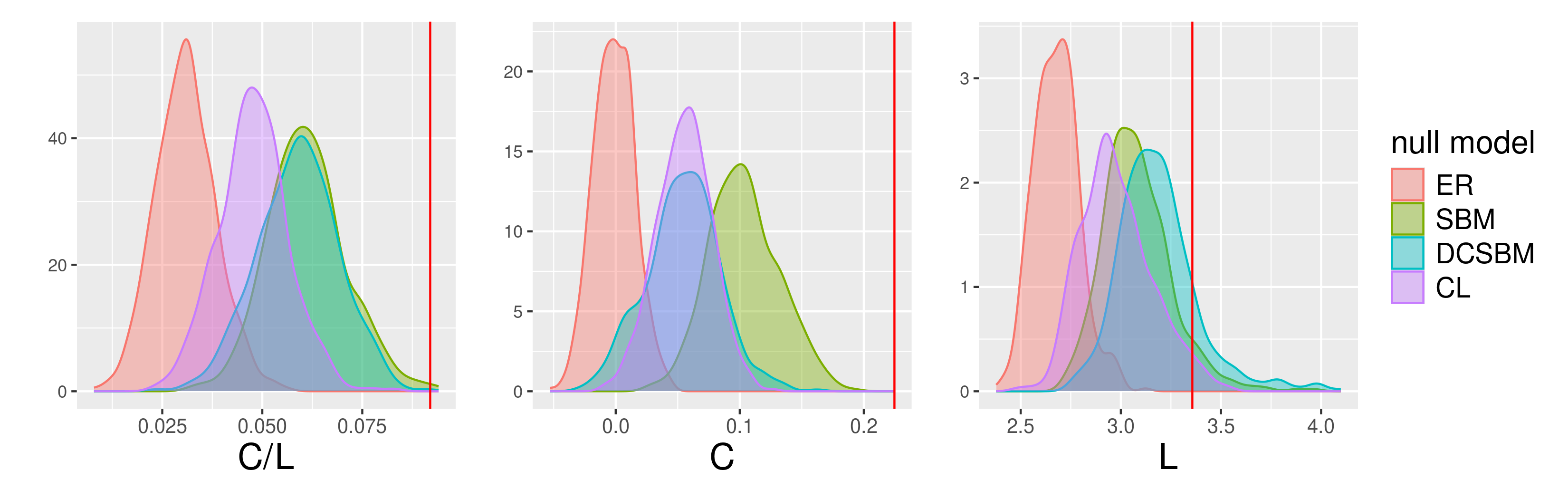}
        \caption{Dolphins \cite{lusseau2003bottlenose}}
    \end{subfigure}
    \begin{subfigure}[b]{\textwidth}
        \centering
        \includegraphics[width=\linewidth, height=3.2cm]{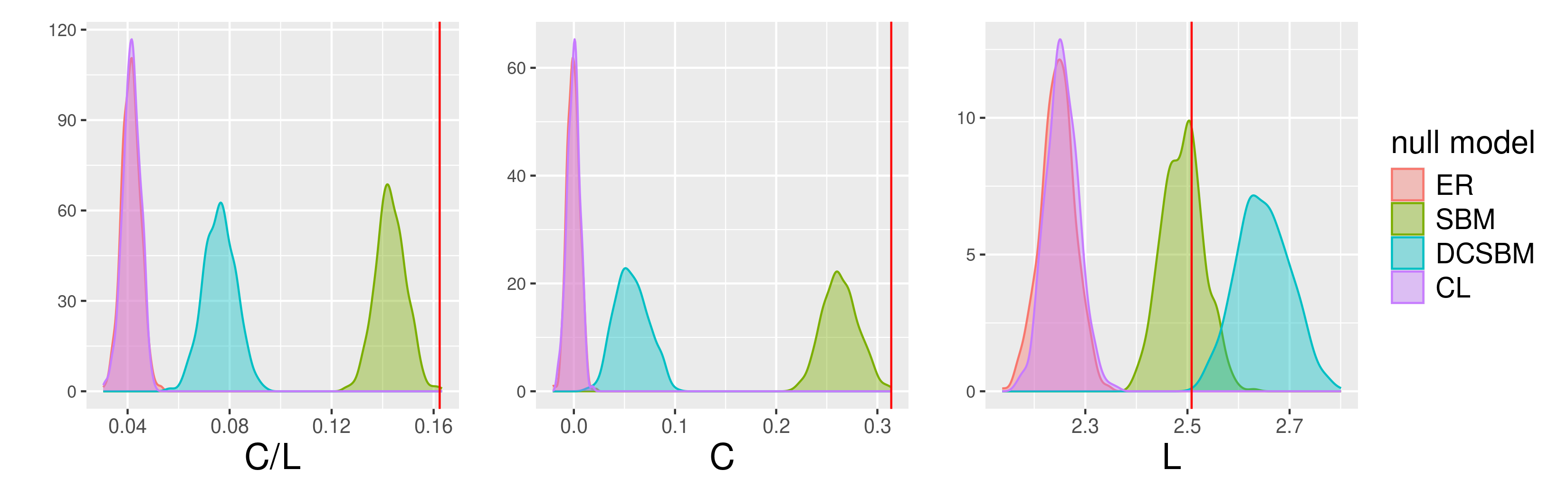}
        \caption{Football \cite{girvan2002community}}
    \end{subfigure}
    \begin{subfigure}[b]{\textwidth}
        \centering
        \includegraphics[width=\linewidth, height=3.2cm]{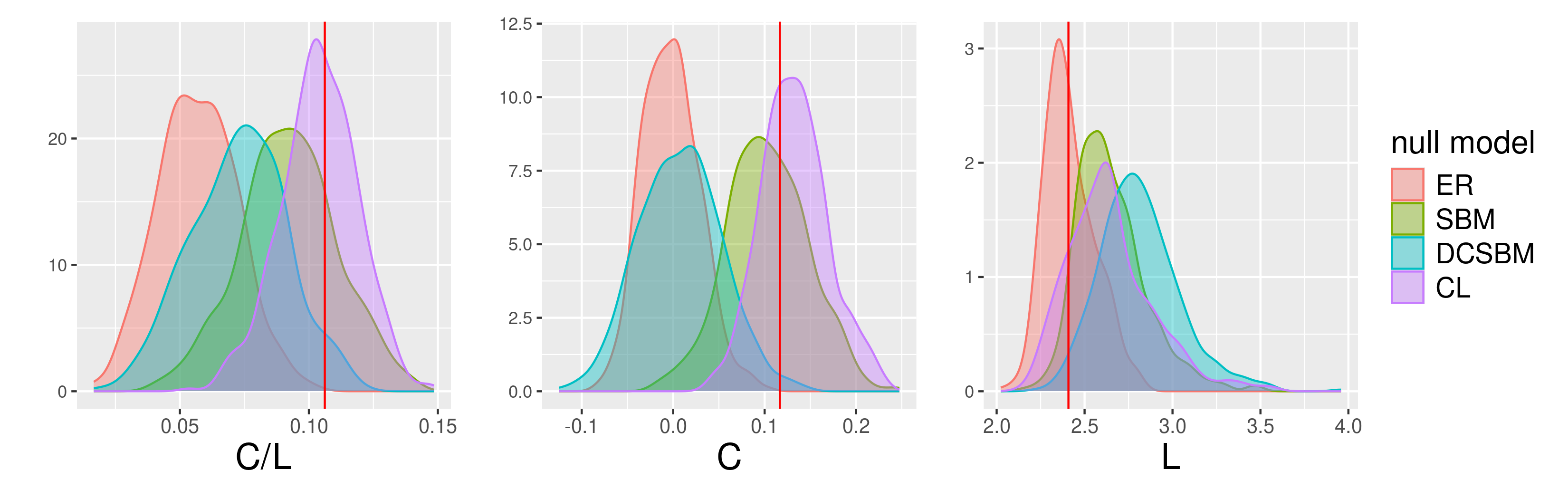}
        \caption{Karate Club network \cite{zachary1977information}}
    \end{subfigure}
    \begin{subfigure}[b]{\textwidth}
        \centering
        \includegraphics[width=\linewidth, height=3.2cm]{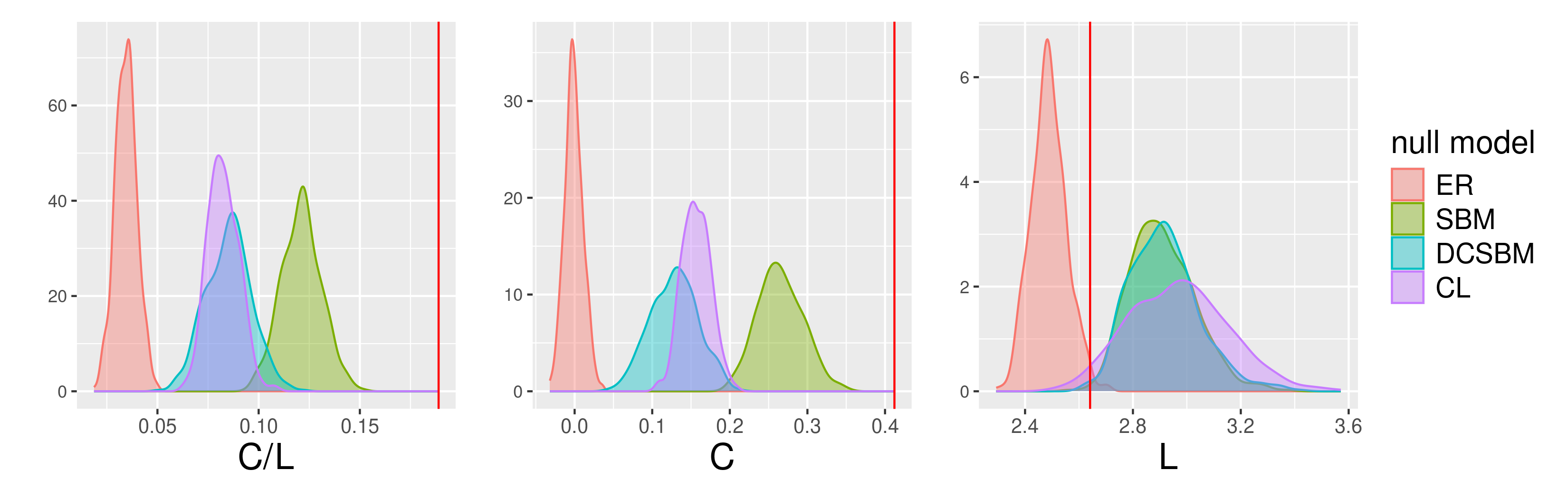}
        \caption{Les Miserables \cite{knuth1993stanford}}
    \end{subfigure}
    \caption{\textbf{Bootstrap tests on real-world networks.} The first figure in each row is the empirical distributions of the small-world coefficient, second shows empirical distributions for the clustering coefficient and the average path length. 500 simulations were used to generate the distributions for ER, SBM, DCSBM and CL null models. The red line in both figures is the observed test-statistic.}
   \label{fig:realworld1}
\end{figure*}

\begin{figure*}
   \centering
    \begin{subfigure}[b]{\textwidth}
        \centering
        \includegraphics[width=\linewidth,height=3.2cm]{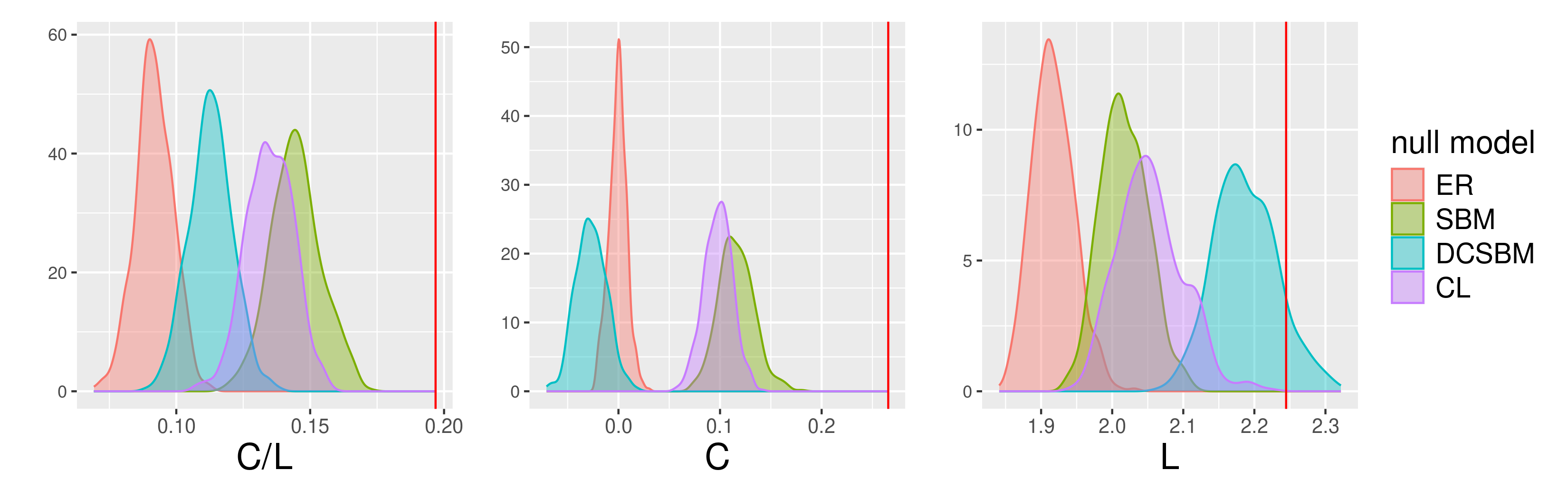}
        \caption{Macaque Cortex \cite{kaiser2006nonoptimal}}
    \end{subfigure}
    \vskip\baselineskip
    \begin{subfigure}[b]{\textwidth}
        \centering
        \includegraphics[width=\linewidth,height=3.2cm]{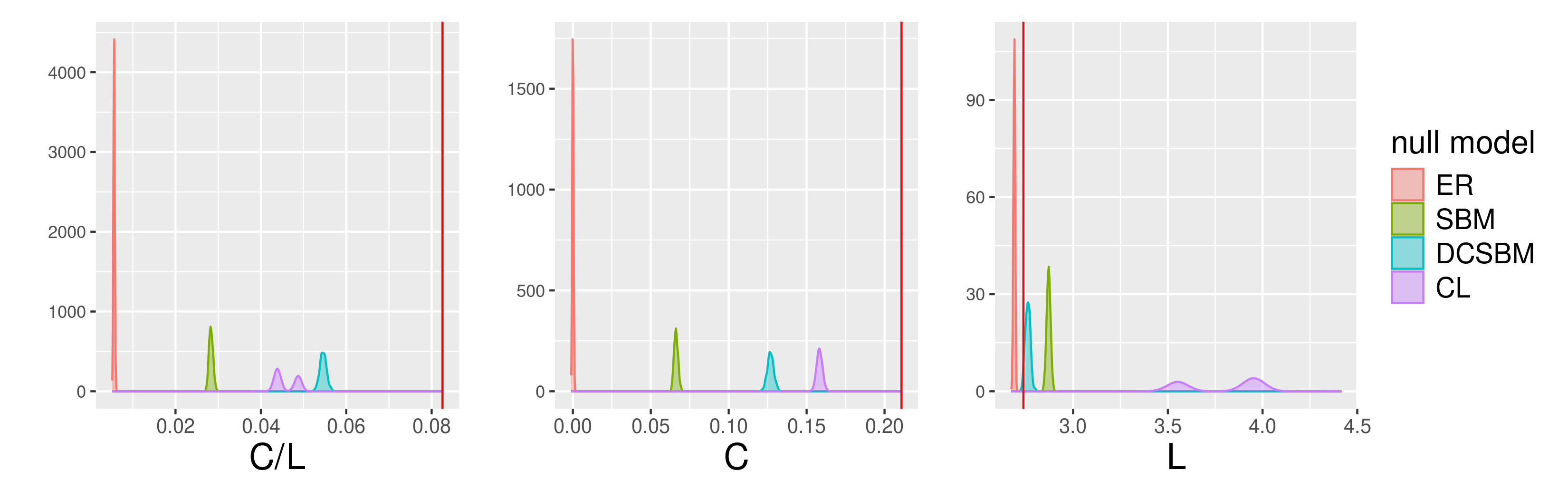}
        \caption{Political Blogs \cite{adamic05}}
    \end{subfigure}
    \begin{subfigure}[b]{\textwidth}
        \centering
        \includegraphics[width=\linewidth,height=3.2cm]{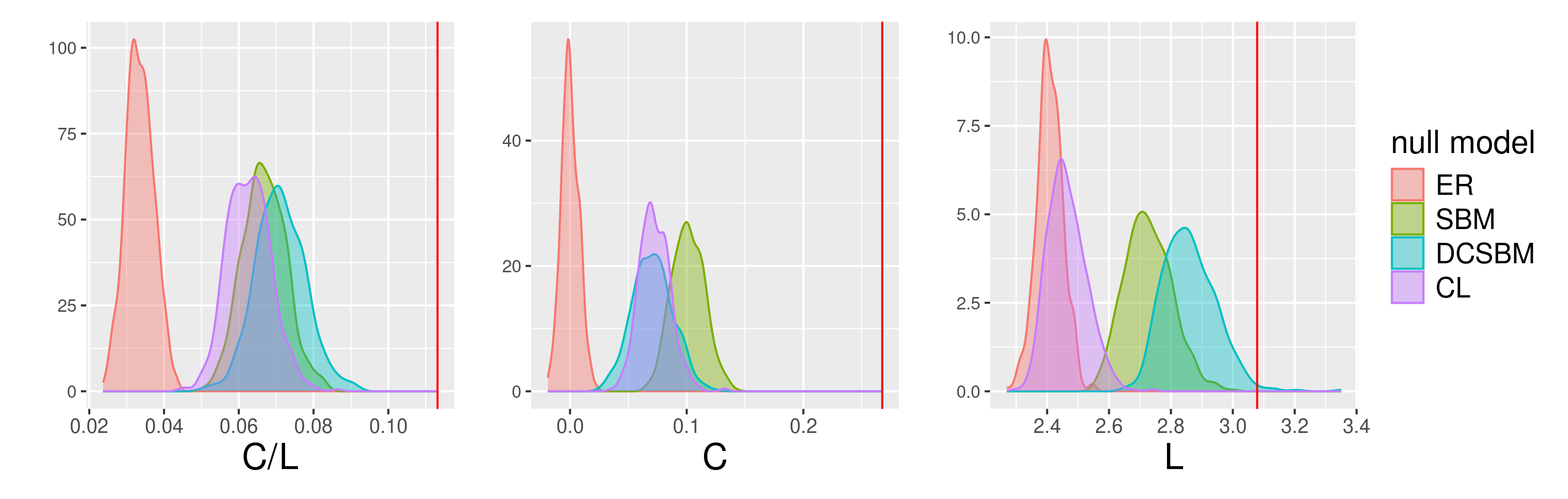}
        \caption{Political Books (V.Kreb (\url{www.orgnet.com}))}
    \end{subfigure}
    \begin{subfigure}[b]{\textwidth}
        \centering
        \includegraphics[width=\linewidth,height=3.2cm]{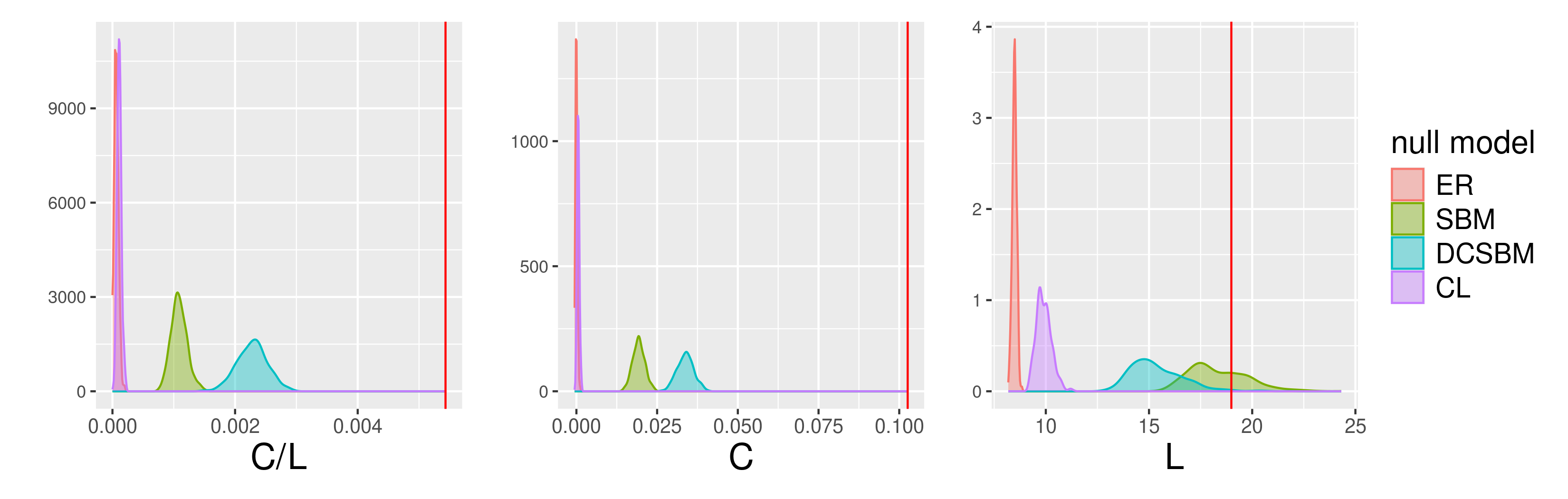}
        \caption{Power Grid \cite{watts1998collective}}
    \end{subfigure}
    \begin{subfigure}[b]{\textwidth}
        \centering
        \includegraphics[width=\linewidth,height=3.2cm]{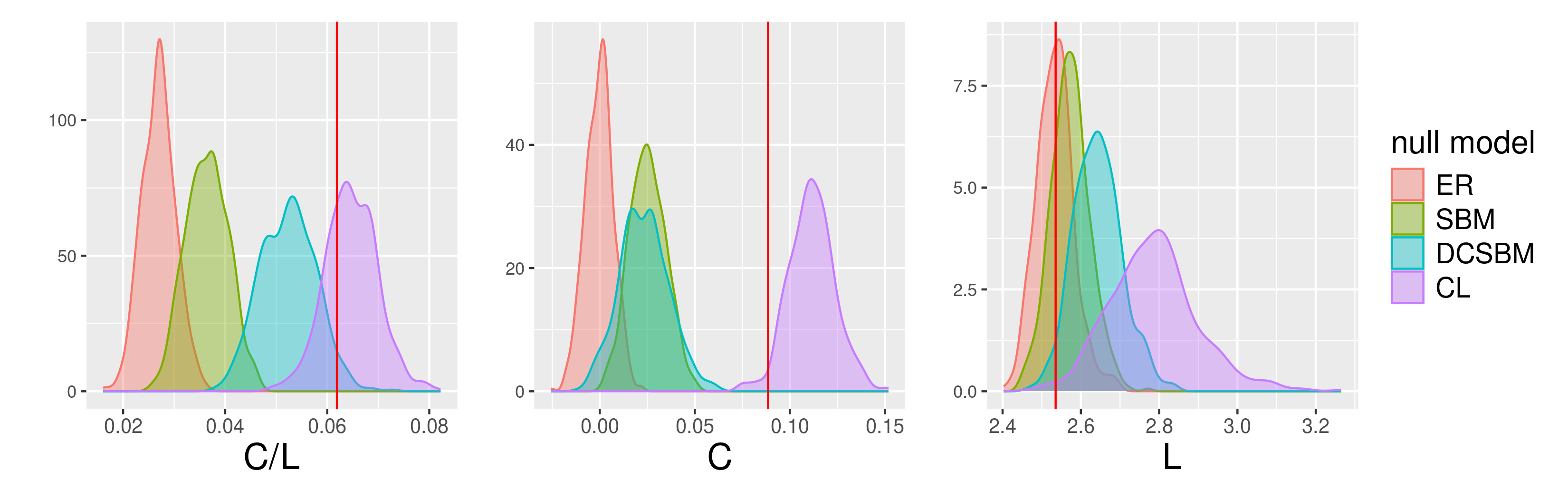}
        \caption{Word Adjacency \cite{newman2006finding}}
    \end{subfigure}
    \caption{\textbf{Bootstrap tests on real-world networks continued.}}
    \label{fig:realworld2}
\end{figure*}

We apply the \textit{bootstrap detection method} for small world property to several real-world networks using the above-mentioned four null models, namely ER, SBM, DCSBM, and CL. For each null model, we generate 500 networks to derive empirical distributions of $C$ and $L$ using parameters learned from a given real-world network. We present these empirical distributions for 10 real network datasets in the right columns of Figures ~\ref{fig:realworld1} and ~\ref{fig:realworld2}. The observed values of $C$ and $L$ are indicated in the Figures with a red colored vertical line. In Table ~\ref{table:realworld}, we further present p-values associated with the two components of our intersection test statistic in Equation \ref{test_statistic}, as well as the overall decision from our test. In the table ~\ref{table:realworld}, the column for $C$ depicts  $\mathbf{p_{C}}$, which is the proportion of simulated networks which have a \textit{higher} clustering coefficient, and the column for $L$ depicts $\mathbf{p_{L}}$, which is the proportion of simulated networks which have a \textit{lower} average path length. The Decision column presents the verdict from the intersection test, which rejects the null hypothesis to conclude that a given network is small-world, \emph{if the null hypothesis for both $C$ and $L$ are rejected}. The test for $C$ rejects the null if $\mathbf{p_{C}} < 0.05$, indicating that the given network has a significantly higher clustering coefficient compared to the null model. The test for $L$ rejects the null if $\mathbf{p_{L}} < 0.99$, indicating the given network has a comparable path length to the null model. Therefore the test for $L$ fails to reject if almost all the simulated networks have lower $L$ compared to the one observed. Finally, the left columns of Figure \ref{fig:realworld1} and \ref{fig:realworld2}
further depicts the empirical distribution of the test statistic $\mathbf{C/L}$ along with its observed value.

Several interesting features emerge from our results. From Figure \ref{fig:realworld1}, the C-elegans and Les Miserables networks are small world under all four null models. The Karate club network is not a small world under SBM and CL null models because the clustering coefficient $C$ is not significantly higher than what the two null models predict, despite $L$ being within the distribution of $L$ from all the null models. Therefore, the Karate club network's high clustering coefficient can be well explained by either community structure or degree heterogeneity. On the other hand, the Football network is not a small world under ER and CL null models, and the Dolphin network is not a small world under the ER model because the average path length $L$ is not within the distribution of $L$ from the null model. In both networks, the clustering coefficient $C$ is higher than what any of the null models would predict. For both networks the average path length is high enough that a ER random graph model cannot explain it. However, such an average path length can be well predicted by models that include community structure and/or degree heterogeneity. 

In Figure \ref{fig:realworld2}, none of the 5 networks is small world under all four null models. The Macaque Cortex and Political books networks are small world only under DCSBM null model. For the other three null models, the distribution of $L$ values is completely in the left hand side of the observed $L$ value.  The power grid, political blogs, and political books networks have very high observed values of $C$ which is higher than what any of the null models would predict. Therefore, in terms of clustering, the networks cannot be well approximated by any of the null models and require models with additional features. However, $L$ is comparable to only SBM and DCSBM null models for power grid network, SBM, DCSBM and CL null models for political blogs and DCSBM for political books networks. The word adjacencies network has a $C$ which is lower than the distribution of $C$ from the CL model and therefore it is not small world under the CL model. The network is small world under all other null models. Only the CL model can explain both the high clustering coefficient and the low average path length for this network.

Overall it appears that many networks are able to ``pass" (i.e., reject) the clustering coefficient test for most of the null models, but ``fail" (i.e., fail to reject) the average path length rest for some null models. Clearly the more complex null models, namely, CL and DCSBM consistently predicts higher average path length than the ER random graph model and are able to model the observed path lengths in many cases. Therefore many networks are small world only under those models and not under the SBM and ER models. This is an useful finding in terms of modeling choice for real-world network data.

Further, as the left columns of the Figures \ref{fig:realworld1} and \ref{fig:realworld2} show, the results with the traditional small world coefficient is identical to the result one would obtain from the clustering coefficient test alone. Consequently, \textit{using the traditional coefficient fails to take into account the average path length of the observed networks}. This is contradictory to the philosophical spirit of the small world phenomenon - clustering structure despite a small average path length. Consequently, with ER null model, the traditional small world coefficient declares all networks under test as small world. With SBM, the metric only detects the Karate Club network as not being a small world; with CL it detects Karate and word adjacencies as not small world, while with DCSBM it again detects all networks as small world. In comparison, results with our proposed methodology bring out the consequences of various modeling choices and help distinguish small-world property from community structure and degree heterogeneity. The results from the asymptotic test on these datasets are presented in the Appendix B.

\section{Limitations and Conclusions}
We have developed a hypothesis testing framework for detecting the small world property of a network by defining suitable null and alternative families of models and hypotheses. The test is an intersection of two tests on average path length and clustering coefficient, which is \textit{rejected} (network is designated \textit{small-world}) if both the tests are rejected simultaneously. Our empirical results on several real networks bring out nuances associated with the small-world property. Separately testing for average path length and clustering coefficient has two advantages over the existing small-world coefficient. First, it removes the undue influence of clustering on the small-world coefficient and distinguishes the \textit{property} from the \textit{coefficient}. Second, the analyst is able to ascertain which of the two requirements a network does not satisfy. The superimposed Newman-Watts type small-world models have been proposed as \textit{plausible models} of \textit{small-world property} generalizing the Newman-Watts and Watts-Strogatz models. Using these new tools we are able to  analyze the properties of real networks in greater detail.

In this paper, we have proposed and empirically studied a parametric bootstrap strategy without a formal proof of its validity. Recent years have seen substantial growth in theoretical results for bootstrapping network data, with notable contributions from \cite{bhattacharyya2015subsampling}, \cite{green2022bootstrapping}, \cite{lundesubsampling}, and \cite{levin2019bootstrapping}. However, none of these existing works on network bootstrapping include the average path length, which makes it challenging to prove theoretical results for our method. We consider it an important direction for future work to develop a theory for the proposed parametric bootstrap method.

\paragraph{Ackowledgement:}  This work was supported in part by  National Science Foundation grants DMS-1830547 and DMS-2413327, and the NIH grant 7R01LM013309.

\bibliographystyle{plainnat}

\bibliography{triads.bib,Networks.bib,vc.bib,cluster.bib,GraphTheory.bib,triads1.bib,sw.bib}

\section{Appendix A: Proofs}

\subsection{Proof of Proposition \ref{prop1}}
The calculations for the ER null and superimposed models are similar to those in \cite{barrat2000properties} (see also \cite{newman2018networks}). Note, for $\beta=0$,  $E[3T] = 3 (\frac{2\delta}{n-1})^3$. Next, for $\beta = 1$, we have $3T = 3n \binom{\delta}{2}/\dbinom{n}{3}$ (deterministically). 
Now for a $\beta \neq \{0,1\}$, we need to consider triangles from 3 sources: (a) the random graph component, (b) the ring lattice component, and (c) the incidental triangles. The incidental triangles are generated in two ways - (i) one edge from the random graph and 2 edges from the ring lattice, and (ii) 1 edge from the lattice and 2 edges from the random graph. Throughout the analysis we are assuming that if there are multiple triangles ($T$) or connected triples ($S$) among a set of 3 vertices, then we will count them multiple times in our global counts of $T$ and $S$ structures. The ring lattice component (b) generates $n \binom{\delta \beta}{2}$ triangles. The random graph component is expected to (a) generate $ \binom{n}{3}(\frac{2\delta(1-\beta)}{n-1})^3$ triangles. Finally the expected incidental triangles can be enumerated as follows. The expected number of incidental triangles of type (i) is $n\binom{2 \delta \beta}{2}\frac{2\delta(1-\beta)}{n-1}$, while the expected number of incidental triangles of type (ii) is $n.2 \delta\beta (n-2)(\frac{2\delta(1-\beta)}{n-1})^2$. Note that while counting triangles, we count a triple having 2 triangles if the triple has a triangle and at least one of the edges is a multi-edge. Combining the above expected values and multiplying by 3, we get $E[3T]$. Further, as $n \to \infty$, the expression for $T$ is asymptotically equivalent to
\[
\dbinom{n}{3}E[T] \asymp \frac{n \delta^2\beta^2}{2} + \frac{8 \delta^3 (1-\beta)^3}{6} + 4 \delta^3\beta^2(1-\beta) + 8 \delta^3 \beta(1-\beta)^2.
\]
We can similarly compute the expected number of connected triples that goes in the denominator. Once again, for $\beta=0$, we have $E[S]=3(\frac{2\delta}{n-1})^2$, and for $\beta=1$, we have $S = n\dbinom{2\delta}{2}/\dbinom{n}{3}$ (deterministically). 
For $0<\beta<1$, we have  3 sources for connected triples (S): The random graph component (a) is expected to generate $3\dbinom{n}{3}(\frac{2\delta(1-\beta)}{n-1})^2$ connected triples. The ring lattice component (b) generates $n \dbinom{2\delta \beta}{2}$ connected triples. Finally the number of  incidental connected triples (c) are expected to be $2n \delta \beta (n-2)\frac{2\delta(1-\beta)}{n-1}$. This is because for each of the $n\delta \beta$ edges in the ring lattice, an edge from the random graph can connect one of the ends of the edge to any of the $n-2$ remaining nodes. Note that we count a triple to have 2 $S$ structures if there is an $S$ structure and at least one of the edges is a multi-edge. Combining the above we arrive at an expression for $E[S]$.  As $n \to \infty$, then we have
\[
\dbinom{n}{3} E[S] \asymp 2 n \delta^2\beta^2 + 2n \delta^2 (1-\beta)^2 + 4 n \delta^2\beta(1-\beta) \asymp 2n\delta^2.
\]

Now, we note from second-order bivariate Taylor series expansion that
\[
E[C] = \frac{E[3T]}{E[S]} - \frac{Cov(3T,S)}{(E[S])^2} + \frac{Var(S)E[3T]}{E[S]^3} + E[R].
\]
For $\beta=0$, approximating $E[C] \asymp \frac{E[3T]}{E[S]}$, we have $E[C] \asymp \frac{2\delta}{(n-1)} \to 0$, as $n \to \infty$. This approximation is valid, since for $n \to \infty$, we have $Cov(3T,S) = O(\delta^4/n^6), E(S) =O(\delta^2/n^2), E[3T] = O(\delta^3/n^3)$, and $Var[S]=O(\delta^3/n^5)$, and consequently, the second order terms $\frac{Cov(3T,S)}{(E[S])^2} =O(\frac{1}{n^2})$, and $\frac{Var(S)E[3T]}{E[S]^3} = O(\frac{1}{n^2})
$. One can repeat this calculation to see expectation of the remainder term $E[R]$ is of even smaller asymptotic order. For $\beta=1$, we have $E[C] = \frac{3\delta(\delta-1)}{2\delta(2\delta-1)} \to \frac{3}{4}$ as $n \to \infty$. Finally, for $0 <\beta<1$ we note that whenever $\beta>0$, the expression for $\dbinom{n}{3}E[3T]$ is dominated by $\frac{n\delta^2\beta^2}{2}$, since $\delta = o(n)$ implies that $\delta^3 = o(n\delta^2)$. Therefore, as $n \to \infty$, we have 
\[
E[\frac{3T}{S}] \asymp \frac{E[3T]}{E[S]} \asymp \frac{3n\delta^2\beta^2}{4 n\delta^2 } \to \frac{3}{4}\beta^2.
\]
\subsection{Proof of Proposition \ref{prop2}}
Recall we have defined $\bar{p} =  \sum_{1 \leq i<j \leq n} p_{ij}/\dbinom{n}{2}$. We note that when $\beta=0$, $E[3T]$ for a graph from inhomogeneous random graph model
can be written as 
\begin{align*}
    E[3T] & =\sum_{i}\sum_{ j\neq i} \sum_{k \neq (i,j)} p_{ij}p_{jk}p_{ik} /\dbinom{n}{3} \\
    & \leq \frac{1}{2}\sum_i \sum_{ j\neq i} p_{ij} (n-2)p_{\max}^2 /\dbinom{n}{3} = \frac{6n(n-1)(n-2)}{2n(n-1)(n-2)}\bar{p}p_{\max}^2 = 3\bar{p}p_{\max}^2.
\end{align*}

On the other hand, a lower bound on  $E[S]$ can be found as follows
\begin{align*}
  E[S] & = \sum_{i}\sum_{ j\neq i} p_{ij} (\sum_{k \neq (i,j)} p_{jk} + \sum_{k \neq (i,j)} p_{ik}) /\dbinom{n}{3}\\
  & \geq \frac{6n(n-1)(n-2)}{2n(n-1)(n-2)}\bar{p}p_{\min} = 3 \bar{p}p_{\min} 
\end{align*}

Then as $n \to \infty$, the ratio, $C$ is asymptotically upper bounded by $\frac{p_{\max}^2}{p_{\min}} \asymp \frac{2\delta_{\max}^2}{\delta_{\min}n} .$

For $0<\beta<1$, the random variable $3T$ is lower bounded by three times the number of triangles in the (deterministic) ring lattice  component of the graph. Further note that, $\delta_{\max} \geq \delta \geq \delta_{\min}$. Therefore $\dbinom{n}{3}E[3T]$ is lower bounded by  $\frac{3n \delta^2\beta^2}{2} \geq \frac{3n \delta \delta_{\min}\beta^2}{2}  $.  Therefore, following the calculation for $\dbinom{n}{3}E[S]$ as described in the proof of Proposition 1, we note that the components are asymptotically upper bounded by $2n\delta \delta_{\max}\beta^2$, $2n\delta \delta_{\max}(1-\beta)^2$, and $4n\delta \delta_{\max}\beta(1-\beta)$ respectively. Hence combining the three, we have $\dbinom{n}{3}E[S] \lesssim 2n\delta \delta_{\max}.$ Taking the ratio of the two expectations and first-order Taylor series expansion leads to the proof of the proposition.

\subsection{Proof of Theorem \ref{C-dist}} 

From the asymptotic order of various terms of $T$ and $S$, we have, in the regime $1/n <<p<<1/\sqrt{n}$, the dominant term in $T-p^3$ is $R_3 + pR_2$. Consequently,
\[
\sqrt{\dbinom{n}{3}} \frac{(T-p^3)}{\sqrt{p^3(1-p)^3 + 3p^4 (1-p)^2}} \overset{D}{\to} N (0,1),
\]
and 
\[
\frac{S}{3p^2} = 1 + O_p(\frac{1}{n\sqrt{p}})= 1+o_p(1).
\]
Then we have by Slutsky's theorem
\begin{align*}
   \sqrt{\frac{n^3}{6p^3(1+2p)(1-p)^2}}p^3\frac{(T/p^3 -1)}{S/3p^2} \overset{D}{\to} N(0,1). 
\end{align*}
Next we note that
\[
\sqrt{\frac{n^3}{6p^3(1+2p)(1-p)^2}}p^3 (\frac{1}{S/3p^2}-1) = O_p(\sqrt{n^3p^3}.\frac{1}{n\sqrt{p}}) =O_p(\sqrt{np^2}) = o_p(1),
\]
by the condition on the growth rate of $p$. This implies
\[
\sqrt{\frac{n^3p}{6(1+2p)(1-p)^2}}(C -p) \overset{D}{\to} N (0,1).
\]
This completes the proof for part (1) of the theorem.

Next in the regime $p >> 1/\sqrt{n}$,
we compute the joint asymptotic distribution as follows. Note from central limit theorem,
\[
\sqrt{n(n-1)}(\hat{p}-p) \overset{D}{\to} N(0, 2p(1-p)).
\]
Then we have the following joint asymptotic normality
\begin{align*}
\sqrt{\frac{n(n-1)}{2p(1-p)}} \begin{pmatrix}
    T/(3p^2)-p/3\\
    S/(6p)-p/2
\end{pmatrix}
& =  \sqrt{\frac{n(n-1)}{2p(1-p)}} \begin{pmatrix}
    1\\
    1
\end{pmatrix} (\hat{p}-p) \\
& \overset{D}{\to} 
\mathcal{MVN}_2 \left(\begin{pmatrix}
    0 \\
    0
\end{pmatrix},
\begin{pmatrix}
    1 & 1 \\
    1 & 1
\end{pmatrix}\right).
\end{align*}
We note that for fixed $p$, the limiting covariance matrix becomes
\[
\sqrt{n(n-1)} 
\begin{pmatrix}
    T-p^3\\
    S-3p^2
\end{pmatrix} \overset{D}{\to} 
\mathcal{MVN}_2 (\begin{pmatrix}
    0 \\
    0
\end{pmatrix},  18p(1-p)\begin{pmatrix}
    p^4 & 2p^3 \\
    2p^3 & 4p^2
\end{pmatrix}.
\]
It is well-known the limiting covariance matrix of joint subgraph statistics for $T$ and $S$ is a rank 1 matrix \cite{reinert2010random}. For fixed $p$, the above result was obtained by \cite{nowicki1989asymptotic}.

Consider the function for the random variables $[X_1, X_2]^T = [T/(3p^2), S/(6p)]^T$,
\[
g([X_1,X_2]^T) =  \frac{3p}{2}\left(\frac{X_1}{X_2}\right) .
\]
Clearly this is a continuous function at $\mu = [p/3,p/2]^T$. We note the function $g(\mu)$ is $\mathcal{R}^2 \to \mathcal{R}$, and is given by
\[
g(\mu) = p.
\]
Then from the multivariate delta method, we have
\begin{align*}
    & \sqrt{\frac{n(n-1)}{2p(1-p)}}   \left (g(T/(3p^2), S/(6p)) - g(p/3,p/2) \right)\\
    & = \sqrt{\frac{n(n-1)}{2p(1-p)}} 
 [D(g(\mu))] 
\begin{pmatrix}
    T/(3p^2)-p/3\\
    S/(6p)-p/2
\end{pmatrix} + o_p(1) \\
    & \to N_2 \left(0, [D(g(\mu))]\begin{pmatrix}
    1 & 1 \\
    1 & 1
\end{pmatrix}) [D(g(\mu))]^T\right), 
\end{align*}
where $D(g(\mu))$ is the gradient of $g(X)$ evaluated at $\mu$.

Therefore,
\begin{align*}
   [D(g(\mu))] = \frac{3p}{2}\begin{pmatrix}
       \frac{1}{X_2} \\
       \frac{-X_1}{X_2^2}
   \end{pmatrix}|_{\mu} = \frac{3p}{2} \begin{pmatrix}
       2/p \\
      -4/(3p)
   \end{pmatrix} = \begin{pmatrix}
       3 \\
      -2
   \end{pmatrix}. 
\end{align*}
and consequently,
\begin{align*}
    & [D(g(\mu))]  \Sigma_1 [D(g(\mu))]^T  \\
    & = [3,-2]\begin{pmatrix}
    1 & 1 \\
    1 & 1
    \end{pmatrix}
     \begin{pmatrix}
     3\\
     -2 
    \end{pmatrix}
    = 1.
\end{align*}  

Therefore, 
\begin{align*}
& \sqrt{\frac{n(n-1)}{2p(1-p)}}\left( C - p\right)  \overset{D}{\to} N \left(0,1\right)
\end{align*}
 
This completes the proof for the second part of the theorem.

\subsection{Proof of Theorem \ref{C-power}}
We first consider $\beta=0$. Let $p_0$ denotes the true population parameter for the ER model, i.e., $A \sim ER(n,p_0)$. Let $C_A$ denote the observed clustering coefficient of the graph $A$. Define $K_{1,\alpha}(p)$ as the $100*(1-\alpha)$th upper quantile of the distribution of $C$ under the ER model with parameter $p$. First consider $p >> \frac{1}{\sqrt{n}}$. Then, 
\begin{equation}
K_{1,\alpha}(p) =\left(p + Z_{\alpha} \sqrt{\frac{2p(1-p)}{n(n-1)}}\right),
\end{equation}
where $Z_{\alpha}$ is the $100*(1-\alpha)$th upper quantile of the standard normal distribution (a constant as a function of $n$).
Let $K_{1,\alpha}(\hat{p})$ be the corresponding $100*(1-\alpha)$th upper quantile that we calculate by replacing $p$ with the estimated value $\hat{p}$ in $K_{1,\alpha}(p)$. 
By part (2) of Theorem \ref{C-dist}, 
\[
P(C_A > K_{1,\alpha}(p_0)) \to \alpha.
\]
Moreover, from central limit theorem $\hat{p}-p_0 = O_p(\frac{\sqrt{p_0}}{n})$.
We further note that
\[
K_{1,\alpha}(\hat{p}) - K_{1,\alpha}(p_0) = (\hat{p}-p_0) + Z_{\alpha} \left(\sqrt{\frac{2\hat{p}(1-\hat{p})}{n(n-1)}} - \sqrt{\frac{2p_0(1-p_0)}{n(n-1)}}\right).
\]
From Taylor series expansion, it is clear that
\begin{align*}
\left(\sqrt{\frac{2\hat{p}(1-\hat{p})}{n(n-1)}} - \sqrt{\frac{2p_0(1-p_0)}{n(n-1)}}\right) & = \frac{(1-2p_0)(\hat{p}-p_0)}{\sqrt{2n(n-1)p_0(1-p_0)}} + O_p\left(\frac{1}{n^3\sqrt{p}}\right)\\
& = O_p\left(\frac{1}{n^2}\right) + O_p\left(\frac{1}{n^3\sqrt{p}}\right) =  O_p\left(\frac{1}{n^2}\right).
\end{align*}
Therefore,
\[
K_{1,\alpha}(\hat{p}) - K_{1,\alpha}(p_0) = O_p\left(\frac{\sqrt{p_0}}{n}\right)= o_p(1).
\]

Similarly, for $\frac{1}{n}<<p<<\frac{1}{\sqrt{n}}$,
\begin{equation}
K_{1,\alpha}(p) =\left(p + Z_{\alpha} \sqrt{\frac{6(1+2p)(1-p^2)}{n^3p}}\right).
\end{equation}
As in the previous case, part (1) of Theorem \ref{C-dist} guarantees that
\[
P(C_A > K_{1,\alpha}(p_0)) \to \alpha.
\]
We can also verify from Taylor expansion that,
\[
\sqrt{\frac{6(1+2\hat{p})(1-\hat{p}^2)}{n^3\hat{p}}} - \sqrt{\frac{6(1+2p_0)(1-p_0^2)}{n^3p_0}} = O_p\left(\frac{1}{\sqrt{p_0^3n^3}}\right)O_p\left(\frac{\sqrt{p_0}}{n}\right),
\]
and $\hat{p}-p_0 =O_p(\frac{\sqrt{p_0}}{n}) $, as stated before. Then,
\[
K_{1,\alpha}(\hat{p}) - K_{1,\alpha}(p_0) =O_p\left(\frac{\sqrt{p_0}}{n}\right) + O_p\left(\frac{1}{\sqrt{p_0^3n^3}}\right)O_p\left(\frac{\sqrt{p_0}}{n}\right) = o_p(1).
\]

Therefore, combining,  for any $p>>\frac{1}{n}$, 
\[
P(C_A>K_{1,\alpha}(\hat{p})) \to \alpha, \text{ when } A \sim ER (n,p),
\]
as $n \to \infty$. This completes the proof of the first part of the theorem.

Next we need to show that
\[
P(C_A>K_{1,\alpha}(\hat{p})) \rightarrow 1, \text{ when } A \sim NW (n,p,\beta) \text{ and } n\rightarrow \infty.
\]

We start proving this by defining the following two graphs.
 Let $A_R$ be the ring lattice component of the graph, i.e, obtained by removing the ER edges from the graph. In our notation, $A_R \sim RL(n,2\beta \delta)$ for some $0<\beta<1$, however this is a deterministic graph not a probabilistic one.\\
Let $A_E$ be the Erdos Renyi component of the graph, i.e, obtained by removing the Ring lattice edges from the graph. In our notation, $A_E \sim ER(n,\frac{(1-\beta) 2\delta}{n-1})$ for some $0<\beta<1$.

Now since the number of triangles in $A$ must be higher than just the ring lattice portion of the graph $A_R$, we can say
\[
C_A = \frac{T_A}{S_A} \geq \frac{T_{A_R}}{S_A}.
\]

Next we further note that,
\[
K_{1,\alpha}(p_0) = p_0 + o(p_0).
\]
This holds since in the case of $p_0 >>\frac{1}{\sqrt{n}}$, we have, 
\[
K_{1,\alpha}(p_0) = p_0 + O\left(\frac{\sqrt{p_0}}{n}\right)=p_0+o(p_0),
\]
and in the case of $\frac{1}{n}<<p_0 <<\frac{1}{\sqrt{n}}$, we have
\[
K_{1,\alpha}(p_0) = p_0 + O\left(\frac{1}{\sqrt{n^3p_0}}\right)=p_0+o(p_0),
\]
Therefore combining the two,
\[
K_{1,\alpha}(\hat{p}) = K_{1,\alpha}(p_0) + o_p(1) = (1+o(1))p_0 +o_p(1),
\]
for any $p_0>>\frac{1}{\sqrt{n}}$.

Therefore, the desired result follows if we can prove that
 \begin{equation}
 P\left(\frac{T_{A_R}}{S_A} \geq C_n \frac{2\delta}{n}\right) \rightarrow 1,
 \label{eqSA}
 \end{equation}
 for a sequence $C_n \to \infty$ at any rate.
 We note that the quantity $T_{A_R}$ is deterministic and calculate
 \[
 \dbinom{n}{3}T_{A_R} = n \dbinom{\delta \beta}{2} \asymp  n \delta^2\beta^2. 
 \]
 Then the result in \ref{eqSA} is  equivalent to
 \begin{equation}
 P\left(\dbinom{n}{3}S_A \leq c_2 \frac{n^2 \delta \beta^2 }{C_n}\right) \rightarrow 1,
 \label{eqSA1}
 \end{equation}
  where $c_2 $ is a constant not dependent $n$.
  
We note $S_A$ consists of three parts:
\[
\dbinom{n}{3} S_A = \sum_{i,j,k} \{ (S_{A_E})_{ijk} + (S_{A_R})_{ijk} + (E_{A_E})_{ij}(E_{A_R})_{jk}\}.
\]
The middle term is a deterministic quantity and is upper bounded by 
\[
\sum_{i,j,k} (S_{A_R})_{ijk} = \frac{n.2\delta \beta(2\delta \beta-1)}{2} \leq c_4n\delta^2\beta^2.
\]
By assumption of $\frac{2\delta C_n}{n} \rightarrow 0$, (i.e., the graph is not too dense), we have 
\[\sum_{i,j,k} (S_{A_R})_{ijk}  < c_2 \frac{n^2 \delta \beta^2 }{C_n}.
\]
Next we derive an upper bound for the first term which holds with high probability. For notational convenience, define $S_{A_E}=\sum_{i,j,k}(S_{A_E})_{ijk}$. Then note the expectation and variance of $S_{A_E}$ is \cite{janson2011random},
\begin{align*}
E[S_{A_E}] & =  \frac{n(n-1)(n-2)}{2}\frac{4\delta^2(1-\beta)^2}{(n-1)^2}  \asymp 2n\delta^2(1-\beta)^2\\
V[S_{A_E}] & \asymp n \delta^3 (1-\beta)^3.
\end{align*}
Then from Chebyshev's inequality
\[
P(|S_{A_E} - E[S_{A_E}]| \geq  E[S_{A_E}]) \leq \frac{V[S_{A_E}]}{(E[S_{A_E}])^2} \leq \frac{c_5n\delta^3(1-\beta)^3}{n^2\delta^4(1-\beta)^4}  = \frac{c_5}{n\delta(1-\beta)}, 
\]
for some constant $c_5$.
Therefore,
\[
P(S_{A_E} \leq c_6  n\delta^2(1-\beta)^2) \geq 1 - \frac{c_5}{n\delta(1-\beta)},
\]
where $c_5, c_6$ are constants independent of $n$.
By our assumption of $\frac{\delta C_n}{n} \rightarrow 0$, 
\[
c_6  n\delta^2(1-\beta)^2)< c_7 \frac{n^2 \delta \beta^2 }{C_n}.
\]

Finally, we tackle the third term. Note that the quantities $(E_{A_R})_{jk}$ are deterministic and we precisely know there are 
$n\delta \beta$ such edges. These incidental structures are formed if there is an ER edge in either end of the RL edge. Let $(d_{E})_i$ denote the degree of $i$th node in the graph $A_E$ and $d_{\max} = \max_i (d_{E})_i$ is the maximum degree of a node in $A_E$.
Since $(d_{E})_i$ is sum of i.i.d. Bernoulli random variables, from Bernstein inequality we get
\begin{align*}
P\bigg((d_{E})_i  \geq (1+ c_8)2\delta(1-\beta)\bigg) \leq \exp \left(-\frac{\frac{1}{2}c_8^2 4\delta^2(1-\beta)^2}{(n-1)\frac{2\delta (1-\beta)}{n-1} (1-\frac{2\delta (1-\beta)}{n-1} ) +\frac{1}{3}c_8 2\delta(1-\beta)}\right).
\end{align*}
 Taking an union bound over all $n$ vertices, we have
 \[
 P(d_{\max}\geq (1+ c_8)2 \delta(1-\beta)) \leq \exp(-c_9 (\delta(1-\beta) -\log n)) \to 0,
 \]
 by the assumption that $\delta>> \log n$.
Again for notational convenience, define $S_{A_{ER}} = \sum_{ijk} (E_{A_E})_{ij}(E_{A_R})_{jk}$. Therefore we can bound the total number of incidental ``V" shaped structures as
\[
P\left(S_{A_{ER}}\geq  n \delta \beta (n-2)\frac{2\delta (1-\beta)}{n-1} \right) \leq P\left(d_{\max}\geq (1+ c_8)\delta(1-\beta)\right) 
\]
and using the bound for the right hand side above we have
\[
P\left( S_{A_{ER}}\leq 2 n \delta \beta (n-2)\frac{\delta (1-\beta)}{n-1} \right) \geq 1- o(1).
\]
 Next, combining the three results we have
 \begin{equation}
 P\left( \dbinom{n}{3} S_A \leq c_2 \frac{n^2 \delta \beta^2 }{C_n}\right) \geq 1-o(1) \rightarrow 1.
 \end{equation}

\subsection{Proof of Theorem \ref{L-power}}
We start with some definitions.
For any pair of nodes $(u,v)$, let  $d(u,v)$ be the geodesic distance between $u$ and $v$.
$\Gamma_i(v)$ is defined as the set of vertices at distance $i$ from vertex $v$, that is,
    $ \Gamma_i(v) = \{u: d(u,v) = i\}$,
    and $|\Gamma_i(v)|$ is the number of vertices in the set $\Gamma_i(v)$.
    
\textbf{Proof of Part 1:}
When $A\sim ER(n,p)$, our proof strategy is to collect the union of the events where $L > K_2$ can happen, and show that the sum of their probabilities go to zero.
Define the following events:
\begin{enumerate}
    \item[E1:] The graph $G$ is not connected.
    
    It is well-known that $P[E1] \rightarrow 0 $ when $\frac{np}{\log(n)} \rightarrow \infty$, which is our assumption.
This result established by \cite{erdos1959random} is one of the most celebrated results in random graph theory.
We skip the proof in the interest of space.

    \item[E2:] $\{\hat{p} > p (1 + \frac{1}{\sqrt{n \log(n)}})\}$.
    
    We know that, for any $\epsilon \in (0,1)$, by Chernoff's inequality,
    $$
    P[\hat{p} \geq (1+\epsilon) p] \le \exp\left(-\frac{\epsilon^2 {n \choose 2}p}{3}\right).
    $$
    Put $\epsilon = \frac{1}{\sqrt{n \log(n)}}$.
    Then, since ${\epsilon^2 {n \choose 2}p} = \frac{(n-1)p}{2\log(n)} \rightarrow \infty$ by assumption,
    the probability on the right hand side goes to zero, which implies $P[E_2] \rightarrow 0$.
    
    (Note that we do not have to consider the case $\hat{p} < p$, since that makes $K_2$ larger than its population version, and therefore the right-tail probability is even lower.)
    
    \item[E3:] There is at least one vertex $v$ and some $i \in (1, \frac{2 \log(n)}{3 \log(np)})$ such that 
    $
    |\Gamma_i(v)| \le \frac{1}{2}(np)^i
    $.

    Here we apply Lemma 8 of \cite{chung2001diameter}, which states that: when $p \geq \frac{c \log(n)}{n}$ for some $c>2$, then for any fixed vertex $v$, any fixed $i \in (1, \frac{2 \log(n)}{3 \log(np)})$, and any $\epsilon > 0$,
    $$
P[|\Gamma_i(v)| \le  (1 - \sqrt{2/c} - \epsilon)(np)^i] = o\left(\frac{1}{n}\right).
$$
Fix any $\epsilon > 0$.
Since $\frac{np}{\log(n)} \rightarrow \infty$, we can use $c=\frac{2}{\epsilon^2}$, which implies that
$$
P[|\Gamma_i(v)| \le  (1-2\epsilon)(np)^i] = o\left(\frac{1}{n}\right).
$$
Taking union over $v$,
$$
P\left[\cup_{v \in (1, \ldots, n)} \left\{|\Gamma_i(v)| \le  \frac{1}{2}(np)^i \right\}\right]  = o(1).
$$
Therefore, $P[E_3] \rightarrow 0$.

    \item[E4:] There exists a pair of vertices $(u,v)$, integers $k_1, k_2$ such that 
    $$
    |\Gamma_{k_1}(u)||\Gamma_{k_1}(v)|p > (2+\epsilon) \log(n)
    $$
    for some $\epsilon>0$, but $ d(u,v) > k_1 + k_2 + 1$.

Suppose such a pair exists. There can be two cases.
    
    Case 1: $\Gamma_{k_1}(u) \cap \Gamma_{k_2}(v)$ is not null.
    Then, there is a path of length less than or equal to $k_1+k_2$ from $u$ to $v$, which means $d(u,v) \le k_1 + k_2$.
    
    Case 2: $\Gamma_{k_1}(u) \cap \Gamma_{k_2}(v)$ is null.
    Let's compute the probability that there is no edge between $\Gamma_{k_1}(u)$ and $\Gamma_{k_2}(u)$.
    This probability is given by
    $$
    (1-p)^{|\Gamma_{k_1}(u)||\Gamma_{k_1}(v)|}.
    $$
Note that $e^{-p} \geq (1-p)$. Therefore,
\begin{align*}
    (1-p)^{|\Gamma_{k_1}(u)||\Gamma_{k_1}(v)|}
     \le 
    \exp(-p|\Gamma_{k_1}(u)||\Gamma_{k_1}(v)|)\le 
    \exp(-(2+\epsilon) \log(n))
    = \frac{1}{n^{2+\epsilon}}.
    \end{align*}
Therefore,
$$
P[\cup_{uv} \{\text{ No edge between }\Gamma_{k_1}(u) \text{ and }\Gamma_{k_1}(u)\}]
\le 
\frac{1}{n^{\epsilon}}
\rightarrow
0.
$$
Therefore, $P[E_4] \le  \frac{1}{n^{\epsilon}} \rightarrow 0$.
\end{enumerate}

Now, armed with the fact that $P[E1 \cup E2 \cup E3 \cup E4] = 0$, we proceed to complete the proof.
Fix any $\epsilon > 0$, and choose
$$
k_1 = \left\lceil \frac{\log(\sqrt{2(1+\epsilon)n\log(n)/(1-2\epsilon)})}{\log(np)} \right\rceil,
$$
$$
k_2 = \left\lceil \frac{\log({2(1+\epsilon)n\log(n)/(1-2\epsilon)^2})}{\log(np)}-k_1-1 \right\rceil.
$$
Then $k_1, k_2 \in (1, \frac{2 \log(n)}{3 \log(np)})$, \footnote{See the proof of Theorem 2 of \cite{chung2001diameter} for details} and therefore,
$$
|\Gamma_{k_1}(u)||\Gamma_{k_2}(v)| 
\geq
(1-2\epsilon)(np)^{k_1} \times (1-2\epsilon)(np)^{k_2}
\geq
2(1+\epsilon) \log(n)
$$
with probability greater than $1-P[E3]$.
Therefore, from the result on E4, we can say that with probability $1-P[E3 \cup E4]$, 
the path length between any two vertices is less than or equal to $k_1+k_2+1$.
This implies that with probability $1-P[E3 \cup E4]$, the average path length
is less than or equal to $k_1+k_2+1$.
Thus, we obtain
$$
P\left[
L \le \left\lceil \frac{\log({2(1+\epsilon)n\log(n)/(1-2\epsilon)^2})}{\log(np)}\right\rceil
\right]
\rightarrow 
1.
$$

We can now conclude that, for any $\epsilon' >0$,
\begin{equation}
\label{K2pop}
    P\left[L > \left\lceil \frac{\log({2(1+\epsilon')n\log(n)})}{\log(np)}\right\rceil\right] \rightarrow 0.
\end{equation}
However, $\lceil x \rceil$ could be anything from $x$ to $x+1$.
Therefore, to be abundantly conservative, we use
\begin{align*}
\left\lceil \frac{\log({2(1+\epsilon')n\log(n)})}{\log(np)}\right\rceil 
& \le
\frac{\log({2(1+\epsilon')n\log(n)})}{\log(np)} + 1\\
& =
\frac{\log({2(1+\epsilon')n^2p\log(n)})}{\log(np)}
\le
\frac{(2+\delta)\log(n)}{\log(np)}
\end{align*}
for any $\delta>0$.
Finally, we have to adjust for the fact that
$
\hat{p} \le p (1 + \frac{1}{\sqrt{n \log(n)}})$,
where we use the Taylor series approximation
$
\log(1+x) \approx x
$
for the denominator.
This gives us the final bound,
$$
K_2 = \frac{(2+\epsilon)\log(n)}{\log(n\hat{p})},
$$
for any $\epsilon>0$.

\textbf{Proof of Part 2:}\\
When $A \sim NW(n,p,\beta)$, consider the network $A'$ obtained by removing the ring lattice edges.
Note that $A' \sim ER(n,(1-\beta)p)$, so it follows from Equation \eqref{K2pop} that
$$
P\left[L(A') > \frac{(2+\epsilon') \log(n)}{\log((1-\beta)np)}\right] \rightarrow 0
$$
for any $\epsilon'>0$.
Next, we prove that 
$$
P\left[K_2 > \frac{(2+\epsilon') \log(n)}{\log((1-\beta)np)}\right]  \rightarrow 1
$$
as $n \rightarrow \infty$.
To see this,
fix some $\epsilon \in (0,1)$, and let
$
K_2 = \frac{(2+\epsilon)\log(n)}{\log(n\hat{p})}$.
Let 
$\epsilon' = \epsilon/2$, and let 
$K_2' = \frac{(2+\epsilon') \log(n)}{\log((1-\beta)np)}$.
Then,
$$
\frac{K_2}{K_2'}
= 
\frac{(2+\epsilon) \log((1-\beta)np)}
{(2+\epsilon/2) \log(n\hat{p})}
= 
\frac{(2+\epsilon)}
{(2+\epsilon/2)}
\frac{\log(np) + \log(1-\beta)}
{\log(np) + \log(\hat{p}/p)}.
$$
Therefore, from the result on E2 from Part 1,
\begin{align*}
\frac{K_2}{K_2'}
& =
\frac{(2+\epsilon)}
{(2+\epsilon/2)}
\frac{\log(np) + \log(1-\beta)}
{\log(np) + \log(1 + \frac{1}{\sqrt{n \log(n)}})}\\
\geq
& \left(1+\frac{\epsilon}{6}\right)
\frac{\log(np) + \log(1-\beta)}
{\log(np) + \log(1 + \frac{1}{\sqrt{n \log(n)}})}
\end{align*}
with probability $1-P[E2]$.
In the final expression, the first part is greater than 1, and the second part converges to 1 as $n \rightarrow \infty$, so we can choose $\epsilon$ to ensure that the product is greater than 1.
Thus, 
$$
P\left[L(A') > K_2\right] 
\le
P\left[L(A') > K_2'\right]
+
P[K_2 < K_2']
\rightarrow
0.
$$
Clearly,  $L(A) \le L(A')$, since $A$ has more edges than $A'$, and every additional edge has a non-decreasing effect on the average path length.
Therefore, we have proved part 2.

\textbf{Proof of Part 3:}
For a ring lattice, we have
$$
L  \approx \frac{n}{2n\hat{p}}  \approx \frac{1}{2{p}}, \text{ and }
K_2 \approx \frac{(2+ \epsilon) \log(n)}{\log(np)}
$$
for large enough $n$ and small $\epsilon > 0$.

Therefore, it suffices to prove that for some $\epsilon > 0$,
 \begin{align*}
    &\frac{1}{2{p}} > \frac{(2+ \epsilon) \log(n)}{\log(np)}\\
    & \Leftrightarrow 
    \log(n) + \log(p) > 2(2+\epsilon) p \log(n) \\
    & \Leftrightarrow 
    \log(n)(1-2(2+\epsilon)p) + \log(p) > 0,
 \end{align*}
 which is true since $p < 1/4$ and for large enough $n$.

\subsection{Proof of Theorem \ref{betapower}}

\textbf{Power analysis as $\beta$ approaches 0:}

We let $\beta$ be a function of $n$, say, $\beta= \frac{h}{n^{l}}$, for some $h, l>0$. Note that cut-off for rejecting the clustering coefficient does not change with $\beta$, as should be expected for a test. Therefore we need to prove that $\dbinom{n}{3}S_A$ is upper bounded by $c_2 \frac{n^2 \delta \beta^2}{C_n}$ for some constant $c_2$ with high probability. The deterministic quantity $S_{A_R}$ is still upper bounded by this quantity for any value of $\beta>0$. For $S_{A_E}$, the random variable is upper bounded by a constant times the expectation happens with high probability for any $\beta$ that is small. However, in order for the upper bound be smaller than  $c_2 \frac{n^2 \delta \beta^2}{C_n}$, we get a restriction that
\[
n \delta^2 \leq c_8 \frac{n^2 \delta \beta^2}{C_n} \Rightarrow \beta > c_9 \sqrt{\frac{\delta C_n}{n}}. 
\]
If we assume $p \asymp \frac{n^{\epsilon}}{n}$, then we obtain $\beta \gtrsim n^{-1/2+\epsilon},$ for some $\epsilon>0$. 

By comparing the upper bound for $S_{A_{ER}}$ with $C_2 \frac{n^2 \delta \beta^2}{\log n}$, we get another restriction on the growth rate of $\beta$, namely,  
\[
\beta \gtrsim \frac{\delta \log n}{n}. 
\]
Similarly, if we assume $p \asymp \frac{\log n}{n}$, then we obtain $\beta \gtrsim n^{-1+\epsilon},$ for some $\epsilon>0$. Therefore we can let $\beta = O(\frac{1}{n^{1/2-\epsilon}})$, and still have power tend to 1.

\textbf{Power analysis as $\beta$ approaches 1:}
Suppose $1-\beta \geq \frac{1}{n^l}$ where $l = \frac{\epsilon}{12} \frac{\log (np)}{\log n}$.
We retrace the steps from the proof of part 3 until the definition of $K_2'$.
Next, note that
\begin{align*}
    \frac{K_2}{K_2'}
& = 
\frac{(2+\epsilon) \log((1-\beta)np)}
{(2+\epsilon/2) \log(n\hat{p})}\\
&\geq 
\left(1+\frac{\epsilon}{6}\right)
\frac{\log(np) + \log(1-\beta)}
{\log(np) + \log(1 + \frac{1}{\sqrt{n \log(n)}})}\\
&\geq 
\left(1+\frac{\epsilon}{6}\right)
\left(1-\frac{\epsilon}{12}\right)
\frac{\log(np)}
{\log(np) + \log(1 + \frac{1}{\sqrt{n \log(n)}})}.
\end{align*}
As before, note that $\left(1+\frac{\epsilon}{6}\right)
\left(1-\frac{\epsilon}{12}\right) \geq 1$, and $\frac{\log(np)}
{\log(np) + \log(1 + \frac{1}{\sqrt{n \log(n)}})}$ converges to 1 as $n \rightarrow \infty$, so we can choose $\epsilon$ to ensure that the product is greater than 1.
The remaining steps follow from the proof of part 2.

\subsection{Proof of Theorem \ref{C-inhomo}}
We first obtain asymptotic order of the term $S-E[S]$ following ideas in \cite{gao2017testing2}. We note
\[
E[(S-E[S])^2] = O\left(\frac{\theta_{\max}^3}{n^2}\right).
\]
Since convergence in MSE implies convergence in probability,
\[
S= E[S] + O_p\left(\frac{\theta_{\max}^{1.5}}{n}\right),
\]
Since $E[S] \asymp 3 \theta_{\max}^2$, this implies
\[
\frac{S}{E[S]} = 1 + O_p\left(\frac{1}{n\theta_{max}^{1/2}}\right)= o_p(1).
\]
A modification of the limit result in \cite{gao2017testing2} also yields
   \[
    \frac{\sqrt{\dbinom{n}{3}} (T-E[T])}{\sqrt{\Theta_T}}  \overset{D}{\to} N(0,1).
    \]
Applying Slutsky's theorem we have 
   \[
    \frac{\sqrt{\dbinom{n}{3}} E[T](T/E[T]-1)}{\sqrt{\Theta_T} S/E[S]}  \overset{D}{\to} N(0,1).
    \]
   Now we note that the deterministic quantities, $E[T]=O(\theta_{\max}^3)$, and $\Theta_T = \Omega(\theta_{\min}^3)$, such that $\frac{1}{\sqrt{\Theta_T}}= O(\frac{1}{\theta_{\min}^{1.5}})$. Then we have 
    \begin{align*}
       \frac{\sqrt{\dbinom{n}{3}} E[T](E[S]/S-1)}{\sqrt{\Theta_T} } & =O(n^{1.5}) O(\theta_{\max}^3) O(\frac{1}{\theta_{\min}^{1.5}}) O_p(\frac{1}{n\theta_{\min}^{1/2}}) \\
       & = O_p(\sqrt{n}\frac{\theta_{\max}^3}{\theta_{\min}^2})=o_p(1),
    \end{align*}
    due to our assumption that $\theta_{\max} \asymp \theta_{\min} << \frac{1}{\sqrt{n}}$.

Therefore, we obtain the desired result.

\subsection{Proof of Theorem \ref{C-power-inhomo}}
We follow the structure of the proof of Theorem \ref{C-power}. Note when $\beta=0$, the asymptotic normality result in Theorem \ref{C-inhomo} implies that $P(C > K_{1,\alpha}(\theta)) \to \alpha$ and assumption A.2, states that $K_{1,\alpha}(\hat{\theta}) = K_{1,\alpha}(\theta) + o_p(1)$. Combining these two results we have the asymptotic level
\[
P(C > K_{1,\alpha}(\hat{\theta})) \to \alpha.
\]

When $\beta >0$, similarly to the proof of Theorem \ref{C-power}, we define two auxiliary graphs from the observed graph $A$. The graph $A_R$ as before is obtained by removing the inhomogeneous random graph edges from the superimposed model. Then $A_R$ is a deterministic ring lattice with $n$ vertices and degree $2 \beta \delta$. The graph $A_E$ denotes the inhomogeneous random graph component obtained by removing the ring lattices edges from the graph, with $n$ vertices and expected density $\tilde{\theta}=\frac{(1-\beta)2\delta}{n-1}$.

Since our assumptions imply $K_{\alpha}(\hat{\theta}) =(1+o_p(1))\frac{t(\theta)}{s(\theta)}$, then to show asymptotic power converges to $1$, we need to prove that
 \begin{equation}
 P\left(\frac{T_{A_R}}{S_A} \geq C_n \frac{t(\theta)}{s(\theta)}\right) \rightarrow 1,
 \end{equation}
 for a sequence $C_n \to \infty$ at any rate.
Following the same argument in the proof of Theorem \ref{C-power}, this is equivalent to,
 \begin{equation}
 P\left(\dbinom{n}{3}S_A \leq c_2 \frac{n \delta^2 \beta^2 }{C_n \frac{t(\theta)}{s(\theta)}}\right) \rightarrow 1.
 \end{equation}
Using similar notations as in the proof of Theorem \ref{C-power},
We note $S_A$ consists of three parts, namely, $S$ structures from ring lattice, $S$ structures from inhomogeneous random graph component and incidental structures formed when there is an inhomogeneous random graph edge attached to either side of a ring lattice edge. 
\[
\dbinom{n}{3} S_A = \sum_{i,j,k} \{ (S_{A_E})_{ijk} + (S_{A_R})_{ijk} + (E_{A_E})_{ij}(E_{A_R})_{jk}\}.
\]
The middle term is upper bounded by 
$\sum_{i,j,k} (S_{A_R})_{ijk} \leq c_4n\delta^2\beta^2 < c_2 \frac{n \delta^2 \beta^2 }{C_n \frac{t(\theta)}{s(\theta)}},$
by assumption of $C_n\frac{t(\theta)}{s(\theta)} \rightarrow 0$, 

As before, defining $S_{A_E}=\sum_{i,j,k}(S_{A_E})_{ijk}$ the expectation and variance of $S_{A_E}$ is,
\begin{align*}
E[S_{A_E}] & \gtrsim  n^3\theta_{\min}^2(1-\beta)^2, \quad \text{ and } E[S_{A_E}] \lesssim  n^3\theta_{\max}^2(1-\beta)^2\\
V[S_{A_E}] & \lesssim n^4 \theta_{\max}^3 (1-\beta)^3.
\end{align*}
Then from Chebyshev's inequality
\[
P(|S_{A_E} - E[S_{A_E}]| \geq  E[S_{A_E}]) \leq \frac{V[S_{A_E}]}{(E[S_{A_E}])^2} \leq \frac{c_5n^4 \theta_{\max}^3(1-\beta)^3}{n^6\theta_{\min}^4(1-\beta)^4}  = \frac{c_5}{n^2\frac{\theta_{\min}^4}{\theta_{\max}^3}(1-\beta)}, 
\]
for some constant $c_5$.
Since $\theta_{\min} \asymp \theta_{\max}>>\frac{1}{n}$,
\[
P(S_{A_E} \leq c_6  n^3\theta_{\max}^2(1-\beta)^2) \geq 1 - o(1).
\]
By our assumption of $C_n\frac{t(\theta)}{s(\theta)} \rightarrow 0$, and the fact that $n^2\theta_{\max}^2 \asymp \delta^2$,
\[
c_6  n^3\theta_{\max}^2(1-\beta)^2< c_7 \frac{n \delta^2 \beta^2 }{C_n \frac{t(\theta)}{s(\theta)}}.
\]

For the third term, we proceed as follows. Recall, $(d_{E})_i$ denotes the degree of $i$th node in the graph $A_E$ and $d_{\max} = \max_i (d_{E})_i$ is the maximum degree of a node in $A_E$.
From Corollary A.1.10 of~\cite{alon2004probabilistic}, we have the following proposition,
\begin{proposition}
For independent Bernoulli random variables $X_u \sim Bern(p_u), u=1,\ldots,n$ and $p=\frac{1}{n}\sum_{u}p_{u}$, we have
\[
P(\sum_{u} (X_u -p_u) \geq a) \leq \exp (a- (a+ pn) \log (1+ a/pn)).
\]
\end{proposition}
Applying this to the sum $(d_{E})_i$, we note that 
\begin{align*}
P((d_{E})_i -2\delta \geq c_8 2 \delta) \leq \exp (c_8 2 \delta- (c_8 2 \delta+ 2 \delta ) \log (1+ c_8 )).
\end{align*}
 Taking an union bound over all $n$ vertices, we have
 \[
 P(d_{\max}\geq (1+ c_8)2\delta(1-\beta)) \leq \exp(-c_9 (2\delta +\log n)) \to 0,
 \]
 as long as $\delta >>\log n$.
We use the notation $S_{A_{ER}} = \sum_{ijk} (E_{A_E})_{ij}(E_{A_R})_{jk}$ and bound the total number of incidental ``V" shaped structures as
\[
P\left(S_{A_{ER}}\geq 2 n \delta \beta (n-2)\frac{\delta (1-\beta)}{n-1} \right) \leq P\left(d_{\max}\geq (1+ c_8)2\delta(1-\beta)\right) = o(1). 
\]
We note $n \delta \beta (n-2)\frac{\delta (1-\beta)}{n-1} < c_2 \frac{n \delta^2 \beta^2 }{C_n\frac{t(\theta)}{s(\theta)}} $, by our assumptions.
Then combining the three results we have
 \begin{equation}
 P\left( \dbinom{n}{3} S_A \leq c_2 \frac{n \delta^2 \beta^2 }{C_n\frac{t(\theta)}{s(\theta)}}\right) \geq 1-o(1) \rightarrow 1.
 \end{equation}

\section{Appendix B: Additional Simulations and results}

\subsection{Simulation for finite sample behavior of results in Propositions 1 and 2}

\begin{figure*}
\centering
\begin{subfigure}{0.45 \linewidth}
\includegraphics[width=\linewidth]{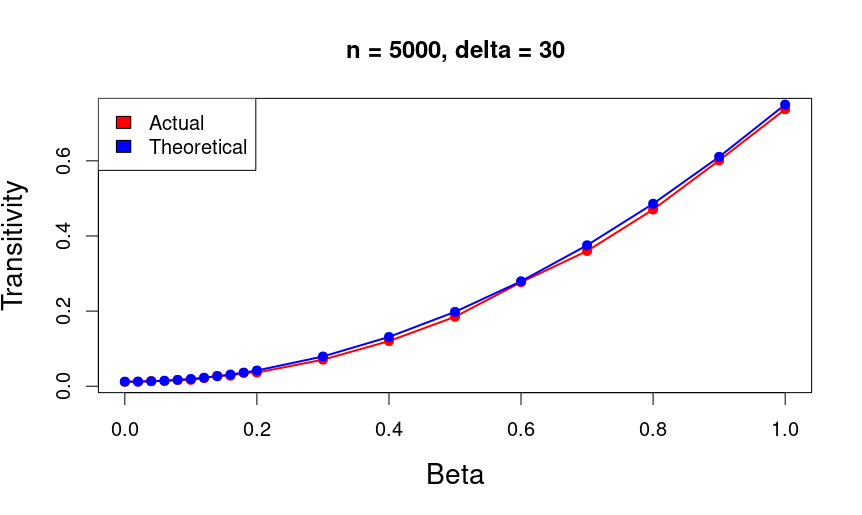}
\end{subfigure}%
\begin{subfigure}{0.45 \linewidth}
\includegraphics[width=\linewidth]{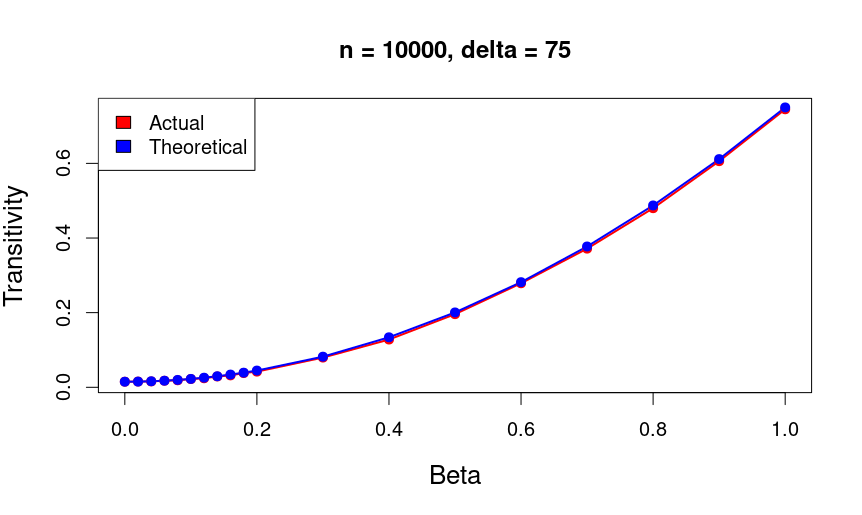}
\end{subfigure}
\caption{Theoretical asymptotic expectation and observed sample mean of C over 100 networks.}
\label{Csim}
\end{figure*}

 We perform a simulation to verify the asymptotic expression for the expected value of $C$ in NW-ER superimposed model given in Proposition 1. We consider two scenarios, one with $n=5000$ and  $\delta = 30$ and another with $n=10000$, and  $\delta=75$ and generate 100 networks from the NW-ER superimposed model with increasing $\beta$ (Figure \ref{Csim}). We note that for $n=10000$ and $\delta = 75$, such that the conditions of the proposition are reasonably satisfied, the theoretical result is very close to the sample average of $C$ over the 100 networks.

\begin{figure*}
\centering
\begin{subfigure}{0.45 \linewidth}
\includegraphics[width=\linewidth]{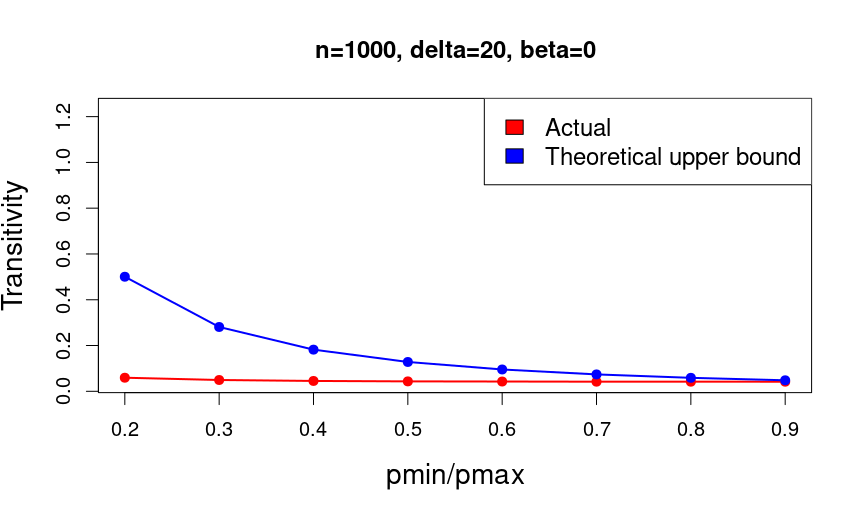}
\end{subfigure}%
\begin{subfigure}{0.45 \linewidth}
\includegraphics[width=\linewidth]{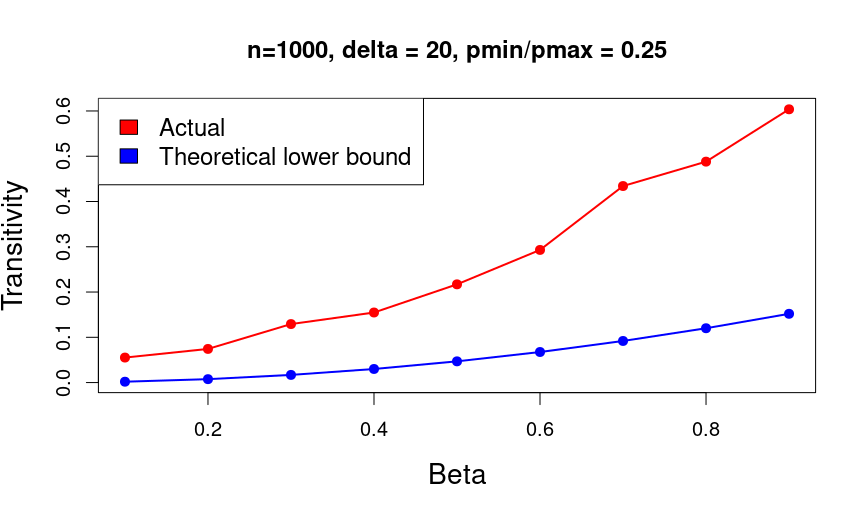}
\end{subfigure}
\begin{center}
    (a) \hspace{175pt} (b)
\end{center}
\begin{subfigure}{0.45 \linewidth}
\includegraphics[width=\linewidth]{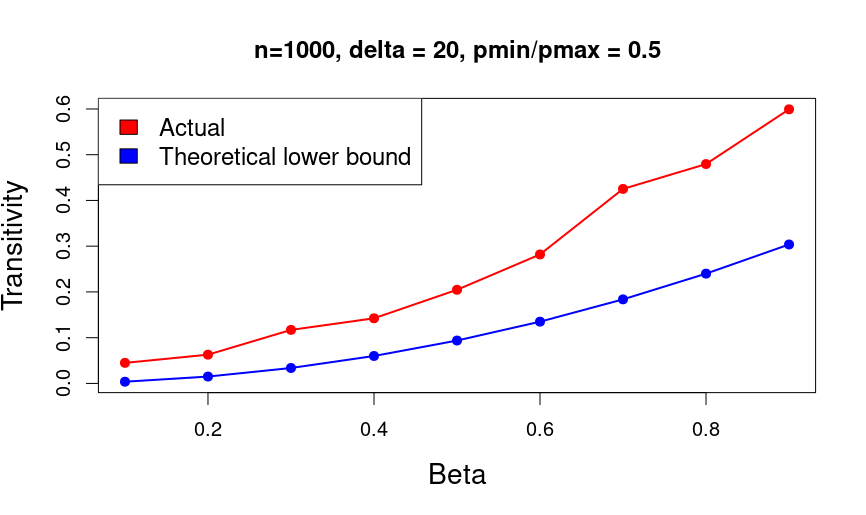}
\end{subfigure}
\begin{subfigure}{0.45 \linewidth}
\includegraphics[width=\linewidth]{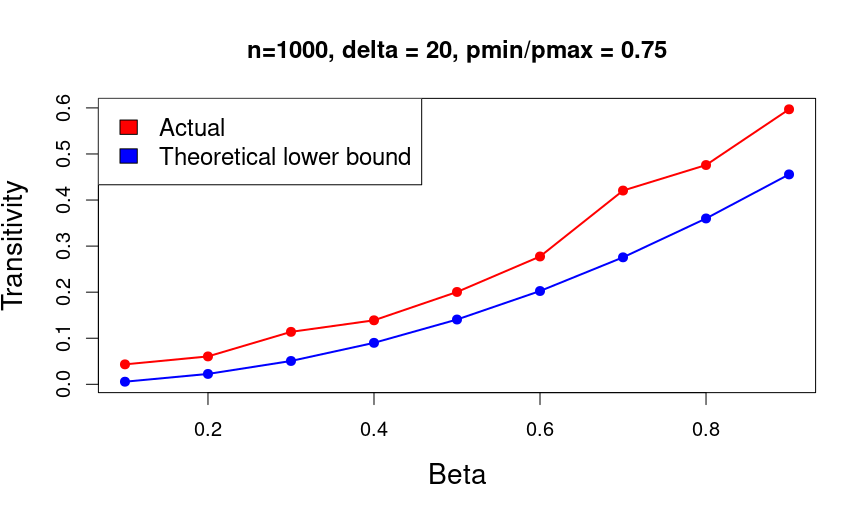}
\end{subfigure}
\begin{center}
    (c) \hspace{175pt} (d)
\end{center}
\caption{(a) Comparison of observed mean of $C$ and theoretical upper bound from Proposition 2 for inhomogeneous random graph with increasing ratio of $p_{\min}$ and $p_{\max}$, (b), (c), (d) Comparison of observed mean of $C$ and theoretical lower bound from Proposition 2 for superimposed inhomogeneous random graph model with increasing $\beta$ for 3 different cases of the ratio of $p_{\min}$ and $p_{\max}$. In each case the observed mean of $C$ is computed over 50 simulated networks.}
\label{prop2sim}
\end{figure*}

To empirically verify the bounds in the proposition 2, we conduct a simulation with $n=1000$ and average degree $\delta = 20$. Figure \ref{prop2sim} presents the results with 50 simulated networks. In Figure \ref{prop2sim} (a), we simulate networks from the inhomogeneous random graph model (therefore $\beta=0$) with different values of the ratio of $p_{\min}$ and $p_{\max}$. Clearly the theoretical upper bound computed in part (a) of the proposition is above the actual average of the $C$ values in this case, with the difference decreasing as the ratio of $p_{\min}$ to $p_{\max}$ increases. In Figure \ref{prop2sim} (b), (c), and (d), we present results of the observed average of $C$ from the NW superimposed inhomogeneous random graph model with varying ratio of $p_{\min}$ and $p_{\max}$ and mixing proportion $\beta$. In each case, we observe the theoretically computed lower bound is below the observed average, with the gap decreasing as the ratio of $p_{\min}$ to $p_{\max}$ increases. 

\subsection{Comparing community detection methods on real datasets}

\begin{figure*}[h!]
   \centering
    \begin{subfigure}[b]{\textwidth}
        \centering
\includegraphics[width=0.9\linewidth, height=3.5cm]{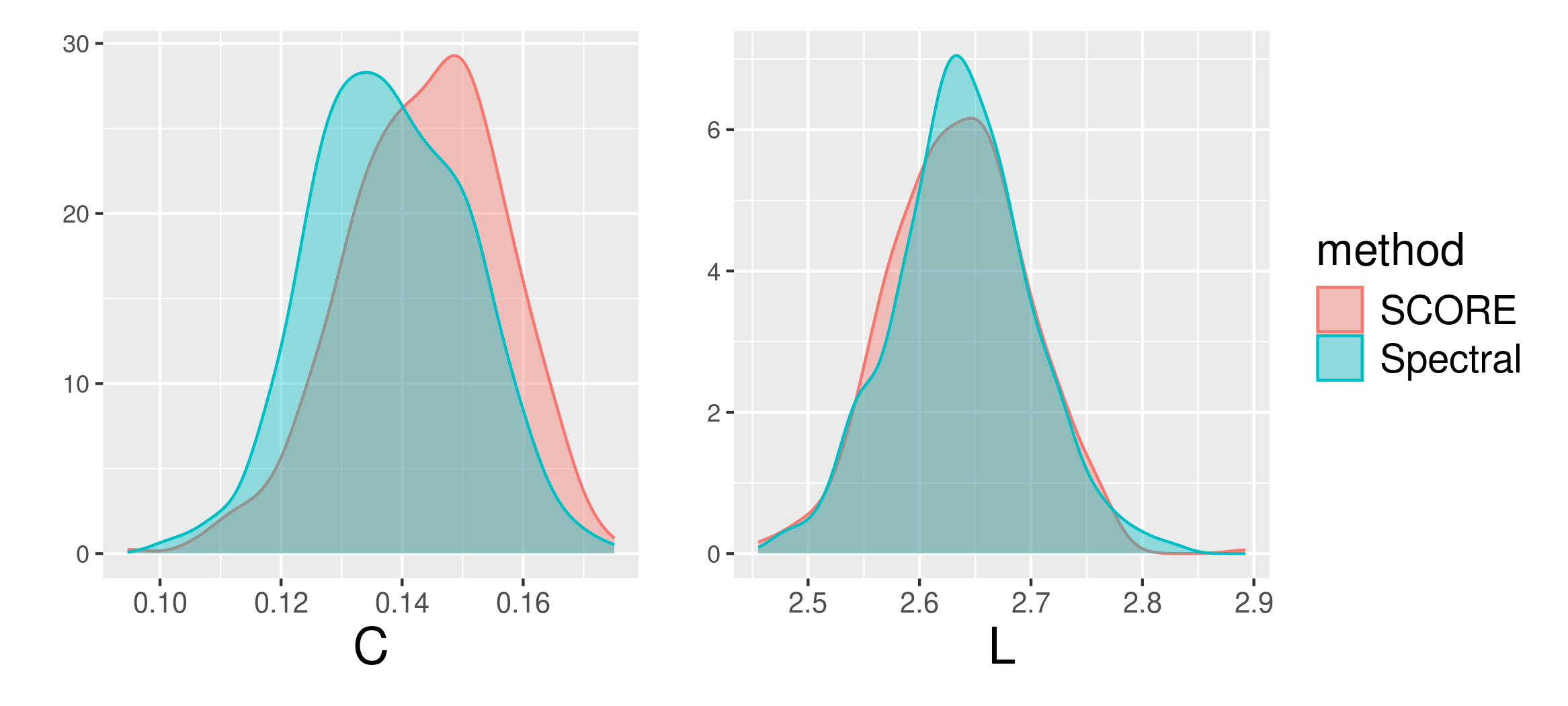}
        \caption{Dolphins \cite{lusseau2003bottlenose}}
    \end{subfigure}
    \vskip\baselineskip
    
    \begin{subfigure}[b]{\textwidth}
        \centering
        \includegraphics[width=0.9\linewidth, height=3.5cm]{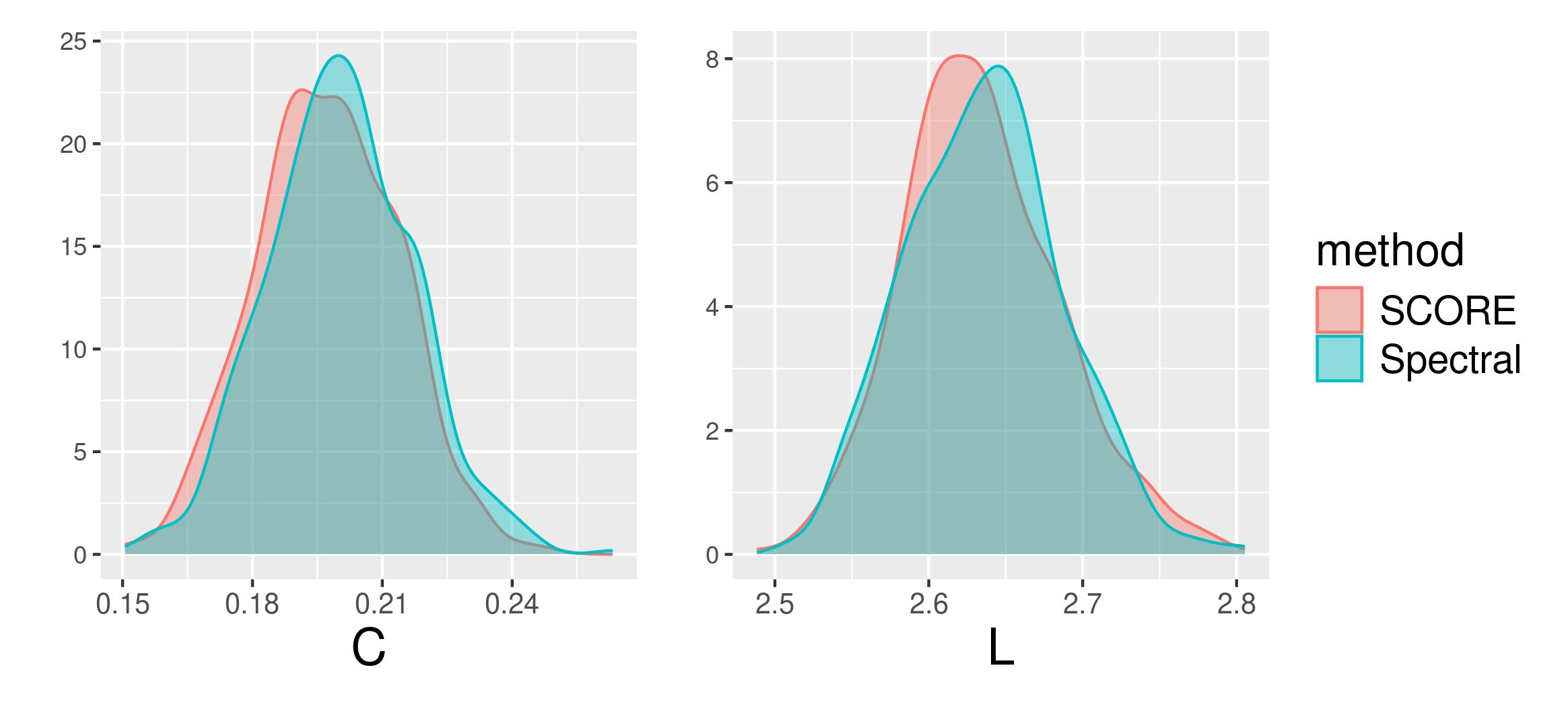}
       \caption{Football \cite{girvan2002community}}
    \end{subfigure}
    \begin{subfigure}[b]{\textwidth}
        \centering
        \includegraphics[width=0.9\linewidth, height=3.5cm]{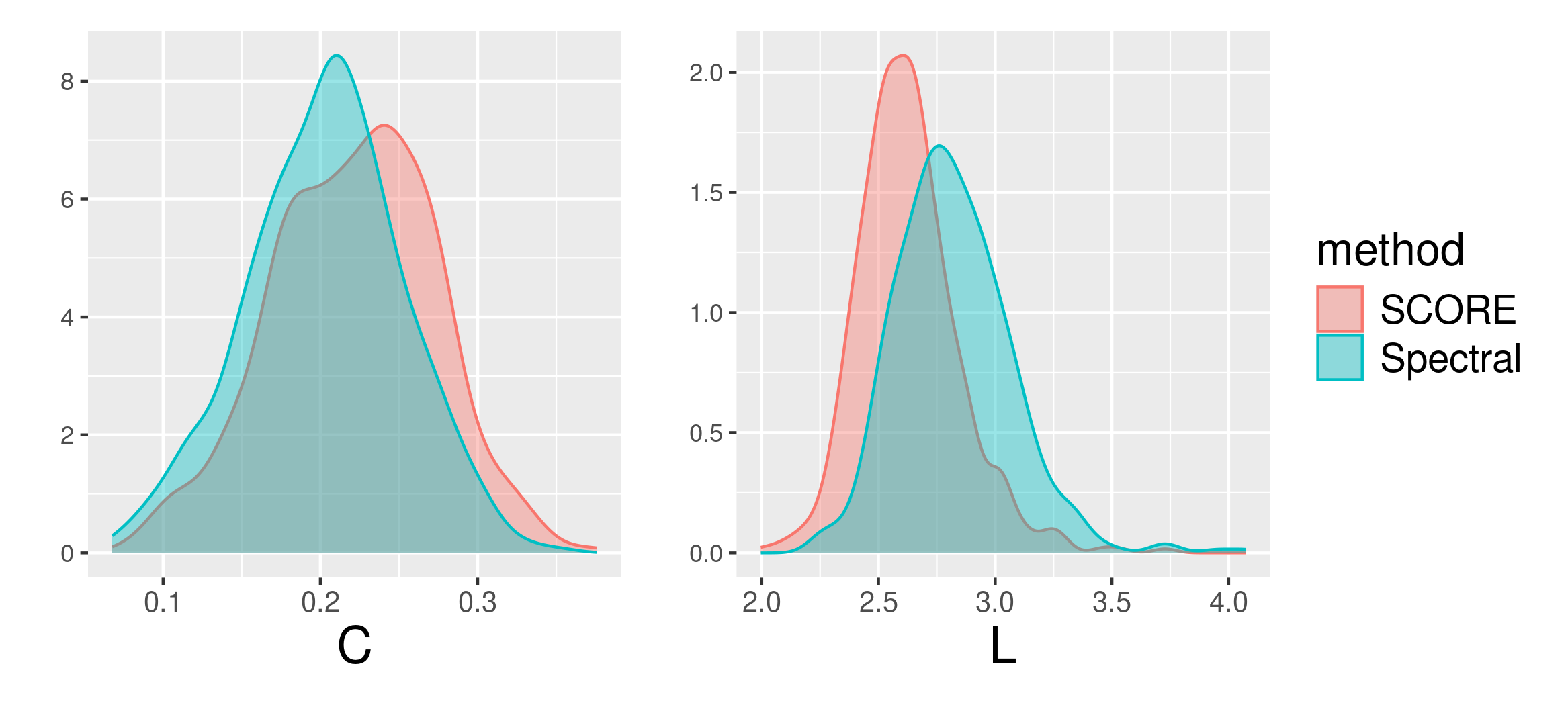}
        \caption{Karate Club network \cite{zachary1977information}}
    \end{subfigure}
    \begin{subfigure}[b]{\textwidth}
        \centering
        \includegraphics[width=0.9\linewidth, height=3.5cm]{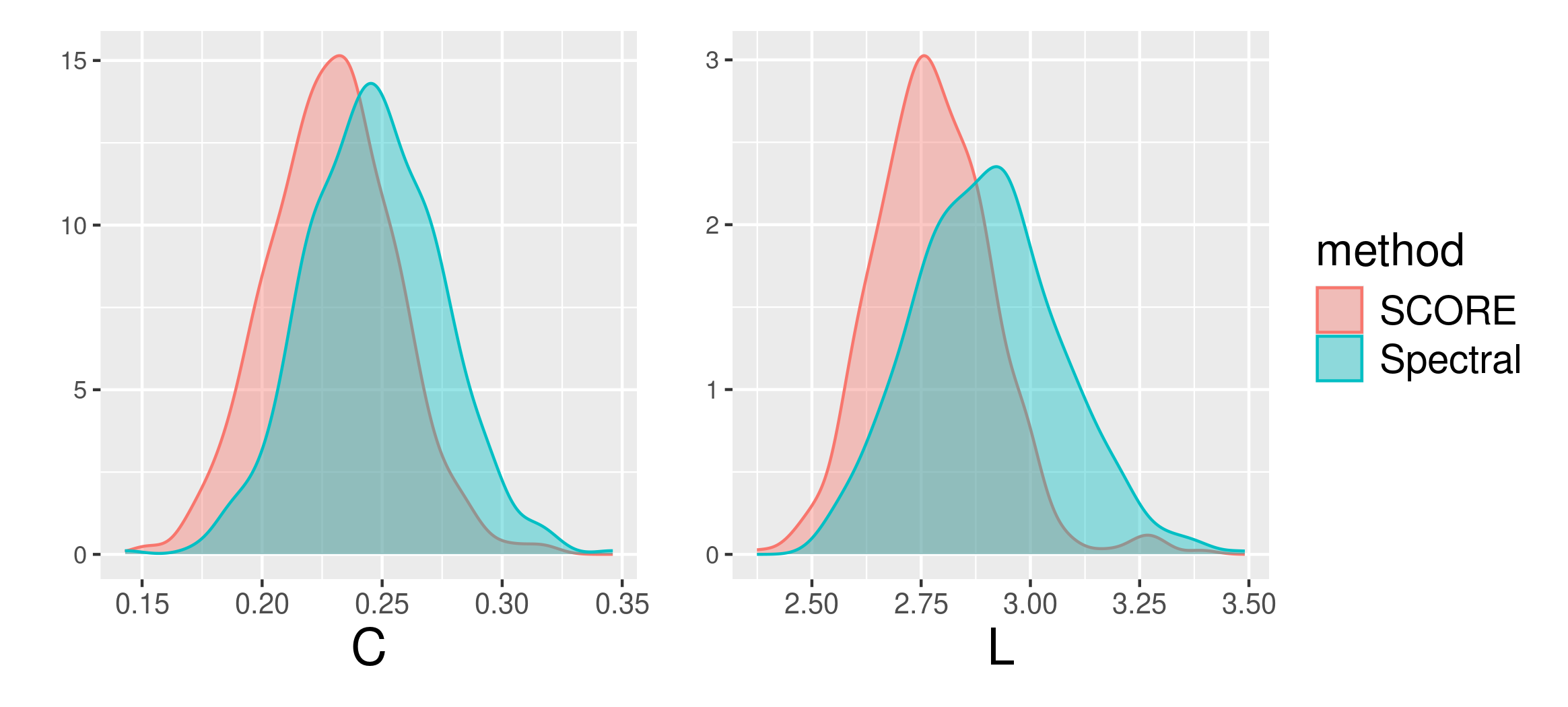}
       \caption{Les Miserables \cite{knuth1993stanford}}
    \end{subfigure}
    \begin{subfigure}[b]{\textwidth}
        \centering
        \includegraphics[width=0.9\linewidth, height=3.5cm]{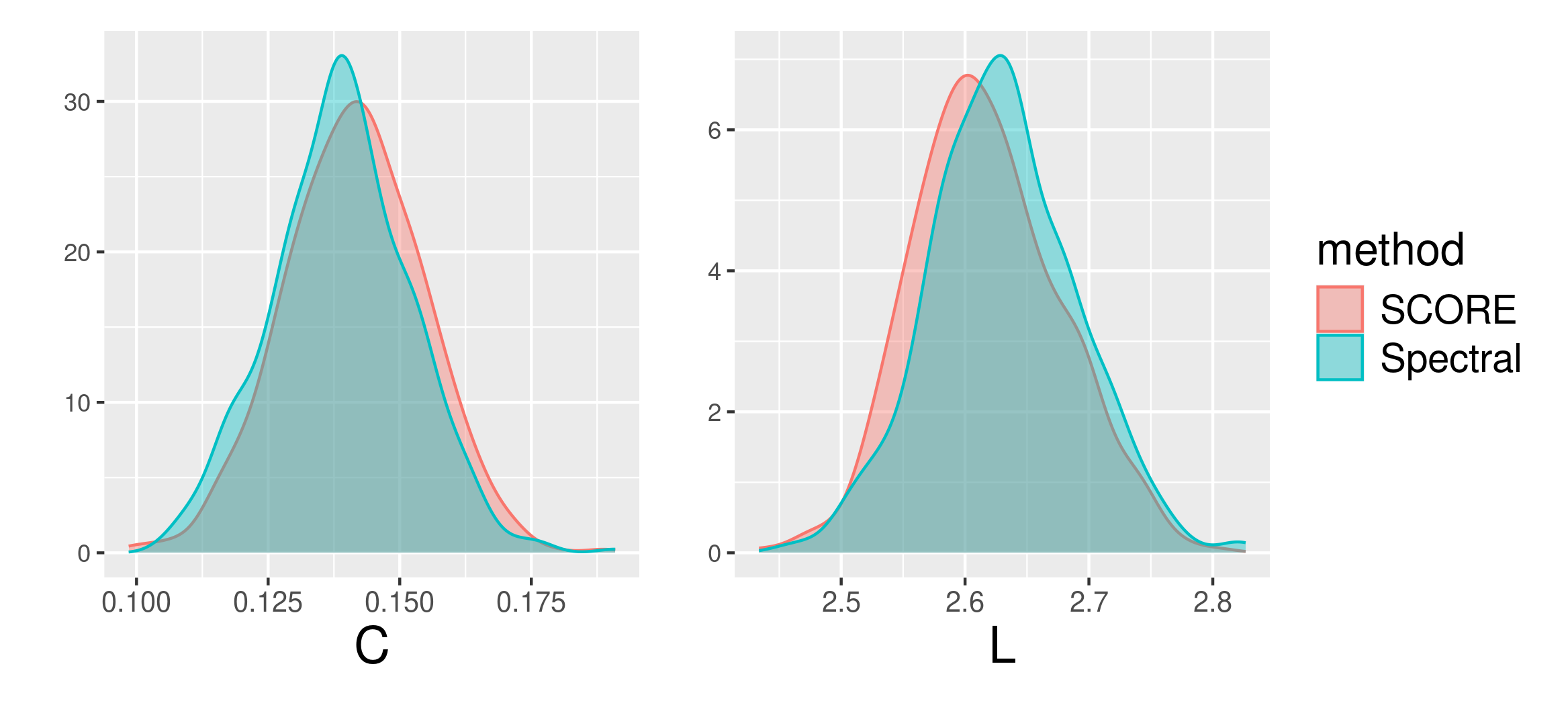}
        \caption{Word Adjacency \cite{newman2006finding}}
    \end{subfigure}
    \caption{Comparison of community detection methods}
   \label{fig:comm_det_comp}
\end{figure*}

We perform a simulation to compare the effect of using different community detection methods on the bootstrap test under a DCSBM null. In figure \ref{fig:comm_det_comp} we generate 500 networks to obtain the sampling distributions for both $C$ and $L$ for five real world networks using SCORE \cite{jin2022mixedmembershipestimationsocial} and spectral methods for community detection. We find that the sampling distributions appear to be similar, indicating the choice of community detection method does not have a large impact on the test.

\subsection{Results from the asymptotic test on real datasets}

\begin{table}[h!]
    \centering
    \begin{tabular}{c|c|c|c|c|c}
    \hline
         Network & $C$ & $K_C$ & $L$ & $K_l$ & Decision \\
         \hline
          C Elegans & 0.132 & 0.039 & 2.455 & 4.255 & Reject \\
          Dolphin & 0.225 & 0.034  & 3.357 & 4.99 & Reject \\
          Football & 0.314 & 0.012 & 2.508 & 3.995 & Reject \\
          Karate & 0.117 & 0.068 & 2.408 & 4.541 & Reject \\
          Les Miserables & 0.414 & 0.024 & 2.641 & 4.573 & Reject \\ 
          Macaque Cortex & 0.266 & 0.019 & 2.245 & 3.374 & Reject \\
          Political Blogs & 0.211 & 0.0006 & 2.737 & 4.696 & Reject \\
          Political Books & 0.267 & 0.015 & 3.079 & 4.354 & Reject \\
          Power Grid & 0.103 & 0.0008 & 18.989 & 17.324 & Fail to Reject \\
          Word Adjacencies & 0.089 & 0.015 & 2.536 & 4.636 & Reject \\ 
          \hline
    \end{tabular}
    \caption{Results from the asymptotic small world test for the ER null mode. A network is small world if the observed $C$ is greater than $K_C$ and the observed $L$ is less than $K_L$}
    \label{tab:ER_asymp}
\end{table}

Finally, we present the results from the asymptotic test with ER null model in Table \ref{tab:ER_asymp}. In each case we present the cutoff values for rejection for $K_C$ and $K_L$. We will call a network small world if the observed value of $C$ is greater than the cutoff $K_C$ and the observed value of $L$ is less than the cutoff $K_L$. The cutoff $K_C$ is the theoretical $95th$ quantile of the asymptotic distribution of $C$ derived in  Theorem 2, while the cutoff $K_L$ is the value $K_2$ defined in Theorem 3. 
We note that for all the networks (except for the Power grid network, the observed value of $L$ is less than the small-world cutoff of twice the average expected path length  The observed value of $C$ is universally higher than the cutoff. 

An interesting point we note is that while many networks are not deemed to be  small-world in Table for bootstrap tests because their observed average path lengths are higher than even the high quantiles of the distribution of average path length, they are deemed small-world in Table \ref{tab:ER_asymp}. This is because their observed path lengths are roughly within twice the expected average path length under the ER model, which is effectively the cutoff the asymptotic test uses. This observation is in line with our observation in the simulation showing the distribution of $L$ is highly concentrated around its mean.
\end{document}